 \documentstyle[amsart12,bezier]{amsart}

 \newcount\beziercnt
\makeatletter

\def\diagram{\leftwidth=\z@ \rightwidth=\z@ \topheight=\z@
\botheight=\z@ \setbox\@picbox\hbox\bgroup}

\def\enddiagram{\egroup\wd\@picbox\rightwidth\unitlength
\ht\@picbox\topheight\unitlength \dp\@picbox\botheight\unitlength
\hskip\leftwidth\unitlength\box\@picbox}

\def\bfig{\begin{diagram}}
\def\efig{\end{diagram}}
\newcount\wideness \newcount\leftwidth \newcount\rightwidth
\newcount\highness \newcount\topheight \newcount\botheight

\def\ratchet#1#2{\ifnum#1<#2 \global #1=#2 \fi}

\def\putbox(#1,#2)#3{
\horsize{\wideness}{#3} \divide\wideness by 2
{\advance\wideness by #1 \ratchet{\rightwidth}{\wideness}}
{\advance\wideness by -#1 \ratchet{\leftwidth}{\wideness}}
\vertsize{\highness}{#3} \divide\highness by 2
{\advance\highness by #2 \ratchet{\topheight}{\highness}}
{\advance\highness by -#2 \ratchet{\botheight}{\highness}}
\put(#1,#2){\makebox(0,0){$#3$}}}

\def\putlbox(#1,#2)#3{
\horsize{\wideness}{#3}
{\advance\wideness by #1 \ratchet{\rightwidth}{\wideness}}
{\ratchet{\leftwidth}{-#1}}
\vertsize{\highness}{#3} \divide\highness by 2
{\advance\highness by #2 \ratchet{\topheight}{\highness}}
{\advance\highness by -#2 \ratchet{\botheight}{\highness}}
\put(#1,#2){\makebox(0,0)[l]{$#3$}}}

\def\putrbox(#1,#2)#3{
\horsize{\wideness}{#3}
{\ratchet{\rightwidth}{#1}}
{\advance\wideness by -#1 \ratchet{\leftwidth}{\wideness}}
\vertsize{\highness}{#3} \divide\highness by 2
{\advance\highness by #2 \ratchet{\topheight}{\highness}}
{\advance\highness by -#2 \ratchet{\botheight}{\highness}}
\put(#1,#2){\makebox(0,0)[r]{$#3$}}}

\def\adjust[#1]{}

\newcount \coefa
\newcount \coefb
\newcount \coefc
\newcount\tempcounta
\newcount\tempcountb
\newcount\tempcountc
\newcount\tempcountd
\newcount\xext
\newcount\yext
\newcount\xoff
\newcount\yoff
\newcount\gap
\newcount\arrowtypea
\newcount\arrowtypeb
\newcount\arrowtypec
\newcount\arrowtyped
\newcount\arrowtypee
\newcount\height
\newcount\width
\newcount\xpos
\newcount\ypos
\newcount\run
\newcount\rise
\newcount\arrowlength
\newcount\halflength
\newcount\arrowtype
\newdimen\tempdimen
\newdimen\xlen
\newdimen\ylen
\newsavebox{\tempboxa}
\newsavebox{\tempboxb}
\newsavebox{\tempboxc}

\newdimen\w@dth

\def\setw@dth#1#2{\setbox\z@\hbox{$#1$}\w@dth=\wd\z@
\setbox\@ne\hbox{$#2$}\ifnum\w@dth<\wd\@ne \w@dth=\wd\@ne \fi
\advance\w@dth by 1.2em}

\def\t@^#1_#2{\def\n@one{#1}\def\n@two{#2}\mathrel{\setw@dth{#1}{#2}
\mathop{\hbox to \w@dth{\rightarrowfill}}\limits
\ifx\n@one\empty\else ^{\box\z@}\fi
\ifx\n@two\empty\else _{\box\@ne}\fi}}
\def\t@@^#1{\@ifnextchar_ {\t@^{#1}}{\t@^{#1}_{}}}
\def\to{\@ifnextchar^ {\t@@}{\t@@^{}}}

\def\t@left^#1_#2{\def\n@one{#1}\def\n@two{#2}\mathrel{\setw@dth{#1}{#2}
\mathop{\hbox to \w@dth{\leftarrowfill}}\limits
\ifx\n@one\empty\else ^{\box\z@}\fi
\ifx\n@two\empty\else _{\box\@ne}\fi}}
\def\t@@left^#1{\@ifnextchar_ {\t@left^{#1}}{\t@left^{#1}_{}}}
\def\toleft{\@ifnextchar^ {\t@@left}{\t@@left^{}}}

\def\two@^#1_#2{\def\n@one{#1}\def\n@two{#2}\mathrel{\setw@dth{#1}{#2}
\mathop{\vcenter{\hbox to \w@dth{\rightarrowfill}\kern-1.7ex
                 \hbox to \w@dth{\rightarrowfill}}
       }\limits
\ifx\n@one\empty\else ^{\box\z@}\fi
\ifx\n@two\empty\else _{\box\@ne}\fi}}
\def\tw@@^#1{\@ifnextchar_ {\two@^{#1}}{\two@^{#1}_{}}}
\def\two{\@ifnextchar^ {\tw@@}{\tw@@^{}}}

\def\tofr@^#1_#2{\def\n@one{#1}\def\n@two{#2}\mathrel{\setw@dth{#1}{#2}
\mathop{\vcenter{\hbox to \w@dth{\rightarrowfill}\kern-1.7ex
                 \hbox to \w@dth{\leftarrowfill}}
       }\limits
\ifx\n@one\empty\else ^{\box\z@}\fi
\ifx\n@two\empty\else _{\box\@ne}\fi}}
\def\t@fr@^#1{\@ifnextchar_ {\tofr@^{#1}}{\tofr@^{#1}_{}}}
\def\tofro{\@ifnextchar^ {\t@fr@}{\t@fr@^{}}}

\def\mon{\mathop{\m@th\hbox to
      14.6\P@{\lasyb\char'51\hskip-2.1\P@$\arrext$\hss
$\mathord\rightarrow$}}\limits}
\def\leftmono{\mathrel{\m@th\hbox to
14.6\P@{$\mathord\leftarrow$\hss$\arrext$\hskip-2.1\P@\lasyb\char'50
}}\limits}
\mathchardef\arrext="0200

\def\settypes(#1,#2,#3){\arrowtypea#1 \arrowtypeb#2 \arrowtypec#3}
\def\settoheight#1#2{\setbox\@tempboxa\hbox{#2}#1\ht\@tempboxa\relax}
\def\settodepth#1#2{\setbox\@tempboxa\hbox{#2}#1\dp\@tempboxa\relax}
\def\settokens[#1`#2`#3`#4]{
     \def\tokena{#1}\def\tokenb{#2}\def\tokenc{#3}\def\tokend{#4}}
\def\setsqparms[#1`#2`#3`#4;#5`#6]{
\arrowtypea #1
\arrowtypeb #2
\arrowtypec #3
\arrowtyped #4
\width #5
\height #6
}
\def\setpos(#1,#2){\xpos=#1 \ypos#2}

\def\settriparms[#1`#2`#3;#4]{\settripairparms[#1`#2`#3`1`1;#4]}

\def\settripairparms[#1`#2`#3`#4`#5;#6]{
\arrowtypea #1
\arrowtypeb #2
\arrowtypec #3
\arrowtyped #4
\arrowtypee #5
\width #6
\height #6
}

\def\resetparms{\settripairparms[1`1`1`1`1;500]\width 500}

\resetparms

\def\mvector(#1,#2)#3{
\put(0,0){\vector(#1,#2){#3}}
\put(0,0){\vector(#1,#2){26}}
}
\def\evector(#1,#2)#3{{
\arrowlength #3
\put(0,0){\vector(#1,#2){\arrowlength}}
\advance \arrowlength by-30
\put(0,0){\vector(#1,#2){\arrowlength}}
}}

\def\horsize#1#2{
\settowidth{\tempdimen}{$#2$}
#1=\tempdimen
\divide #1 by\unitlength
}

\def\vertsize#1#2{
\settoheight{\tempdimen}{$#2$}
#1=\tempdimen
\settodepth{\tempdimen}{$#2$}
\advance #1 by\tempdimen
\divide #1 by\unitlength
}

\def\putvector(#1,#2)(#3,#4)#5#6{{
\ifnum3<\arrowtype
\putdashvector(#1,#2)(#3,#4)#5\arrowtype
\else
\ifnum\arrowtype<-3
\putdashvector(#1,#2)(#3,#4)#5\arrowtype
\else
\xpos=#1
\ypos=#2
\run=#3
\rise=#4
\arrowlength=#5
\ifnum \arrowtype<0
    \ifnum \run=0
        \advance \ypos by-\arrowlength
    \else
        \tempcounta \arrowlength
        \multiply \tempcounta by\rise
        \divide \tempcounta by\run
        \ifnum\run>0
            \advance \xpos by\arrowlength
            \advance \ypos by\tempcounta
        \else
            \advance \xpos by-\arrowlength
            \advance \ypos by-\tempcounta
        \fi
    \fi
    \multiply \arrowtype by-1
    \multiply \rise by-1
    \multiply \run by-1
\fi
\ifcase \arrowtype
\or \put(\xpos,\ypos){\vector(\run,\rise){\arrowlength}}
\or \put(\xpos,\ypos){\mvector(\run,\rise)\arrowlength}
\or \put(\xpos,\ypos){\evector(\run,\rise){\arrowlength}}
\fi\fi\fi
}}

\def\putsplitvector(#1,#2)#3#4{
\xpos #1
\ypos #2
\arrowtype #4
\halflength #3
\arrowlength #3
\gap 140
\advance \halflength by-\gap
\divide \halflength by2
\ifnum\arrowtype>0
   \ifcase \arrowtype
   \or \put(\xpos,\ypos){\line(0,-1){\halflength}}
       \advance\ypos by-\halflength
       \advance\ypos by-\gap
       \put(\xpos,\ypos){\vector(0,-1){\halflength}}
   \or \put(\xpos,\ypos){\line(0,-1)\halflength}
       \put(\xpos,\ypos){\vector(0,-1)3}
       \advance\ypos by-\halflength
       \advance\ypos by-\gap
       \put(\xpos,\ypos){\vector(0,-1){\halflength}}
   \or \put(\xpos,\ypos){\line(0,-1)\halflength}
       \advance\ypos by-\halflength
       \advance\ypos by-\gap
       \put(\xpos,\ypos){\evector(0,-1){\halflength}}
   \fi
\else \arrowtype=-\arrowtype
   \ifcase\arrowtype
   \or \advance \ypos by-\arrowlength
       \put(\xpos,\ypos){\line(0,1){\halflength}}
       \advance\ypos by\halflength
       \advance\ypos by\gap
       \put(\xpos,\ypos){\vector(0,1){\halflength}}
   \or \advance \ypos by-\arrowlength
       \put(\xpos,\ypos){\line(0,1)\halflength}
       \put(\xpos,\ypos){\vector(0,1)3}
       \advance\ypos by\halflength
       \advance\ypos by\gap
       \put(\xpos,\ypos){\vector(0,1){\halflength}}
   \or \advance \ypos by-\arrowlength
       \put(\xpos,\ypos){\line(0,1)\halflength}
       \advance\ypos by\halflength
       \advance\ypos by\gap
       \put(\xpos,\ypos){\evector(0,1){\halflength}}
   \fi
\fi
}

\def\putmorphism(#1)(#2,#3)[#4`#5`#6]#7#8#9{{
\run #2
\rise #3
\ifnum\rise=0
  \puthmorphism(#1)[#4`#5`#6]{#7}{#8}#9
\else\ifnum\run=0
  \putvmorphism(#1)[#4`#5`#6]{#7}{#8}#9
\else
\setpos(#1)
\arrowlength #7
\arrowtype #8
\ifnum\run=0
\else\ifnum\rise=0
\else
\ifnum\run>0
    \coefa=1
\else
   \coefa=-1
\fi
\ifnum\arrowtype>0
   \coefb=0
   \coefc=-1
\else
   \coefb=\coefa
   \coefc=1
   \arrowtype=-\arrowtype
\fi
\width=2
\multiply \width by\run
\divide \width by\rise
\ifnum \width<0  \width=-\width\fi
\advance\width by60
\if l#9 \width=-\width\fi
\putbox(\xpos,\ypos){#4}
{\multiply \coefa by\arrowlength
\advance\xpos by\coefa
\multiply \coefa by\rise
\divide \coefa by\run
\advance \ypos by\coefa
\putbox(\xpos,\ypos){#5} }
{\multiply \coefa by\arrowlength
\divide \coefa by2
\advance \xpos by\coefa
\advance \xpos by\width
\multiply \coefa by\rise
\divide \coefa by\run
\advance \ypos by\coefa
\if l#9
   \putrbox(\xpos,\ypos){#6}
\else\if r#9
   \putlbox(\xpos,\ypos){#6}
\fi\fi }
{\multiply \rise by-\coefc
\multiply \run by-\coefc
\multiply \coefb by\arrowlength
\advance \xpos by\coefb
\multiply \coefb by\rise
\divide \coefb by\run
\advance \ypos by\coefb
\multiply \coefc by70
\advance \ypos by\coefc
\multiply \coefc by\run
\divide \coefc by\rise
\advance \xpos by\coefc
\multiply \coefa by140
\multiply \coefa by\run
\divide \coefa by\rise
\advance \arrowlength by\coefa
\ifcase\arrowtype
\or \put(\xpos,\ypos){\vector(\run,\rise){\arrowlength}}
\or \put(\xpos,\ypos){\mvector(\run,\rise){\arrowlength}}
\or \put(\xpos,\ypos){\evector(\run,\rise){\arrowlength}}
\fi}\fi\fi\fi\fi}}

\newcount\numbdashes \newcount\lengthdash \newcount\increment

\def\howmanydashes{
\numbdashes=\arrowlength \lengthdash=40
\divide\numbdashes by \lengthdash
\lengthdash=\arrowlength
\divide\lengthdash by \numbdashes
\increment=\lengthdash
\multiply\lengthdash by 3
\divide\lengthdash by 5
}

\def\putdashvector(#1)(#2,#3)#4#5{
\ifnum#3=0 \putdashhvector(#1){#4}#5
\else
\ifnum#2=0
\putdashvvector(#1){#4}#5\fi\fi}

\def\putdashhvector(#1,#2)#3#4{{
\arrowlength=#3 \howmanydashes
\multiput(#1,#2)(\increment,0){\numbdashes}
{\vrule height .4pt width \lengthdash\unitlength}
\arrowtype=#4 \xpos=#1
\ifnum\arrowtype<0 \advance\arrowtype by 7 \fi
\ifcase\arrowtype
\or \advance\xpos by 10
    \put(\xpos,#2){\vector(-1,0){\lengthdash}}
    \advance\xpos by 40
    \put(\xpos,#2){\vector(-1,0){\lengthdash}}
\or \advance \xpos by 10
    \put(\xpos,#2){\vector(-1,0){\lengthdash}}
    \advance\xpos by  \arrowlength
    \advance\xpos by  -50
    \put(\xpos,#2){\vector(-1,0){\lengthdash}}
\or \advance\xpos by 10
    \put(\xpos,#2){\vector(-1,0){\lengthdash}}
\or \advance\xpos by \arrowlength
    \advance\xpos by -\lengthdash
    \put(\xpos,#2){\vector(1,0){\lengthdash}}
\or {\advance\xpos by 10
    \put(\xpos,#2){\vector(1,0){\lengthdash}}}
    \advance\xpos by \arrowlength
    \advance\xpos by -\lengthdash
    \put(\xpos,#2){\vector(1,0){\lengthdash}}
\or \advance\xpos by \arrowlength
    \advance\xpos by -\lengthdash
    \put(\xpos,#2){\vector(1,0){\lengthdash}}
    \advance\xpos by -40
    \put(\xpos,#2){\vector(1,0){\lengthdash}}
   \fi
}}

\def\putdashvvector(#1,#2)#3#4{{
\arrowlength=#3 \howmanydashes
\ypos=#2 \advance\ypos by -\arrowlength
\multiput(#1,#2)(0,\increment){\numbdashes}
    {\vrule width .4pt height \lengthdash\unitlength}
\arrowtype=#4 \ypos=#2
\ifnum\arrowtype<0 \advance\arrowtype by 7 \fi
\ifcase\arrowtype
\or \advance\ypos by \arrowlength \advance\ypos by -40
    \put(#1,\ypos){\vector(0,1){\lengthdash}}
    \advance\ypos by -40
    \put(#1,\ypos){\vector(0,1){\lengthdash}}
\or \advance\ypos by 10
    \put(#1,\ypos){\vector(0,1){\lengthdash}}
    \advance\ypos by \arrowlength \advance\ypos by -40
    \put(#1,\ypos){\vector(0,1){\lengthdash}}
\or \advance\ypos by \arrowlength \advance\ypos by -40
    \put(#1,\ypos){\vector(0,1){\lengthdash}}
\or \advance\ypos by 10
    \put(#1,\ypos){\vector(0,-1){\lengthdash}}
\or \advance\ypos by 10
    \put(#1,\ypos){\vector(0,-1){\lengthdash}}
    \advance\ypos by \arrowlength \advance\ypos by -40
    \put(#1,\ypos){\vector(0,-1){\lengthdash}}
\or \advance\ypos by 10
    \put(#1,\ypos){\vector(0,-1){\lengthdash}}
    \advance\ypos by 40
    \put(#1,\ypos){\vector(0,-1){\lengthdash}}
\fi
}}

\def\puthmorphism(#1,#2)[#3`#4`#5]#6#7#8{{
\xpos #1
\ypos #2
\width #6
\arrowlength #6
\arrowtype=#7
\putbox(\xpos,\ypos){#3\vphantom{#4}}
{\advance \xpos by\arrowlength
\putbox(\xpos,\ypos){\vphantom{#3}#4}}
\horsize{\tempcounta}{#3}
\horsize{\tempcountb}{#4}
\divide \tempcounta by2
\divide \tempcountb by2
\advance \tempcounta by30
\advance \tempcountb by30
\advance \xpos by\tempcounta
\advance \arrowlength by-\tempcounta
\advance \arrowlength by-\tempcountb
\putvector(\xpos,\ypos)(1,0)\arrowlength\arrowtype
\divide \arrowlength by2
\advance \xpos by\arrowlength
\vertsize{\tempcounta}{#5}
\divide\tempcounta by2
\advance \tempcounta by20
\if a#8
   \advance \ypos by\tempcounta
   \putbox(\xpos,\ypos){#5}
\else
   \advance \ypos by-\tempcounta
   \putbox(\xpos,\ypos){#5}
\fi}}

\def\putvmorphism(#1,#2)[#3`#4`#5]#6#7#8{{
\xpos #1
\ypos #2
\arrowlength #6
\arrowtype #7
\settowidth{\xlen}{$#5$}
\putbox(\xpos,\ypos){#3}
{\advance \ypos by-\arrowlength
\putbox(\xpos,\ypos){#4}}
{\advance\arrowlength by-140
\advance \ypos by-70
\ifdim\xlen>0pt
   \if m#8
      \putsplitvector(\xpos,\ypos)\arrowlength\arrowtype
   \else
   \putvector(\xpos,\ypos)(0,-1)\arrowlength\arrowtype
   \fi
\else
   \putvector(\xpos,\ypos)(0,-1)\arrowlength\arrowtype
\fi}
\ifdim\xlen>0pt
   \divide \arrowlength by2
   \advance\ypos by-\arrowlength
   \if l#8
      \advance \xpos by-40
      \putrbox(\xpos,\ypos){#5}
   \else\if r#8
      \advance \xpos by40
      \putlbox(\xpos,\ypos){#5}
   \else
      \putbox(\xpos,\ypos){#5}
   \fi\fi
\fi
}}

\def\putsquarep<#1>(#2)[#3;#4`#5`#6`#7]{{
\setsqparms[#1]
\setpos(#2)
\settokens[#3]
\puthmorphism(\xpos,\ypos)[\tokenc`\tokend`{#7}]{\width}{\arrowtyped}b
\advance\ypos by \height
\puthmorphism(\xpos,\ypos)[\tokena`\tokenb`{#4}]{\width}{\arrowtypea}a
\putvmorphism(\xpos,\ypos)[``{#5}]{\height}{\arrowtypeb}l
\advance\xpos by \width
\putvmorphism(\xpos,\ypos)[``{#6}]{\height}{\arrowtypec}r
}}

\def\putsquare{\@ifnextchar <{\putsquarep}{\putsquarep
   <\arrowtypea`\arrowtypeb`\arrowtypec`\arrowtyped;\width`\height>}}
\def\square{\@ifnextchar< {\squarep}{\squarep
   <\arrowtypea`\arrowtypeb`\arrowtypec`\arrowtyped;\width`\height>}}

\def\squarep<#1>[#2`#3`#4`#5;#6`#7`#8`#9]{{
\setsqparms[#1]
\diagram
\putsquarep<\arrowtypea`\arrowtypeb`\arrowtypec`
\arrowtyped;\width`\height>
(0,0)[#2`#3`#4`{#5};#6`#7`#8`{#9}]
\enddiagram
}}

\def\putptrianglep<#1>(#2,#3)[#4`#5`#6;#7`#8`#9]{{
\settriparms[#1]
\xpos=#2 \ypos=#3
\advance\ypos by \height
\puthmorphism(\xpos,\ypos)[#4`#5`{#7}]{\height}{\arrowtypea}a
\putvmorphism(\xpos,\ypos)[`#6`{#8}]{\height}{\arrowtypeb}l
\advance\xpos by\height
\putmorphism(\xpos,\ypos)(-1,-1)[``{#9}]{\height}{\arrowtypec}r
}}

\def\putptriangle{\@ifnextchar <{\putptrianglep}{\putptrianglep
   <\arrowtypea`\arrowtypeb`\arrowtypec;\height>}}
\def\ptriangle{\@ifnextchar <{\ptrianglep}{\ptrianglep
   <\arrowtypea`\arrowtypeb`\arrowtypec;\height>}}

\def\ptrianglep<#1>[#2`#3`#4;#5`#6`#7]{{
\settriparms[#1]
\diagram
\putptrianglep<\arrowtypea`\arrowtypeb`
\arrowtypec;\height>
(0,0)[#2`#3`#4;#5`#6`{#7}]
\enddiagram
}}

\def\putqtrianglep<#1>(#2,#3)[#4`#5`#6;#7`#8`#9]{{
\settriparms[#1]
\xpos=#2 \ypos=#3
\advance\ypos by\height
\puthmorphism(\xpos,\ypos)[#4`#5`{#7}]{\height}{\arrowtypea}a
\putmorphism(\xpos,\ypos)(1,-1)[``{#8}]{\height}{\arrowtypeb}l
\advance\xpos by\height
\putvmorphism(\xpos,\ypos)[`#6`{#9}]{\height}{\arrowtypec}r
}}

\def\putqtriangle{\@ifnextchar <{\putqtrianglep}{\putqtrianglep
   <\arrowtypea`\arrowtypeb`\arrowtypec;\height>}}
\def\qtriangle{\@ifnextchar <{\qtrianglep}{\qtrianglep
   <\arrowtypea`\arrowtypeb`\arrowtypec;\height>}}

\def\qtrianglep<#1>[#2`#3`#4;#5`#6`#7]{{
\settriparms[#1]
\width=\height
\diagram
\putqtrianglep<\arrowtypea`\arrowtypeb`
\arrowtypec;\height>
(0,0)[#2`#3`#4;#5`#6`{#7}]
\enddiagram
}}

\def\putdtrianglep<#1>(#2,#3)[#4`#5`#6;#7`#8`#9]{{
\settriparms[#1]
\xpos=#2 \ypos=#3
\puthmorphism(\xpos,\ypos)[#5`#6`{#9}]{\height}{\arrowtypec}b
\advance\xpos by \height \advance\ypos by\height
\putmorphism(\xpos,\ypos)(-1,-1)[``{#7}]{\height}{\arrowtypea}l
\putvmorphism(\xpos,\ypos)[#4``{#8}]{\height}{\arrowtypeb}r
}}

\def\putdtriangle{\@ifnextchar <{\putdtrianglep}{\putdtrianglep
   <\arrowtypea`\arrowtypeb`\arrowtypec;\height>}}
\def\dtriangle{\@ifnextchar <{\dtrianglep}{\dtrianglep
   <\arrowtypea`\arrowtypeb`\arrowtypec;\height>}}

\def\dtrianglep<#1>[#2`#3`#4;#5`#6`#7]{{
\settriparms[#1]
\width=\height
\diagram
\putdtrianglep<\arrowtypea`\arrowtypeb`
\arrowtypec;\height>
(0,0)[#2`#3`#4;#5`#6`{#7}]
\enddiagram
}}

\def\putbtrianglep<#1>(#2,#3)[#4`#5`#6;#7`#8`#9]{{
\settriparms[#1]
\xpos=#2 \ypos=#3
\puthmorphism(\xpos,\ypos)[#5`#6`{#9}]{\height}{\arrowtypec}b
\advance\ypos by\height
\putmorphism(\xpos,\ypos)(1,-1)[``{#8}]{\height}{\arrowtypeb}r
\putvmorphism(\xpos,\ypos)[#4``{#7}]{\height}{\arrowtypea}l
}}

\def\putbtriangle{\@ifnextchar <{\putbtrianglep}{\putbtrianglep
   <\arrowtypea`\arrowtypeb`\arrowtypec;\height>}}
\def\btriangle{\@ifnextchar <{\btrianglep}{\btrianglep
   <\arrowtypea`\arrowtypeb`\arrowtypec;\height>}}

\def\btrianglep<#1>[#2`#3`#4;#5`#6`#7]{{
\settriparms[#1]
\width=\height
\diagram
\putbtrianglep<\arrowtypea`\arrowtypeb`
\arrowtypec;\height>
(0,0)[#2`#3`#4;#5`#6`{#7}]
\enddiagram
}}

\def\putAtrianglep<#1>(#2,#3)[#4`#5`#6;#7`#8`#9]{{
\settriparms[#1]
\xpos=#2 \ypos=#3
{\multiply \height by2
\puthmorphism(\xpos,\ypos)[#5`#6`{#9}]{\height}{\arrowtypec}b}
\advance\xpos by\height \advance\ypos by\height
\putmorphism(\xpos,\ypos)(-1,-1)[#4``{#7}]{\height}{\arrowtypea}l
\putmorphism(\xpos,\ypos)(1,-1)[``{#8}]{\height}{\arrowtypeb}r
}}

\def\putAtriangle{\@ifnextchar <{\putAtrianglep}{\putAtrianglep
   <\arrowtypea`\arrowtypeb`\arrowtypec;\height>}}
\def\Atriangle{\@ifnextchar <{\Atrianglep}{\Atrianglep
   <\arrowtypea`\arrowtypeb`\arrowtypec;\height>}}

\def\Atrianglep<#1>[#2`#3`#4;#5`#6`#7]{{
\settriparms[#1]
\width=\height
\diagram
\putAtrianglep<\arrowtypea`\arrowtypeb`
\arrowtypec;\height>
(0,0)[#2`#3`#4;#5`#6`{#7}]
\enddiagram
}}

\def\putAtrianglepairp<#1>(#2)[#3;#4`#5`#6`#7`#8]{{
\settripairparms[#1]
\setpos(#2)
\settokens[#3]
\puthmorphism(\xpos,\ypos)[\tokenb`\tokenc`{#7}]{\height}{\arrowtyped}b
\advance\xpos by\height
\puthmorphism(\xpos,\ypos)[\phantom{\tokenc}`\tokend`{#8}]
{\height}{\arrowtypee}b
\advance\ypos by\height
\putmorphism(\xpos,\ypos)(-1,-1)[\tokena``{#4}]{\height}{\arrowtypea}l
\putvmorphism(\xpos,\ypos)[``{#5}]{\height}{\arrowtypeb}m
\putmorphism(\xpos,\ypos)(1,-1)[``{#6}]{\height}{\arrowtypec}r
}}

\def\putAtrianglepair{\@ifnextchar
   <{\putAtrianglepairp}{\putAtrianglepairp
   <\arrowtypea`\arrowtypeb`\arrowtypec`\arrowtyped`\arrowtypee;
   \height>}}
\def\Atrianglepair{\@ifnextchar
   <{\Atrianglepairp}{\Atrianglepairp
   <\arrowtypea`\arrowtypeb`\arrowtypec`\arrowtyped`\arrowtypee;
   \height>}}

\def\Atrianglepairp<#1>[#2;#3`#4`#5`#6`#7]{{
\settripairparms[#1]
\settokens[#2]
\width=\height
\diagram
\putAtrianglepairp
<\arrowtypea`\arrowtypeb`\arrowtypec`
\arrowtyped`\arrowtypee;\height>
(0,0)[{#2};#3`#4`#5`#6`{#7}]
\enddiagram
}}

\def\putVtrianglep<#1>(#2,#3)[#4`#5`#6;#7`#8`#9]{{
\settriparms[#1]
\xpos=#2 \ypos=#3
\advance\ypos by\height
{\multiply\height by2
\puthmorphism(\xpos,\ypos)[#4`#5`{#7}]{\height}{\arrowtypea}a}
\putmorphism(\xpos,\ypos)(1,-1)[`#6`{#8}]{\height}{\arrowtypeb}l
\advance\xpos by\height
\advance\xpos by\height
\putmorphism(\xpos,\ypos)(-1,-1)[``{#9}]{\height}{\arrowtypec}r
}}

\def\putVtriangle{\@ifnextchar <{\putVtrianglep}{\putVtrianglep
   <\arrowtypea`\arrowtypeb`\arrowtypec;\height>}}
\def\Vtriangle{\@ifnextchar <{\Vtrianglep}{\Vtrianglep
   <\arrowtypea`\arrowtypeb`\arrowtypec;\height>}}

\def\Vtrianglep<#1>[#2`#3`#4;#5`#6`#7]{{
\settriparms[#1]
\width=\height
\diagram
\putVtrianglep<\arrowtypea`\arrowtypeb`
\arrowtypec;\height>
(0,0)[#2`#3`#4;#5`#6`{#7}]
\enddiagram
}}

\def\putVtrianglepairp<#1>(#2)[#3;#4`#5`#6`#7`#8]{{
\settripairparms[#1]
\setpos(#2)
\settokens[#3]
\advance\ypos by\height
\putmorphism(\xpos,\ypos)(1,-1)[`\tokend`{#6}]{\height}{\arrowtypec}l
\puthmorphism(\xpos,\ypos)[\tokena`\tokenb`{#4}]{\height}{\arrowtypea}a
\advance\xpos by\height
\puthmorphism(\xpos,\ypos)[\phantom{\tokenb}`\tokenc`{#5}]
{\height}{\arrowtypeb}a
\putvmorphism(\xpos,\ypos)[``{#7}]{\height}{\arrowtyped}m
\advance\xpos by\height
\putmorphism(\xpos,\ypos)(-1,-1)[``{#8}]{\height}{\arrowtypee}r
}}

\def\putVtrianglepair{\@ifnextchar
    <{\putVtrianglepairp}{\putVtrianglepairp
    <\arrowtypea`\arrowtypeb`\arrowtypec`\arrowtyped`\arrowtypee;
    \height>}}
\def\Vtrianglepair{\@ifnextchar
    <{\Vtrianglepairp}{\Vtrianglepairp
    <\arrowtypea`\arrowtypeb`\arrowtypec`\arrowtyped`\arrowtypee;
    \height>}}

\def\Vtrianglepairp<#1>[#2;#3`#4`#5`#6`#7]{{
\settripairparms[#1]
\settokens[#2]
\diagram
\putVtrianglepairp
<\arrowtypea`\arrowtypeb`\arrowtypec`
\arrowtyped`\arrowtypee;\height>
(0,0)[{#2};#3`#4`#5`#6`{#7}]
\enddiagram
}}

\def\putCtrianglep<#1>(#2,#3)[#4`#5`#6;#7`#8`#9]{{
\settriparms[#1]
\xpos=#2 \ypos=#3
\advance\ypos by\height
\putmorphism(\xpos,\ypos)(1,-1)[``{#9}]{\height}{\arrowtypec}l
\advance\xpos by\height
\advance\ypos by\height
\putmorphism(\xpos,\ypos)(-1,-1)[#4`#5`{#7}]{\height}{\arrowtypea}l
{\multiply\height by 2
\putvmorphism(\xpos,\ypos)[`#6`{#8}]{\height}{\arrowtypeb}r}
}}

\def\putCtriangle{\@ifnextchar <{\putCtrianglep}{\putCtrianglep
    <\arrowtypea`\arrowtypeb`\arrowtypec;\height>}}
\def\Ctriangle{\@ifnextchar <{\Ctrianglep}{\Ctrianglep
    <\arrowtypea`\arrowtypeb`\arrowtypec;\height>}}

\def\Ctrianglep<#1>[#2`#3`#4;#5`#6`#7]{{
\settriparms[#1]
\width=\height
\diagram
\putCtrianglep<\arrowtypea`\arrowtypeb`
\arrowtypec;\height>
(0,0)[#2`#3`#4;#5`#6`{#7}]
\enddiagram
}}

\def\putDtrianglep<#1>(#2,#3)[#4`#5`#6;#7`#8`#9]{{
\settriparms[#1]
\xpos=#2 \ypos=#3
\advance\xpos by\height \advance\ypos by\height
\putmorphism(\xpos,\ypos)(-1,-1)[``{#9}]{\height}{\arrowtypec}r
\advance\xpos by-\height \advance\ypos by\height
\putmorphism(\xpos,\ypos)(1,-1)[`#5`{#8}]{\height}{\arrowtypeb}r
{\multiply\height by 2
\putvmorphism(\xpos,\ypos)[#4`#6`{#7}]{\height}{\arrowtypea}l}
}}

\def\putDtriangle{\@ifnextchar <{\putDtrianglep}{\putDtrianglep
    <\arrowtypea`\arrowtypeb`\arrowtypec;\height>}}
\def\Dtriangle{\@ifnextchar <{\Dtrianglep}{\Dtrianglep
   <\arrowtypea`\arrowtypeb`\arrowtypec;\height>}}

\def\Dtrianglep<#1>[#2`#3`#4;#5`#6`#7]{{
\settriparms[#1]
\width=\height
\diagram
\putDtrianglep<\arrowtypea`\arrowtypeb`
\arrowtypec;\height>
(0,0)[#2`#3`#4;#5`#6`{#7}]
\enddiagram
}}

\def\setrecparms[#1`#2]{\width=#1 \height=#2}

\def\recursep<#1`#2>[#3;#4`#5`#6`#7`#8]{{
\width=#1 \height=#2
\settokens[#3]
\settowidth{\tempdimen}{$\tokena$}
\ifdim\tempdimen=0pt
  \savebox{\tempboxa}{\hbox{$\tokenb$}}
  \savebox{\tempboxb}{\hbox{$\tokend$}}
  \savebox{\tempboxc}{\hbox{$#6$}}
\else
  \savebox{\tempboxa}{\hbox{$\hbox{$\tokena$}\times\hbox{$\tokenb$}$}}
  \savebox{\tempboxb}{\hbox{$\hbox{$\tokena$}\times\hbox{$\tokend$}$}}
  \savebox{\tempboxc}{\hbox{$\hbox{$\tokena$}\times\hbox{$#6$}$}}
\fi
\ypos=\height
\divide\ypos by 2
\xpos=\ypos
\advance\xpos by \width
\bfig
\putCtrianglep<-1`1`1;\ypos>(0,0)[`\tokenc`;#5`#6`{#7}]
\puthmorphism(\ypos,0)[\tokend`\usebox{\tempboxb}`{#8}]{\width}{-1}b
\puthmorphism(\ypos,\height)[\tokenb`\usebox{\tempboxa}`{#4}]{\width}{-1}a
\advance\ypos by \width
\putvmorphism(\ypos,\height)[``\usebox{\tempboxc}]{\height}1r
\efig
}}

\def\recurse{\@ifnextchar <{\recursep}{\recursep<\width`\height>}}

\def\puttwohmorphisms(#1,#2)[#3`#4;#5`#6]#7#8#9{{
\puthmorphism(#1,#2)[#3`#4`]{#7}0a
\ypos=#2
\advance\ypos by 20
\puthmorphism(#1,\ypos)[\phantom{#3}`\phantom{#4}`#5]{#7}{#8}a
\advance\ypos by -40
\puthmorphism(#1,\ypos)[\phantom{#3}`\phantom{#4}`#6]{#7}{#9}b
}}

\def\puttwovmorphisms(#1,#2)[#3`#4;#5`#6]#7#8#9{{
\putvmorphism(#1,#2)[#3`#4`]{#7}0a
\xpos=#1
\advance\xpos by -20
\putvmorphism(\xpos,#2)[\phantom{#3}`\phantom{#4}`#5]{#7}{#8}l
\advance\xpos by 40
\putvmorphism(\xpos,#2)[\phantom{#3}`\phantom{#4}`#6]{#7}{#9}r
}}

\def\puthcoequalizer(#1)[#2`#3`#4;#5`#6`#7]#8#9{{
\setpos(#1)
\puttwohmorphisms(\xpos,\ypos)[#2`#3;#5`#6]{#8}11
\advance\xpos by #8
\puthmorphism(\xpos,\ypos)[\phantom{#3}`#4`#7]{#8}1{#9}
}}

\def\putvcoequalizer(#1)[#2`#3`#4;#5`#6`#7]#8#9{{
\setpos(#1)
\puttwovmorphisms(\xpos,\ypos)[#2`#3;#5`#6]{#8}11
\advance\ypos by -#8
\putvmorphism(\xpos,\ypos)[\phantom{#3}`#4`#7]{#8}1{#9}
}}

\def\putthreehmorphisms(#1)[#2`#3;#4`#5`#6]#7(#8)#9{{
\setpos(#1) \settypes(#8)
\if a#9
     \vertsize{\tempcounta}{#5}
     \vertsize{\tempcountb}{#6}
     \ifnum \tempcounta<\tempcountb \tempcounta=\tempcountb \fi
\else
     \vertsize{\tempcounta}{#4}
     \vertsize{\tempcountb}{#5}
     \ifnum \tempcounta<\tempcountb \tempcounta=\tempcountb \fi
\fi
\advance \tempcounta by 60
\puthmorphism(\xpos,\ypos)[#2`#3`#5]{#7}{\arrowtypeb}{#9}
\advance\ypos by \tempcounta
\puthmorphism(\xpos,\ypos)[\phantom{#2}`\phantom{#3}`#4]{#7}
        {\arrowtypea}{#9}
\advance\ypos by -\tempcounta \advance\ypos by -\tempcounta
\puthmorphism(\xpos,\ypos)[\phantom{#2}`\phantom{#3}`#6]{#7}
        {\arrowtypec}{#9}}}

\def\setarrowtoks[#1`#2`#3`#4`#5`#6]{
\def\toka{#1}
\def\tokb{#2}
\def\tokc{#3}
\def\tokd{#4}
\def\toke{#5}
\def\tokf{#6}
}
\def\hex{\@ifnextchar <{\hexp}{\hexp<1000`400>}}
\def\hexp<#1`#2>[#3`#4`#5`#6`#7`#8;#9]{
\setarrowtoks[#9]
\yext=#2 \advance \yext by #2
\xext=#1 \advance\xext by \yext
\bfig
\putCtriangle<-1`0`1;#2>(0,0)[`#5`;\tokb``\tokd]
\xext=#1 \yext=#2 \advance \yext by #2
\putsquare<1`0`0`1;\xext`\yext>(#2,0)[#3`#4`#7`#8;\toka```\tokf]
\advance \xext by #2
\putDtriangle<0`1`-1;#2>(\xext,0)[`#6`;`\tokc`\toke]
\efig
}

 \newcount\grcalca
 \newcount\grcalcb
 \newcount\grcalcc
 \newcount\grcalcd
 \newcount\mordiam
 \newcommand\thlines{\thinlines}

 \newcommand{\comult}{
   \thlines
   \put(5,0){\oval(10,10)[t]}
   \put(0,1) {\line(1,0){10}}  }

 \newcommand{\bcomult}[1]{
   \thlines
   \grcalca = #1
   \divide \grcalca by 2
   \put(0,1) {\line(1,0){#1}}
   \put(\grcalca,0){\oval(#1,10)[t]} }

 \newcommand{\mult}{
   \thlines
   \put(0,4) {\line(1,0){10}}
   \put(5,5){\oval(10,10)[b]} }

 \newcommand{\bmult}[1]{
   \thlines
   \grcalca = #1
   \divide \grcalca by 2
   \put(0,4) {\line(1,0){#1}}
   \put(\grcalca,5){\oval(#1,10)[b]} }

 \newcommand{\eval}{
   \thlines
   \put(5,5){\oval(10,10)[b]} }

 \newcommand{\beval}[1]{
   \thlines
   \grcalca = #1
   \divide \grcalca by 2
   \put(\grcalca,5){\oval(#1,10)[b]} }

 \newcommand{\coeval}{
   \thlines
   \put(5,0){\oval(10,10)[t]} }

 \newcommand{\bcoeval}[1]{
   \thlines
   \grcalca = #1
   \divide \grcalca by 2
   \put(\grcalca,0){\oval(#1,10)[t]} }

 \newcommand{\braid}{
   \thlines
   \bezier{\beziercnt}(10,10)(10,7.5)(7,6)
   \bezier{\beziercnt}(0,0)(0,2.5)(3,4)
   \bezier{\beziercnt}(0,10)(0,7.5)(5,5)
   \bezier{\beziercnt}(10,0)(10,2.5)(5,5) }

 \newcommand{\ibraid}{
   \thlines
   \bezier{\beziercnt}(10,10)(10,7.5)(5,5)
   \bezier{\beziercnt}(0,0)(0,2.5)(5,5)
   \bezier{\beziercnt}(0,10)(0,7.5)(3,6)
   \bezier{\beziercnt}(10,0)(10,2.5)(7,4) }

 \newcommand{\dbloverbraid}{
   \thlines
   \bezier{\beziercnt}(20,10)(20,7.5)(17,6)
   \bezier{\beziercnt}(0,0)(0,2.5)(3,4)
   \bezier{\beziercnt}(0,10)(0,7.5)(5,5)
   \bezier{\beziercnt}(10,0)(10,2.5)(5,5)
   \bezier{\beziercnt}(10,10)(10,7.5)(15,5)
   \bezier{\beziercnt}(20,0)(20,2.5)(15,5) }

 \newcommand{\overdblbraid}{
   \thlines
   \bezier{\beziercnt}(10,10)(10,8)(7,7)
   \bezier{\beziercnt}(0,0)(0,3)(3,5)
   \bezier{\beziercnt}(20,10)(20,7)(17,5)
   \bezier{\beziercnt}(10,0)(10,2)(13,3)
   \bezier{\beziercnt}(0,10)(0,8)(10,5)
   \bezier{\beziercnt}(20,0)(20,2)(10,5) }

 \newcommand{\brmor}{
   \thlines
   \bezier{\beziercnt}(10,10)(10,7.5)(7,6)
   \bezier{\beziercnt}(0,0)(0,2.5)(3,4)
   \bezier{\beziercnt}(0,10)(0,7.5)(5,5)
   \bezier{\beziercnt}(10,0)(10,2.5)(5,5)
   \put(5,5) {\circle{5}}}

 \newcommand{\mor}[2]{
   \thlines
   \grcalca = #2
   \advance \grcalca by -\mordiam
   \divide \grcalca by 2
   \put(0,#2) {\line(0,-1){\grcalca}}
   \put(0,0) {\line(0,1){\grcalca}}
   \grcalca = #2
   \divide \grcalca by 2
   \put(0,\grcalca) {\circle{\mordiam}}
   \put(0,\grcalca) {\makebox(0,0){$#1$}} }

  \newcommand{\multmor}[2]{
    \thlines
    \grcalca = #2
    \advance \grcalca by 5
    \put(-2.5,0){\framebox(\grcalca,10){$#1$}}}

 \newcommand{\twist}[2]{
   \thlines
   \grcalca = #1
   \divide \grcalca by 2
   \grcalcb = #2
   \divide \grcalcb by 2
   \grcalcc = #2
   \divide \grcalcc by 3
   \bezier{\beziercnt}(0,0)(0,\grcalcc)(\grcalca,\grcalcb)
   \grcalcd = #2
   \advance \grcalcd by -\grcalcc
   \multiply \grcalcc by 3
   \bezier{\beziercnt}(\grcalca,\grcalcb)(#1,\grcalcd)(#1,#2) }

 \newcommand{\idgr}[1]{
   \thlines
   \put(0,0) {\line(0,1){#1}}}

 \newcommand{\objo}[1]{
   \put(0,1){\makebox(0,0)[b]{$#1$}}}

 \newcommand{\obju}[1]{
   \put(0,-10){\makebox(0,0)[b]{$#1$}}}

 \newcommand{\bgr}[2]{\vskip2ex
   \begin{center}
   \unitlength=0.22ex
   \grcalca = #2
   \advance \grcalca by 7
   \begin{picture}(#1,\grcalca)}

 \newcommand{\egr}{
   \end{picture}{\vskip2.5ex}\end{center} }

 \newcommand{\bsgr}[1]{
   \newsavebox{#1}
   \savebox{#1}(0,0)[bl]}

 \font\Fraktur=eufm10 scaled\magstep1
 \newcommand{\fraktur}[1]{\mbox{\Fraktur #1}}
 \font\Fraktu=eufm7 scaled\magstep1
 \newcommand{\fraktu}[1]{\mbox{\Fraktu #1}}
 \font\Frakt=eufm5 scaled\magstep1
 \newcommand{\frakt}[1]{\mbox{\Frakt #1}}
 \def\fr#1{\mathchoice{\fraktur {#1}}
                        {\fraktur {#1}}
                        {\fraktu {#1}}
                        {\frakt {#1}}  }

 \newcommand{\J}{\fr A}

 \newtheorem{thm}{Theorem}[section]
 \newtheorem{cor}[thm]{Corollary}
 \newtheorem{lma}[thm]{Lemma}
 \newtheorem{prop}[thm]{Proposition}

 \theoremstyle{definition}
 \newtheorem{defn}[thm]{Definition}

 \theoremstyle{remark}

 \def\lim{\varinjlim}
 \newcommand\Hom{{\mathop{\mathrm {Hom}}}}
 
 \newcommand\coend{{\mathop{\mathrm {coend}}}}
 \newcommand\coalg{{\mathop{\mathrm {\hbox{-}coalg}}}}
 \newcommand\rend{{\mathop{\mathrm {end}}}}
 \newcommand\vek{{\mathop{\mathrm {vec}}}}
 \newcommand\Nat{{\mathop{\mathrm {Nat}}}}
 \newcommand\Set{{\mathop{\mathrm {Set}}}}
 
 \newcommand\ad{{\mathop{\mathrm {ad}}}}
 
 \newcommand\Id{{\mathop{\mathrm {Id}}}}
 \newcommand\Ob{{\mathop{\mathrm {Ob}}}}
 \newcommand\Mod{{\mathop{\mathrm {\hbox{-}Mod}}}}
 \newcommand\Comod{{\mathop{\mathrm {\hbox{-}Comod}}}}
 \newcommand\modp{{\mathop{\mathrm {\hbox{-}mod}}}}
 \newcommand\Vek{{\mathop{\mathrm {Vec}}}}
 \newcommand\id{{\mathop{\mathrm {id}}}}
 \newcommand\ev{{\mathop{\mathrm {ev}}}}
 \newcommand\db{{\mathop{\mathrm {db}}}}
 \newcommand\tensor{\otimes}
 \newcommand\tensorhat{\ \widehat\otimes\ }
 \newcommand\wtau{\widetilde \tau}
 \newcommand\iso{\cong}
 \newcommand\A{{\cal A}}
 \newcommand\B{{\cal B}}
 \newcommand\C{{\cal C}}
 \newcommand\D{{\cal D}}
 \newcommand\F{{\cal F}}
 
 \newcommand\M{{\cal M}}
 \newcommand\kl{{\Bbb K}}
 \newcommand\kk{{\scriptstyle \Bbb K}}
 \newcommand\x{{\mbox{-}}}
 \newcommand\X{{\mbox{--}}}
 \newcommand\sst{\scriptstyle}
 \newsymbol\boxtimes 1202

 \def\=>{\longrightarrow}

\begin{document}

 \bsgr{\rightcoadj}{
 \unitlength=0.22ex
 \put(5,35){\bcomult{15}}
 \put(5,25){\mor{z}{10}}
 \put(5,15){\braid}
 \put(5,0){\idgr{15}}
 \put(15,5){\mor{S}{10}}
 \put(15,25){\comult}
 \put(19.5,30){\idgr{5}}
 \put(25,5){\idgr{10}}
 \put(25,15){\mor{z}{10}}
 \put(15,0){\mult}  }

 \bsgr{\halfadj}{
 \unitlength=0.22ex
 \put(0,0){\idgr{5}}
 \put(0,5){\braid}
 \put(0,15){\mor{z}{10}}
 \put(0,25){\bcomult{15}}
 \put(10,0){\mult}
 \put(20,5){\idgr{10}}
 \put(10,15){\multmor{\ad}{10}}  }

 \title[Reconstruction of hidden symmetries] {Reconstruction of hidden
symmetries}

 \author{Bodo Pareigis}

 \address{Mathematisches Institut der Universit\"at M\"unchen\\
Germany}

 \email{pareigis@rz.mathematik.uni-muenchen.de}


 \date{December 1, 1994}

 \maketitle

 \tableofcontents

 \section{Introduction}

 Groups $G$ are often obtained as groups of symmetries (or
automorphisms) of mathematical structures like a vector space (over a
fixed field $\kl$) or two vector spaces together with a linear map
between them or a whole diagram of vector spaces, where a symmetry of
such a diagram is a family of automorphisms one for each vector space
which are compatible with the linear maps of the diagram (a natural
automorphism). This process of constructing the group of symmetries is
a special case of the notion of (Tannaka-Krein) reconstruction.

 Conversely given a group $G$ one considers its representations $G \=>
GL(V)$ in vector spaces $V$ over the field $\kl$. All representations
of $G$ form the category ${}_{{{\kk}}G}\cal M$ of modules, which we
may consider as a huge diagram of vector spaces. The category
${}_{{{\kk}}G}\cal M$ has an additional interesting structure -- the
tensor product $V \tensor W$ of two representations is again a
representation in a canonical way, ${}_{{{\kk}}G}\cal M$ is a monoidal
category. A special consequence of reconstruction theory is the fact
that $G$ may be recovered as the full group of those symmetries of
this huge diagram which are compatible with the tensor product. This
process seems to be the inverse of the first one. In a more general
setting there are, however, subtle deviations. One may reconstruct
much larger groups of symmetries than what one started out with.

 More generally we know that algebras $A$, Lie algebras $\frak g$ and
Hopf algebras $H$ can be reconstructed from their categories of
modules. For the reconstruction of an algebra $A$ one actually needs
not only the category of $A$-modules $A\Mod$ but also the underlying
functor $\omega: A\Mod \=> \Vek$. Then $A$ (as an algebra) can be
reconstructed (up to isomorphism) as $\rend(\omega)$, the end of the
underlying functor. For the reconstruction of a Hopf algebra $H$ one
additionally needs the monoidal structure of $H\Mod$. Then the full
Hopf algebra structure can be reconstructed \cite{DEL, PA3, UL1}.

 This stands in a remarkable contrast to another similar result, the
Morita theorems \cite{BA}, which show that the knowledge of the
category of modules $A\Mod$ of an algebra $A$ does not determine $A$
up to isomorphism.

 As we remarked before the forgetful functor $\omega: A\Mod \=> \Vek$
is essential in the process of reconstruction. In particular one has
to consider representations of the given objects (algebras, groups,
Lie algebras, Hopf algebras) in vector spaces. Representations in
categories of objects with a richer structure like super vector
spaces, $\star$-spaces, graded vector spaces, comodules over Hopf
algebras have a different behavior. Instead of the base category
$\Vek$ we wish to use the category $L\Mod$ of modules over a given
quasitriangular Hopf algebra $L$ (or dually $L\Comod$ the category of
comodules over a coquasitriangular Hopf algebra). We answer the
following question: given a Hopf algebra $H$ in $L\Mod$, can it be
reconstructed from $H\x(L\Mod)$, the underlying functor $\omega:
H\x(L\Mod) \=> L\Mod$ and the monoidal structure? A special case is
the reconstruction of a super algebra
 from its super representations.

 The surprising answer shows that one usually reconstructs a much
bigger object
 from $\omega: H\x(L\Mod) \=> L\Mod$ in $L\Mod$. In the group case
this amounts to additional symmetries which we call hidden symmetries,
in the (Hopf) algebra case the situation is even more complex but we
also talk about hidden symmetries. In certain cases we describe
precisely the additional hidden symmetries by a smash product
decomposition of the reconstructed object.

 We control the process of reconstruction by a control category $\C$
which operates on $\omega: H\x(L\Mod) \=> L\Mod$. With different
choices of the control category $\C$ we obtain different reconstructed
objects and study their properties.

 The second section of this paper is devoted to some basic
notions from the theory of braided monoidal categories $\C$ and the
notion of $\C$-categories. The most interesting examples for $\C$ are
the categories of modules resp. comodules over Hopf algebras with an
additional structure known as a quasitriangular structure resp.
braiding, one example being the category of super vector spaces.

 In the third section we study the general algebraic structure of
reconstructed objects in a braided monoidal category. We have decided
to base our investigations on coalgebras and (right) comodules instead
of algebras and (left) modules, because the fundamental structure
theorem for comodules makes certain constructions in this case much
easier. So we study the coend of a functor $\omega: \B \=> \A$ as the
universal natural transformation $\omega \=> \omega \tensor U$ and
show that such a universal $U \in \A$ carries the structure of a
coalgebra or even a Hopf algebra depending on the properties of
$\omega$. Our techniques allow us to restrict the class of natural
transformations (by the notion of $\C$-morphisms). This process gives
us a family of different universal transformations $\omega \=> \omega
\tensor U_\C$ parametrized by the choice of control category $\C$. It
turns out that some of the structure is connected with coadjoint
coactions, cosmash products and transmutation.

 In the fourth section we show under which conditions a coalgebra $C$
in $\A$ can be reconstructed from the category of $C$-comodules $\A^C$
in $\A$ and the functor $\omega: \A^C \=> \A$. Furthermore we show
that certain universal objects $U_\C$ exist in the case of functors
into the subcategory of rigid (finite-dimensional) objects in $\A$.
Our construction of the objects $U_\C$ is an extension of a known
construction in the case of an unparametrized object $U$.

 Section five presents the main result of this paper: The universal
object $U_\C$ for a functor $\omega: \B \=> \A$ tends to decompose
into a cosmash product of a Hopf algebra with a coalgebra. In
particular we show the following. If $H$ is a braided Hopf algebra
over a field $\kl$, $C$ is an $H$-comodule coalgebra, and $\A =
\Vek^H$ is the braided monoidal category of $H$-comodules, then the
coend of the functor $\omega: \A^C \=> \A$ is the cosmash product $H
\# C$.

 In the Appendix we study certain connections between $\kl$-additive
categories and our notion of $\C$-categories and show in particular
(Theorem \ref{nattransisC}) why there are no hidden symmetries in the
case of representations in ordinary vector spaces.

 We close with an example from representation theory of groups
illuminating our point of view and which will get additional comments
in \ref{symmgroups}.
 We consider representations of a group $G$ in vector spaces over a
field $\kl$, i.e. the category $\M_{\kl G}$. Each element $g \in G$
induces a monoidal automorphism $\varphi_g: \omega \=> \omega$,
$\varphi_g(p) := pg$ where $\omega: \hbox{Mod}\x\kl G \=> \Vek$ is the
forgetful functor. Conversely given any monoidal automorphism
$\varphi: \omega \=> \omega$ there is precisely one $g \in G$ with
$\varphi = \varphi_g$. Thus $G$ can be reconstructed from its
representations.

 We now consider representations of $G$ in super vector spaces over
$\kl$, i.e. the category $\A$ of two-graded vector spaces. They define
a category $\A_{\kl G}$ and a forgetful functor $\omega: \A_{\kl G}
\=> \A$. We may view $\kl G$ as a super Hopf algebra $(\kl G,0)$ in
$\A$ and have $(p_0,p_1)g = (p_0g,p_1g)$ with a suitable $G$-structure
on $P_0$ and $P_1$ separately. Then each element $g \in G$ induces a
monoidal automorphism $\varphi_g: \omega \=> \omega$. For the monoidal
automorphism $\varphi: \omega \=> \omega$ with $\varphi (P_0,P_1)
(p_0,p_1) := (p_0,-p_1)$ there is, however, no $g \in G$ with $\varphi
= \varphi_g$. So in this case the group of symmetries (of monoidal
automorphisms of $\omega$) is a bigger group than the one we started
out with. The given $\varphi$ is an example of a hidden symmetry.

 \section{Braided categories and $\C$-categories}

 Throughout this paper let $\A$ be a {\em monoidal category}, i.e.~a
category together with a bifunctor $\tensor: \A \times \A \=> \A$, a
neutral object $I \in \A$, and natural isomorphisms $\alpha: (P
\tensor Q) \tensor R \=> P \tensor (Q \tensor R)$, $\lambda: I \tensor
P \=> P$, and $\rho: P \tensor I \=> P$, satisfying the well-known
coherence (constraint) conditions. Without loss of generality (by Mac
Lane's coherence theorem \cite{ML} Theorem 15.1) we shall assume that
$\A$ is a strict monoidal category, so that all associativity and unit
isomorphisms are identities. Similarly $\C$ will be a monoidal
category throughout.

 We are mainly interested in the case where $\A$ is the category of
(right) modules $\M_B$ or comodules $\M^B$ over a bialgebra $B$ (over
a field $\kl$) with the canonical monoidal structure. Further examples
are $\Vek$ the category of vector spaces over a field $\kl$, $\vek$
the category of finite-dimensional vector spaces, mod-$B$, the
category of finite-dimensional (over $\kl$) $B$-modules, and
 comod-$B$, the category of finite-dimensional $B$-comodules. Some
interesting topological examples may be found in \cite{YE}.

 \subsection{$\C$-categories}

 We will need the notion of categories, functors, and natural
transformations  ``over'' a monoidal category $\C$, which we call
$\C$-categories, $\C$-functors, and $\C$-morphisms. Many of their
properties have been investigated in \cite{PA1, PA3}. They are built
in analogy to $G$-sets and their morphisms or $R$-modules and their
morphisms.

 \begin{defn}
 A category $\B$ together with a bifunctor $\tensor: \C \times \B \=>
\B$ and coherent\footnote{Whenever we use the term ``coherent'' we
mean that the given natural transformation is coherent also with
respect to the already existing coherent natural transformations, in
this case with $\alpha$, $\lambda$, and $\rho$. The minimal
requirements for coherence are obvious in most cases. We do not
further investigate them.} natural isomorphisms $\beta: (X \tensor Y)
\tensor P \=> X \tensor (Y \tensor P)$ (for $X,Y \in \C, P \in \B$)
and $\pi: I \tensor P \=> P$ will be called a (left)
 $\C${\em -category}. (For the coherence conditions see \cite{SD}.) In
such a context we will call $\C$ a {\em control category}.
 \end{defn}

 Some of our main examples are:

 \subsubsection{}
 A monoidal category $\A$ is an $\A$-category.

 \subsubsection{}
 Let $A$ (with $m_A: A \tensor A \=> A$ and $u_A: I \=> A$) be an {\em
algebra} (a monoid) in $\A$, i.e.~the multiplication $m_A$ is
associative and unital (with unit morphism $u_A$). In the situation
$\A = \M^B$, such an algebra $A$ is called a $B${\em -comodule
algebra}. In the case $\A = \M_B$, such an algebra $A$ is called a
$B${\em -module algebra} \cite{SW}.

 \subsubsection{}
 The category $\B = \A_A$ of (right) $A${\em -modules} $(P,\kappa: P
\tensor A \=> P)$ in $\A$ is a (left) $\A$-category, since $X \tensor
P$ carries the structure of a right $A$-module by $(X \tensor P)
\tensor A \iso X \tensor (P \tensor A) \=> X \tensor P$.

 \subsubsection{}
 A vector space $P$ is in $(\M^B)_A$ if and only if $P$ is a right
$B$-comodule and a right $A$-module such that $\delta(pa) = \sum
p_{(0)}a_{(0)} \tensor p_{(1)}a_{(1)}$, a $B$-$A$-Hopf module. A
vector space $P$ is in $(\M_B)_A$ iff $P$ is a right $B$-module and a
right $A$-module such that $(pa)b = \sum (pb_{(1)})(ab_{(2)})$, i.e.~a
$B \# A$-module.

 \subsubsection{}
 Furthermore let $C$ (with $\Delta_C: C \=> C \tensor C$ and
$\varepsilon_C: C \=> I$) be a coalgebra in $\A$. In the situation $\A
= \M^B$, such a coalgebra $C$ is called a $B${\em -comodule
coalgebra}. In the case $\A = \M_B$, such a coalgebra $C$ is called a
$B${\em -module coalgebra}.

 \subsubsection{} \label{comodCcat}
 The category $\B = \A^C$ of (right) $C${\em -comodules} $(P,\delta: P
\=> P \tensor C)$ in $\A$ is a (left) $\A$-category, since $X \tensor
P$ carries the structure of a right $C$-comodule by $X \tensor P \=> X
\tensor (P \tensor C) \iso (X \tensor P) \tensor C$.

 \subsubsection{} \label{cosmashcomods}
 A vector space $P$ is in $(\M^B)^C$ if and only if $P$ is a right
$B$-comodule and a right $C$-comodule such that
 $$\bfig
 \putmorphism(0, 400)(1, 0)[P`P \tensor B`\sst \delta_B]{900}1a
 \putmorphism(0, 0)(1, 0)[P \tensor C`(P \tensor C) \tensor B`\sst
\delta_B]{900}1a
 \putmorphism(0, 400)(0, -1)[``\sst \delta_C]{400}1r
 \putmorphism(900, 400)(0, -1)[``\sst \delta_C \tensor \id_B]{400}1r
 \efig$$
 commutes, i.e. $P$ is a $B \#^c C$-comodule, a comodule over the
cosmash product. A vector space $P$ is in $(\M_B)^C$ iff $P$ is a
right $B$-module and a right $C$-comodule such that $\delta_C(p)b =
\sum p_{(0)}b_{(1)} \tensor p_{(C,1)}b_{(2)}$.

 \begin{defn}
 Let $\B$ and $\B'$ be $\C$-categories. A functor $\omega: \B \=> \B'$
together with a coherent natural isomorphism $\xi: \omega(X \tensor P)
\=> X \tensor \omega(P)$ is called a $\C${\em -functor}.

 Observe that our assumption on coherence implies in particular that
$\pi' \xi(I,P) = \omega(\pi)$.
 \end{defn}

 \subsubsection{} \label{comodCfunct}
 The identity functor $\id: \A \=> \A$ is an $\A$-functor. Furthermore
the forgetful (underlying) functors $\omega: \A_A \=> \A$ resp.
$\omega: \A^C \=> \A$ are easily seen to be $\A$-functors. If $f: A
\=> A'$ is an algebra morphism in $\A$, then the induced functor
$\omega: \A_{A'} \=> \A_A$ is an $\A$-functor. Similarly if $f: C \=>
C'$ is a coalgebra morphism in $\A$, then the induced functor $\omega:
\A^C \=> \A^{C'}$ is an $\A$-functor.

 \subsubsection{}
 We will use additional $\C$-functors. Let $\omega: \B \=> \A$ be a
$\C$-functor and let $M \in \A$. Then $\omega \tensor M: \B \ni P
\mapsto \omega(P) \tensor M \in \A$ is again a $\C$-functor.

 \begin{defn} \label{Cmorph}
 Let $\B$ and $\B'$ be $\C$-categories and $\omega: \B \=> \B'$ and
$\omega': \B \=> \B'$ be $\C$-functors. A natural transformation
$\varphi: \omega \=> \omega'$ is a $\C${\em -morphism} if the
following diagram commutes
 $$\bfig
 \putmorphism(0, 400)(1, 0)[\omega(X \tensor P)`\omega'(X \tensor
P)`\sst \varphi(X \tensor P)]{900}1a
 \putmorphism(0, 0)(1, 0)[X \tensor \omega(P)`X \tensor
\omega'(P).`\sst X \tensor \varphi(P)]{900}1a
 \putmorphism(0, 400)(0, -1)[``\sst \xi(X,P)]{400}1r
 \putmorphism(900, 400)(0, -1)[``\sst \xi'(X,P)]{400}1r
 \efig$$
 We will denote the set\footnote{There are well known standard methods
to handle the set theoretic difficulties of this construction.} of
natural transformation from $\omega$ to $\omega'$ by
$\Nat(\omega,\omega')$ and the subset of $\C$-morphisms by
$\Nat_\C(\omega,\omega')$.
 \end{defn}

 \subsubsection{} \label{Cexamples}
 These $\C$-morphisms will be of central importance for
reconstruction, so we will give an example. Let $B$ be a bialgebra in
$\Vek$ and $\C = \A := \M_B$.
 Let $A$ be an algebra in $\Vek$. It can be considered as a $B$-module
algebra by the trivial action $ab := a\varepsilon(b)$.
 Let $\B := (\M_B)_A$. Consider the $\C$-functor $\omega: (\M_B)_A \=>
\M_B$ with $\omega(P) = P$, the forgetful functor. Then for any $a \in
A$ the morphism $\varphi_a:\omega \=> \omega$, $\varphi_a(P):
\omega(P) \=> \omega(P)$, $\varphi_a(p) = pa$ is a natural
transformation and in fact a $\C$-morphism. For any $b \in
\mbox{center}(B)$ the morphism $\varphi_b:\omega \=> \omega$,
$\varphi_b(P): \omega(P) \=> \omega(P)$, $\varphi_b(p) = pb$ is a
natural transformation, but in general it is not a $\C$-morphism.
 If $\varphi_b$ is a $\C$-morphism then
 for the special choice $X = B$, $P = B \tensor A$, $x = 1_B$, and $p
= 1_B \tensor 1_A$ we have
 $\sum b_{(1)} \tensor b_{(2)} \tensor 1_A
 = \sum b_{(1)} \tensor b_{(2)} \tensor 1_Ab_{(3)}
 = \sum xb_{(1)} \tensor pb_{(2)}
 = \varphi_b(x \tensor p)
 = x \tensor \varphi_b(p)
 = x \tensor pb
 = 1_B \tensor b \tensor 1_A$,
 and hence $\Delta(b) = 1 \tensor b$ which implies $b = \alpha \cdot
1_B$ ($\alpha \in \kl$). Conversely for $b = \alpha \cdot 1_B$ it is
easy to see that $\varphi_b$ is a $\C$-morphism.

 \subsection{Braided categories}

 For the definition and study of more complicated objects, like
bialgebras and Hopf algebras in $\C$, we assume that the monoidal
category $\C$ is {\em braided} (or a {\em quasitensor} category) with
a natural isomorphism of bifunctors $\sigma_{X,Y}: X \tensor Y \iso Y
\tensor X$, the {\em braiding}, such that $(1_Y \tensor
\sigma_{X,Z})(\sigma_{X,Y} \tensor 1_Z) = \sigma_{X,Y \tensor Z}$ and
$(\sigma_{X,Z} \tensor 1_Y)(1_X \tensor \sigma_{Y,Z}) = \sigma_{X
\tensor Y,Z}$.

 \subsubsection{}
 A {\em quasitriangular structure} or {\em universal $R$-matrix}
\cite{DR1} for a bialgebra $B = (B,m,u,\Delta,\varepsilon)$ in $\Vek$
is an invertible element $R = \sum R_1 \tensor R_2 \in B \tensor B$
such that
 \begin{enumerate}
 \item $\forall b \in B: \tau\Delta(b) = R\Delta(b)R^{-1},$
 \item $(\Delta \tensor 1)(R) = R_{13}R_{23},$
 \item $(1 \tensor \Delta)(R) = R_{13}R_{12}$
 \end{enumerate}
 where $R_{12} = R \tensor 1_B$, $R_{13} = \sum R_1 \tensor 1_B
\tensor R_2$, and $R_{23} = 1_B \tensor R$.

 \subsubsection{}
 A {\em coquasitriangular structure} (\cite{SCH2} Definition 2.4.4 and
\cite{LT}) or {\em braiding} is a convolution-invertible homomorphism
$r: B \tensor B \=> \kl$ such that
 \begin{enumerate}
 \item $m\tau = r * m * r^{-1},$
 \item $r(m \tensor 1) = r^{13}r^{23},$
 \item $r(1 \tensor m) = r^{13}r^{12}.$
 \end{enumerate}

 \subsubsection{}
 If $B$ is quasitriangular then $\M_B$ is a braided monoidal category
with $\sigma_{X,Y}(x \tensor y) = \sum(yR_2 \tensor xR_1)$ \cite{DR1}.

 \subsubsection{}
 If $B$ is coquasitriangular then $\M^B$ is a braided monoidal
category with $\sigma_{X,Y}(x \tensor y) = \sum(y_{(0)} \tensor
x_{(0)})r(x_1 \tensor y_1)$ (\cite{SCH2} Remark 2.4.6; see also the
last paragraph in \cite{PA3}).

 \subsubsection{}
 Here are some observations from \cite{MAJ4} about algebras,
bialgebras and Hopf algebras in braided monoidal categories $\C$. If
$A$ and $A'$ are algebras in $\C$ then so is $A \tensor B$. We use the
graphical calculus \cite{YE} to describe the algebra multiplication as
 \bgr{30}{20}
 \put(0,5){\mult}
 \put(20,5){\mult}
 \put(10,10){\braid}
 \put(5,0){\idgr{5}}
 \put(25,0){\idgr{5}}
 \put(0,10){\idgr{10}}
 \put(30,10){\idgr{10}}
 \put(0,20){\objo{A}}
 \put(10,20){\objo{B}}
 \put(20,20){\objo{A}}
 \put(30,20){\objo{B}}
 \put(5,0){\obju{A}}
 \put(25,0){\obju{B}}
 \egr
 which represents the morphism $(m_A \tensor m_B)(1_A \tensor
\sigma_{B,A} \tensor 1_B)$. One checks that $A \tensor B$ becomes an
algebra with this multiplication.

 This allows us to define a {\em bialgebra} in $\C$ which is an
algebra $(B,m,u)$ and a coalgebra $(B,\Delta,\varepsilon)$ such that
 \bgr{230}{30}
 \put(0,5){\comult}
 \put(30,20){\comult}
 \put(50,20){\comult}
 \put(0,20){\mult}
 \put(30,5){\mult}
 \put(50,5){\mult}
 \put(40,10){\braid}
 \put(0,0){\idgr{5}}
 \put(10,0){\idgr{5}}
 \put(35,0){\idgr{5}}
 \put(55,0){\idgr{5}}
 \put(0,25){\idgr{5}}
 \put(10,25){\idgr{5}}
 \put(35,25){\idgr{5}}
 \put(55,25){\idgr{5}}
 \put(5,10){\idgr{10}}
 \put(30,10){\idgr{10}}
 \put(60,10){\idgr{10}}
 \put(0,30){\objo{B}}
 \put(10,30){\objo{B}}
 \put(35,30){\objo{B}}
 \put(55,30){\objo{B}}
 \put(20,15){\objo{=}}
 \put(0,0){\obju{B}}
 \put(10,0){\obju{B}}
 \put(35,0){\obju{B}}
 \put(55,0){\obju{B}}

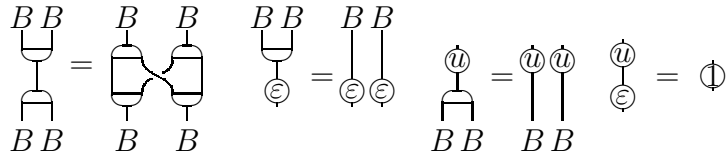
 \put(140,5){\comult}
 \put(80,20){\mult}
 \put(85,5){\mor{\varepsilon}{10}}
 \put(110,5){\mor{\varepsilon}{10}}
 \put(120,5){\mor{\varepsilon}{10}}
 \put(145,15){\mor{u}{10}}
 \put(170,15){\mor{u}{10}}
 \put(180,15){\mor{u}{10}}
 \put(200,17){\mor{u}{10}}
 \put(200,3){\mor{\varepsilon}{10}}
 \put(230,10){\mor{1}{10}}
 \put(80,25){\idgr{5}}
 \put(85,15){\idgr{5}}
 \put(90,25){\idgr{5}}
 \put(110,15){\idgr{15}}
 \put(120,15){\idgr{15}}
 \put(140,0){\idgr{5}}
 \put(145,10){\idgr{5}}
 \put(150,0){\idgr{5}}
 \put(170,0){\idgr{15}}
 \put(180,0){\idgr{15}}
 \put(200,13){\idgr{4}}
 \put(80,30){\objo{B}}
 \put(90,30){\objo{B}}
 \put(110,30){\objo{B}}
 \put(120,30){\objo{B}}
 \put(100,11){\objo{=}}
 \put(160,11){\objo{=}}
 \put(215,11){\objo{=}}
 \put(140,0){\obju{B}}
 \put(150,0){\obju{B}}
 \put(170,0){\obju{B}}
 \put(180,0){\obju{B}}
 \egr
 i.e.~$\Delta m = (m \tensor m)(1 \tensor \sigma \tensor 1)(\Delta
\tensor \Delta)$, $\varepsilon m = \varepsilon \tensor \varepsilon $,
$ \Delta u = u \tensor u$, and $\varepsilon u = 1_\kk$.

 A bialgebra $H$ in $\C$ is a {\em Hopf algebra} with {\em antipode
$S: H \=> H$} in $\C$ if it also satisfies
 \bgr{60}{30}
 \put(0,20){\comult}
 \put(50,20){\comult}
 \put(0,5){\mult}
 \put(50,5){\mult}
 \put(0,10){\mor{S}{10}}
 \put(30,16){\mor{\varepsilon}{10}}
 \put(30,4){\mor{u}{10}}
 \put(60,10){\mor{S}{10}}
 \put(5,0){\idgr{5}}
 \put(5,25){\idgr{5}}
 \put(10,10){\idgr{10}}
 \put(30,0){\idgr{4}}
 \put(30,26){\idgr{4}}
 \put(50,10){\idgr{10}}
 \put(55,0){\idgr{5}}
 \put(55,25){\idgr{5}}
 \put(5,30){\objo{H}}
 \put(30,30){\objo{H}}
 \put(55,30){\objo{H}}
 \put(20,11){\objo{=}}
 \put(40,11){\objo{=}}
 \put(5,0){\obju{H}}
 \put(30,0){\obju{H}}
 \put(55,0){\obju{H}}
 \egr

 A bialgebra $H$ in $\C$ has a {\em twisted antipode $S$} in $\C$ if
it satisfies
 \bgr{60}{40}
 \put(0,30){\comult}
 \put(0,20){\mor{S}{10}}
 \put(5,35){\idgr{5}}
 \put(5,40){\objo{H}}
 \put(10,20){\idgr{10}}
 \put(0,10){\ibraid}
 \put(0,5){\mult}
 \put(5,0){\idgr{5}}
 \put(5,0){\obju{H}}

 \put(20,15){\objo{=}}

 \put(30,26){\mor{\varepsilon}{10}}
 \put(30,36){\idgr{4}}
 \put(30,40){\objo{H}}
 \put(30,4){\mor{u}{10}}
 \put(30,0){\idgr{4}}
 \put(30,0){\obju{H}}

 \put(40,15){\objo{=}}

 \put(50,30){\comult}
 \put(50,20){\idgr{10}}
 \put(55,35){\idgr{5}}
 \put(55,40){\objo{H}}
 \put(60,20){\mor{S}{10}}
 \put(50,10){\braid}
 \put(50,5){\mult}
 \put(55,0){\idgr{5}}
 \put(55,0){\obju{H}}
 \egr

 The notion of a braided bialgebra in $\C$ is somewhat more subtle and
has been studied in \cite{MAJ2}.

 \subsection{$\C$-monoidal categories}

 Now let $\C$ be a braided monoidal category.

 \begin{defn}
 Let $\B$, $\B'$, and $\B''$ be $\C$-categories. A bifunctor $\omega:
\B \times \B' \=> \B''$ together with natural isomorphisms coherent
with the $\C$-structures on $\B$, $\B'$, and $\B''$,
 \begin{enumerate}
 \item $\xi_{X,P,Q}: \omega(X \tensor P,Q) \=> X \tensor \omega(P,Q)$,
 \item $\tau_{P,X,Q}: \omega(P,X \tensor Q) \=> X \tensor
\omega(P,Q)$, and
 \item $\wtau_{X,P,Q}: X \tensor \omega(P,Q) \=> \omega(P,X \tensor
Q)$
 \end{enumerate}
 is called a $\C${\em -bifunctor}, if the following diagrams commute
  $$\bfig
 \putmorphism(0, 400)(1, 0)[\omega(X \tensor P,Y \tensor Q)`Y \tensor
\omega(X \tensor P,Q)`\sst \tau_{X \tensor P,Y,Q}]{1400}1a
 \putmorphism(1400, 400)(1, 0)[\phantom{Y \tensor \omega(X \tensor
P,Q)}`Y \tensor X \tensor \omega(P,Q)`\sst 1 \tensor \xi ]{1200}1a
 \putmorphism(0, 0)(1, 0)[X \tensor \omega(P,Y \tensor Q)`X \tensor Y
\tensor \omega(P,Q)`\sst 1 \tensor \tau_{P,Y,Q}]{1400}1a
 \putmorphism(0, 400)(0, -1)[``\sst \xi]{400}1l
 \putmorphism(2600, 400)(-3, -1)[``\sst \ \sigma_{X,Y} \tensor
1_{\omega(P,Q)} ]{1200}{-1}r
 \efig$$
 $$\bfig
 \putmorphism(0, 400)(1, 0)[\omega(P,X \tensor Y \tensor Q)`X \tensor
Y \tensor \omega(P,Q)`\sst \tau_{P,X \tensor Y,Q}]{1600}1a
 \putmorphism(0, 400)(2, -1)[`X \tensor \omega(P,Y \tensor Q).`\sst
\tau_{P,X,Y \tensor Q} ]{800}1l
 \putmorphism(1600, 400)(-2, -1)[``\sst \ 1_X \tensor
\tau_{P,Y,Q}]{800}{-1}r
 \efig$$
 $$\bfig
 \putmorphism(0, 400)(1, 0)[X \tensor Y \tensor \omega(P,Q)`\omega(P,X
\tensor Y \tensor Q)`\sst \widetilde \tau_{X \tensor Y,P,Q}]{1600}1a
 \putmorphism(0, 400)(2, -1)[`X \tensor \omega(P,Y \tensor Q).`\sst
1_X \tensor \widetilde \tau_{Y,P,Q} ]{800}1l
 \putmorphism(1600, 400)(-2, -1)[``\sst \widetilde \tau_{X,P,Y \tensor
Q} ]{800}{-1}r
 \efig$$
 and
  $$\bfig
 \putmorphism(0, 400)(1, 0)[Y \tensor X \tensor \omega(P,Q)`Y \tensor
\omega(X \tensor P,Q)`\sst 1 \tensor \xi^{-1}]{1200}1a
 \putmorphism(1200, 400)(1, 0)[\phantom{Y \tensor \omega(X \tensor
P,Q)}` \omega(X \tensor P,Y \tensor Q)`\sst \widetilde \tau_{Y,X
\tensor P,Q} ]{1400}1a
 \putmorphism(1200, 0)(1, 0)[X \tensor Y \tensor \omega(P,Q)`X \tensor
\omega(P,Y \tensor Q) `\sst 1 \tensor \widetilde \tau_{Y,P,Q}]{1400}1a
 \putmorphism(2600, 400)(0, -1)[``\sst \xi^{-1}]{400}{-1}r
 \putmorphism(0, 400)(3, -1)[``\sst \sigma_{X,Y} \tensor
1_{\omega(P,Q)} \ ]{1200}1l
 \efig$$
 (suppressing the coherence isomorphisms $\alpha$ and $\beta$ from
Definition 2.1, i.e. going to the strict case.) The corresponding
braid diagrams are
 \bgr{365}{20}
 \put(50,0){\braid}
 \put(60,10){\braid}
 \put(0,10){\dbloverbraid}
 \put(0,0){\idgr{10}}
 \put(10,0){\idgr{10}}
 \put(20,0){\idgr{10}}
 \put(30,0){\idgr{20}}
 \put(50,10){\idgr{10}}
 \put(70,0){\idgr{10}}
 \put(80,0){\idgr{20}}
 \put(0,20){\objo{X}}
 \put(10,20){\objo{P}}
 \put(20,20){\objo{Y}}
 \put(30,20){\objo{Q}}
 \put(50,20){\objo{X}}
 \put(60,20){\objo{P}}
 \put(70,20){\objo{Y}}
 \put(80,20){\objo{Q}}
 \put(0,0){\obju{Y}}
 \put(10,0){\obju{X}}
 \put(20,0){\obju{P}}
 \put(30,0){\obju{Q}}
 \put(50,0){\obju{Y}}
 \put(60,0){\obju{X}}
 \put(70,0){\obju{P}}
 \put(80,0){\obju{Q}}
 \put(40,5){\objo{=}}

 \put(145,10){\braid}
 \put(155,0){\braid}
 \put(095,10){\overdblbraid}
 \put(095,0){\idgr{10}}
 \put(105,0){\idgr{10}}
 \put(115,0){\idgr{10}}
 \put(125,0){\idgr{20}}
 \put(145,0){\idgr{10}}
 \put(165,10){\idgr{10}}
 \put(175,0){\idgr{20}}
 \put(095,20){\objo{P}}
 \put(105,20){\objo{X}}
 \put(115,20){\objo{Y}}
 \put(125,20){\objo{Q}}
 \put(145,20){\objo{P}}
 \put(155,20){\objo{X}}
 \put(165,20){\objo{Y}}
 \put(175,20){\objo{Q}}
 \put(135,5){\objo{=}}
 \put(095,0){\obju{X}}
 \put(105,0){\obju{Y}}
 \put(115,0){\obju{P}}
 \put(125,0){\obju{Q}}
 \put(145,0){\obju{X}}
 \put(155,0){\obju{Y}}
 \put(165,0){\obju{P}}
 \put(175,0){\obju{Q}}

 \put(240,0){\braid}
 \put(250,10){\braid}
 \put(190,10){\dbloverbraid}
 \put(190,0){\idgr{10}}
 \put(200,0){\idgr{10}}
 \put(210,0){\idgr{10}}
 \put(220,0){\idgr{20}}
 \put(240,10){\idgr{10}}
 \put(260,0){\idgr{10}}
 \put(270,0){\idgr{20}}
 \put(190,20){\objo{X}}
 \put(200,20){\objo{Y}}
 \put(210,20){\objo{P}}
 \put(220,20){\objo{Q}}
 \put(240,20){\objo{X}}
 \put(250,20){\objo{Y}}
 \put(260,20){\objo{P}}
 \put(270,20){\objo{Q}}
 \put(190,0){\obju{P}}
 \put(200,0){\obju{X}}
 \put(210,0){\obju{Y}}
 \put(220,0){\obju{Q}}
 \put(240,0){\obju{P}}
 \put(250,0){\obju{X}}
 \put(260,0){\obju{Y}}
 \put(270,0){\obju{Q}}
 \put(230,5){\objo{=}}

 \put(335,10){\braid}
 \put(345,0){\braid}
 \put(285,10){\overdblbraid}
 \put(285,0){\idgr{10}}
 \put(295,0){\idgr{10}}
 \put(305,0){\idgr{10}}
 \put(315,0){\idgr{20}}
 \put(335,0){\idgr{10}}
 \put(355,10){\idgr{10}}
 \put(365,0){\idgr{20}}
 \put(285,20){\objo{Y}}
 \put(295,20){\objo{X}}
 \put(305,20){\objo{P}}
 \put(315,20){\objo{Q}}
 \put(335,20){\objo{Y}}
 \put(345,20){\objo{X}}
 \put(355,20){\objo{P}}
 \put(365,20){\objo{Q}}
 \put(325,5){\objo{=}}
 \put(285,0){\obju{X}}
 \put(295,0){\obju{P}}
 \put(305,0){\obju{Y}}
 \put(315,0){\obju{Q}}
 \put(335,0){\obju{X}}
 \put(345,0){\obju{P}}
 \put(355,0){\obju{Y}}
 \put(365,0){\obju{Q}}
 \egr

 Observe that $\widetilde \tau_{X,P,Q}$ is not the inverse of
$\tau_{P,X,Q}$. Both morphisms are associated with $\sigma$ in the
control category, so in braid diagrams they will be represented by a
braid with the same orientation as $\sigma$.
 \end{defn}

 \subsubsection{}
 If $\A = \C$ is a braided monoidal category, $\B$ is an $\A$-category
and $\omega: \B \=> \A$ is an $\A$-functor, then the bifunctor $\omega
\tensor \omega: \B \times \B \=> \A$ given by $(\omega \tensor
\omega)(P,Q) = \omega(P) \tensor \omega(Q)$ is an $\A$-bifunctor. The
bifunctor $(\omega \tensor \omega \tensor M)(P,Q) = \omega(P) \tensor
\omega(Q) \tensor M$ for $M \in \A$ is also an $\A$-bifunctor.

 \subsubsection{}
 In a similar way define a $\C$-multifunctor property for
multifunctors $\omega: \B_1 \times \ldots \times \B_n \=> \B$. In
particular functors of the form $\omega \tensor \ldots \tensor \omega
= \omega^n: \B \times \ldots \times \B \=> \A$ and $\omega \tensor
\ldots \tensor \omega \tensor M = \omega^n \tensor M : \B \times
\ldots \times \B \=> \A$ are $\A$-multifunctors, if $\omega: \B \=>
\A$ is an $\A$-functor and $M \in \A$.

 \begin{defn}
 Let $\B$, $\B'$, and $\B''$ be $\C$-categories and $\omega, \omega':
\B \times \B' \=> \B''$ be $\C$-bifunctors. A natural transformation
$\varphi: \omega \=> \omega'$ is a $\C${\em -bimorphism}, if the
following diagrams commute
 $$\bfig
 \putmorphism(0, 400)(1, 0)[\omega(X \tensor P,Q)`\omega'(X \tensor
P,Q)`\sst \varphi(X \tensor P,Q)]{1600}1a
 \putmorphism(0, 0)(1, 0)[X \tensor \omega(P,Q)`X \tensor
\omega'(P,Q)`\sst X \tensor \varphi(P,Q)]{1600}1a
 \putmorphism(0, 400)(0, -1)[``\sst \xi ]{400}1r
 \putmorphism(1600, 400)(0, -1)[``\sst \xi']{400}1r
 \efig$$
 $$\bfig
 \putmorphism(0, 400)(1, 0)[\omega(P,Y \tensor Q)`\omega'(P,Y \tensor
Q)`\sst \varphi(P,Y \tensor Q)]{1600}1a
 \putmorphism(0, 0)(1, 0)[Y \tensor \omega(P,Q)`Y \tensor
\omega'(P,Q)`\sst Y \tensor \varphi(P,Q)]{1600}1a
 \putmorphism(0, 400)(0, -1)[``\sst \tau_{P,Y,Q}]{400}1l
 \putmorphism(1600, 400)(0, -1)[``\sst \tau'_{P,Y,Q}]{400}1r
 \efig$$
 $$\bfig
 \putmorphism(0, 0)(1, 0)[\omega(P,Y \tensor Q)`\omega'(P,Y \tensor
Q)`\sst \varphi(P,Y \tensor Q)]{1600}1a
 \putmorphism(0, 400)(1, 0)[Y \tensor \omega(P,Q)`Y \tensor
\omega'(P,Q)`\sst Y \tensor \varphi(P,Q)]{1600}1a
 \putmorphism(0, 400)(0, -1)[``\sst \widetilde \tau_{Y,P,Q}]{400}1l
 \putmorphism(1600, 400)(0, -1)[``\sst \widetilde
\tau'_{Y,P,Q}]{400}1r
 \efig$$
 Let $\Nat_\C(\omega, \omega' )$ denote the set of $\C$-bimorphisms.
 \end{defn}

 For multifunctors $\omega, \omega': \B_1 \times \ldots \times \B_n
\=> \B$ we proceed in a similar way. A natural transformation of
multifunctors $\varphi: \omega  \=> \omega'$ is called a
 $\C$-multimorphism, if commutative diagrams as above hold for all
variables.

 \begin{defn}
 Let $\A$ be a $\C$-category with a coherent structure of a monoidal
category with tensor product $P \tensorhat Q$. If $\tensorhat: \A
\times \A \=> \A$ is a coherent $\C$-bifunctor, then $\A$ is called a
$\C${\em -monoidal category}. In particular the structural morphisms
$\alpha$, $\lambda$, and $\rho$ for $\A$ are $\C$-morphisms in each
variable from $\A$.

 If we go to the strict case we assume $\alpha_\C$, $\lambda_\C$,
$\rho_\C$, $\beta$, $\pi$, $\alpha_\A$, $\lambda_\A$, $\rho_\A$, and
$\xi$ to be identities. Then the necessary equalities for the strict
case are
 $$\begin{array}{l}
 \tau(I,X,P) = \id,\\
 \wtau((X,I,P) = \id,\\
 \tau(P,X,Q) \tensorhat 1_R = \tau(P,X,Q \tensorhat R),\\
 \wtau(X,P,Q \tensorhat R) = \wtau(X,P,Q) \tensorhat 1_R,\\
 (\tau(P,X,Q) \tensorhat 1_R)(1_P \tensorhat \tau(Q,X,R)) = \tau(P
\tensorhat Q,X,R),\\
 (1_P \tensorhat \wtau(X,Q,R))(\wtau(X,P,Q) \tensorhat 1_R) =
\wtau(X,P \tensorhat Q,R).
 \end{array}$$
 \end{defn}

 Observe that $\C$ is a $\C$-monoidal category (since $\C$ is
braided).

 \begin{defn}
 Let $\A$ and $\B$ be monoidal categories. A {\em monoidal functor} is
a functor $\omega:\A \=> \B$ together with coherent natural
isomorphisms $\upsilon: \omega(P \tensorhat Q) \iso \omega(P)
\tensorhat \omega(Q)$ and $\varsigma: \omega(I_\A) \iso I_\B$.

 If $\A$ and $\B$ are $\C$-monoidal categories, $\omega: \A \=> \B$ is
a monoidal functor and a $\C$-functor, and $\upsilon: \omega(P
\tensorhat Q) \iso \omega(P) \tensorhat \omega(Q)$ is a
 $\C$-bimorphism, then $\omega$ is called a $\C${\em -monoidal
functor}.

 Let $\omega, \omega': \A \=> \B$ be $\C$-monoidal functors. A natural
transformation $\varphi: \omega \=> \omega'$ is a $\C$-{\em monoidal
morphism}, if $\varphi$ is a $\C$-morphism and monoidal.
 \end{defn}

 \begin{defn}
 Let $\A$ be a $\C$-monoidal category together with a braiding $\sigma
= \sigma_\A: P \tensorhat Q \=> Q \tensorhat P$. We call $\A$ a
$\C${\em -braided} $\C$-monoidal category, if the braid morphisms in
both categories are coherent w.r.t. the braid group, in particular if
 $$\bfig
 \putmorphism(0, 400)(1, 0)[P \tensorhat (X \tensor Q)`(X \tensor Q)
\tensorhat P`\sst \sigma]{1000}1a
 \putmorphism(0, 0)(1, 0)[X \tensor (P \tensorhat Q)`X \tensor (Q
\tensorhat P)`\sst 1 \tensor \sigma]{1000}1a
 \putmorphism(0, 400)(0, -1)[``\sst \tau]{400}1l
 \putmorphism(1000, 400)(0, -1)[``\sst \xi]{400}1r
 \efig$$
 and
 $$\bfig
 \putmorphism(0, 0)(1, 0)[(X \tensor P) \tensorhat Q`Q \tensorhat (X
\tensor P)`\sst \sigma]{1000}1a
 \putmorphism(0, 400)(1, 0)[X \tensor (P \tensorhat Q)`X \tensor (Q
\tensorhat P)`\sst 1 \tensor \sigma]{1000}1a
 \putmorphism(0, 400)(0, -1)[``\sst \xi^{-1}]{400}1l
 \putmorphism(1000, 400)(0, -1)[``\sst \widetilde \tau]{400}1r
 \efig$$
 commute. Observe, however, that the diagrams
 $$\bfig
 \putmorphism(0, 400)(1, 0)[(X \tensor P) \tensorhat Q`Q \tensorhat (X
\tensor P)`\sst \sigma]{1000}1a
 \putmorphism(0, 0)(1, 0)[X \tensor (P \tensorhat Q)`X \tensor (Q
\tensorhat P)`\sst 1 \tensor \sigma]{1000}1a
 \putmorphism(0, 400)(0, -1)[``\sst \xi]{400}1l
 \putmorphism(1000, 400)(0, -1)[``\sst \tau]{400}1r
 \efig$$
 and
 $$\bfig
 \putmorphism(0, 0)(1, 0)[P \tensorhat (X \tensor Q)`(X \tensor Q)
\tensorhat P`\sst \sigma]{1000}1a
 \putmorphism(0, 400)(1, 0)[X \tensor (P \tensorhat Q)`X \tensor (Q
\tensorhat P)`\sst 1 \tensor \sigma]{1000}1a
 \putmorphism(0, 400)(0, -1)[``\sst \widetilde \tau]{400}1l
 \putmorphism(1000, 400)(0, -1)[``\sst \xi^{-1}]{400}1r
 \efig$$
 do not necessarily commute since their braid diagrams are
 \bgr{160}{30}
 \put(10,10){\braid}
 \put(40,0){\braid}
 \put(40,10){\braid}
 \put(50,20){\braid}
 \put(0,0){\idgr{30}}
 \put(10,0){\idgr{10}}
 \put(10,20){\idgr{10}}
 \put(20,0){\idgr{10}}
 \put(20,20){\idgr{10}}
 \put(40,20){\idgr{10}}
 \put(60,0){\idgr{20}}
 \put(0,30){\objo{X}}
 \put(10,30){\objo{P}}
 \put(20,30){\objo{Q}}
 \put(40,30){\objo{X}}
 \put(50,30){\objo{P}}
 \put(60,30){\objo{Q}}
 \put(0,0){\obju{X}}
 \put(10,0){\obju{Q}}
 \put(20,0){\obju{P}}
 \put(40,0){\obju{X}}
 \put(50,0){\obju{Q}}
 \put(60,0){\obju{P}}
 \put(30,15){\obju{=}}
 \put(30,25){\obju{?}}
 \put(80,20){\obju{\mbox{and}}}
 \put(110,10){\braid}
 \put(140,10){\braid}
 \put(140,20){\braid}
 \put(150,0){\braid}
 \put(100,0){\idgr{30}}
 \put(110,0){\idgr{10}}
 \put(110,20){\idgr{10}}
 \put(120,0){\idgr{10}}
 \put(120,20){\idgr{10}}
 \put(140,0){\idgr{10}}
 \put(160,10){\idgr{20}}
 \put(100,30){\objo{X}}
 \put(110,30){\objo{P}}
 \put(120,30){\objo{Q}}
 \put(140,30){\objo{X}}
 \put(150,30){\objo{P}}
 \put(160,30){\objo{Q}}
 \put(100,0){\obju{X}}
 \put(110,0){\obju{Q}}
 \put(120,0){\obju{P}}
 \put(140,0){\obju{X}}
 \put(150,0){\obju{Q}}
 \put(160,0){\obju{P}}
 \put(130,15){\obju{=}}
 \put(130,25){\obju{?}}
 \egr
 In principle arbitrary tensor products of objects from $\C$ and from
$\A$ can be formed and twisted by elements of the braid group with the
exception of tensor factors from $\C$ appearing on the far right of a
tensor product containing tensor factors from $\A$.
 \end{defn}

 \begin{defn} \label{central}
 An object $P$ in a $\C$-monoidal category $\A$ is called $\C$-{\em
central}, if
 $$ (X \tensor P \tensor Q \buildrel \wtau(X,P,Q) \over
\longrightarrow P \tensor X \tensor Q
 \buildrel \tau(P,X,Q) \over \longrightarrow X \tensor P \tensor Q) =
\id$$
 holds for all $X \in \C$ and $Q \in \A$.
 \end{defn}

 \begin{thm}\label{comodcats}
 Let $\A$ be a $\C$-braided $\C$-monoidal category. Let $B$ be a
 $\C$-central bialgebra in $\A$, $C$ be a coalgebra in $\A$ and $z: C
\=> B$ be a coalgebra morphism. Then
 \begin{enumerate}
 \item $\A^C$ is a $\C$-category;
 \item $\A^B$ is a $\C$-monoidal category;
 \item $\omega:= \A^z : \A^C \=> \A^B $ is a $\C$-functor;
 \item the forgetful functor $\omega: \A^C \=> \A$ is a $\C$-functor;
 \item if $C$ is a $\C$-central bialgebra and $z: C \=> B$ is a
bialgebra morphism then $\omega:= \A^z : \A^C \=> \A^B $ is a
 $\C$-monoidal functor;
 \item the forgetful functor $\omega: \A^B \=> \A$ is a $\C$-monoidal
functor.
 \end{enumerate}
 \end{thm}

 \begin{pf}
 (1) similar to \ref{comodCcat}.

 (2)  A little calculation shows that $\A^B$ is a monoidal category
(\cite{MAJ4} Prop. 2.5) with the comultiplication on the tensor
product given by $(1_P \tensor 1_Q \tensor m_B)(1_P \tensor
\sigma_{B,Q} \tensor 1_B)(\delta_P \tensor \delta_Q): P \tensor Q \=>
P \tensor Q \tensor B$.  $\A^B$ is also a $\C$-category by (1). The
natural transformation $\xi: (X \tensor P) \tensorhat Q \=> X \tensor
(P \tensorhat Q)$ is compatible with the comultiplication with
$B$ from the right. So it is in $\A^B$. The natural transformation
$\tau: P \tensorhat (X \tensor Q) \iso X \tensor (P \tensorhat Q)$
satisfies
 \bgr{100}{45}
 \put(10,20){\multmor{\delta}{10}}
 \put(30,20){\multmor{\delta}{10}}
 \put(60,30){\multmor{\delta}{10}}
 \put(90,20){\multmor{\delta}{10}}
 \put(30,5){\mult}
 \put(90,5){\mult}
 \put(0,35){\braid}
 \put(20,10){\braid}
 \put(60,10){\braid}
 \put(70,20){\braid}
 \put(80,10){\braid}
 \put(15,30){\twist{-5}{5}}
 \put(0,0){\idgr{35}}
 \put(10,0){\idgr{20}}
 \put(20,0){\idgr{10}}
 \put(35,0){\idgr{5}}
 \put(35,30){\idgr{15}}
 \put(40,10){\idgr{10}}
 \put(60,0){\idgr{10}}
 \put(60,20){\idgr{10}}
 \put(65,40){\idgr{5}}
 \put(70,0){\idgr{10}}
 \put(80,0){\idgr{10}}
 \put(80,30){\idgr{15}}
 \put(95,0){\idgr{5}}
 \put(95,30){\idgr{15}}
 \put(100,10){\idgr{10}}
 \put(0,45){\objo{P}}
 \put(10,45){\objo{X}}
 \put(35,45){\objo{Q}}
 \put(65,45){\objo{P}}
 \put(80,45){\objo{X}}
 \put(95,45){\objo{Q}}
 \put(50,20){\objo{=}}
 \put(0,0){\obju{X}}
 \put(10,0){\obju{P}}
 \put(20,0){\obju{Q}}
 \put(35,0){\obju{B}}
 \put(60,0){\obju{X}}
 \put(70,0){\obju{P}}
 \put(80,0){\obju{Q}}
 \put(95,0){\obju{B}}
 \egr
 hence it is in $\A^B$, too. Finally the natural transformation
$\wtau: X \tensor (P \tensorhat Q) \iso P \tensorhat (X \tensor Q)$
satisfies
 \bgr{160}{65}
 \put(30,5){\mult}
 \put(90,5){\mult}
 \put(150,5){\mult}
 \put(10,20){\braid}
 \put(20,10){\braid}
 \put(5,55){\braid}
 \put(60,40){\braid}
 \put(70,20){\braid}
 \put(70,30){\braid}
 \put(80,10){\braid}
 \put(120,40){\braid}
 \put(140,10){\braid}
 \put(20,45){\twist{-5}{10}}
 \put(0,30){\multmor{\delta}{10}}
 \put(30,30){\multmor{\delta}{10}}
 \put(70,50){\multmor{\delta}{10}}
 \put(90,50){\multmor{\delta}{10}}
 \put(130,50){\multmor{\delta}{10}}
 \put(150,50){\multmor{\delta}{10}}
 \put(0,0){\idgr{30}}
 \put(5,40){\idgr{15}}
 \put(10,0){\idgr{20}}
 \put(20,0){\idgr{10}}
 \put(20,30){\idgr{15}}
 \put(30,20){\idgr{10}}
 \put(35,0){\idgr{5}}
 \put(35,40){\idgr{25}}
 \put(60,0){\idgr{40}}
 \put(60,50){\idgr{15}}
 \put(70,0){\idgr{20}}
 \put(75,60){\idgr{5}}
 \put(80,0){\idgr{10}}
 \put(80,40){\idgr{10}}
 \put(90,20){\idgr{30}}
 \put(95,0){\idgr{5}}
 \put(95,60){\idgr{5}}
 \put(100,10){\idgr{40}}
 \put(120,0){\idgr{40}}
 \put(120,50){\idgr{15}}
 \put(130,0){\idgr{40}}
 \put(135,60){\idgr{5}}
 \put(140,0){\idgr{10}}
 \put(140,20){\idgr{30}}
 \put(150,20){\idgr{30}}
 \put(155,0){\idgr{5}}
 \put(155,60){\idgr{5}}
 \put(160,10){\idgr{40}}
 \put(40,10){\idgr{20}}
 \put(5,65){\objo{X}}
 \put(15,65){\objo{P}}
 \put(35,65){\objo{Q}}
 \put(60,65){\objo{X}}
 \put(75,65){\objo{P}}
 \put(95,65){\objo{Q}}
 \put(120,65){\objo{X}}
 \put(135,65){\objo{P}}
 \put(155,65){\objo{Q}}
 \put(50,30){\objo{=}}
 \put(110,30){\objo{=}}
 \put(0,0){\obju{P}}
 \put(10,0){\obju{X}}
 \put(20,0){\obju{Q}}
 \put(35,0){\obju{B}}
 \put(60,0){\obju{P}}
 \put(70,0){\obju{X}}
 \put(80,0){\obju{Q}}
 \put(95,0){\obju{B}}
 \put(120,0){\obju{P}}
 \put(130,0){\obju{X}}
 \put(140,0){\obju{Q}}
 \put(157,0){\obju{B.}}
 \egr
 Since the diagrams defining the structure of a $\C$-monoidal category
on $\A^B$ commute in $\A$ and consist of morphisms of $B$-comodules
they also commute as diagrams in $\A^B$. Thus $\A^B$ is a
 $\C$-monoidal category.

 (3) and (4) similar to \ref{comodCfunct}

 (5) Since the tensor products in $\A^C$ and $\A^B$ are induced by the
tensor product in $\A$ the natural transformation $\upsilon: \omega(P
\tensorhat Q) \=> \omega(P) \tensor \omega(Q)$ is the identity which
makes $\omega$ a $\C$-monoidal functor.

 (6) is a special case of (5).
 \end{pf}

 A corresponding result holds by duality for the category $\A_B$ of
modules over a bialgebra $B$ in $\A$.

 \newpage

 \subsection{Rigid categories}

 \subsubsection{}
 Another important categorical notion is that of a (right) dual
object. This is a generalization of finite-dimensional vector spaces.
An object $X \in \C$ is {\em rigid} or has a {\em dual} $(X^*,\ev)$
where $X^* \in \C$ and $\ev: X^* \tensor X \=> I$ is called the {\em
evaluation}, if there is a morphism $\db: I \=> X \tensor X^*$, the
{\em dual basis}, such that\\
 \centerline{$(X \buildrel {\db \tensor 1} \over \longrightarrow X
\tensor X^* \tensor X \buildrel {1 \tensor \ev} \over \longrightarrow
X) = 1_X\ ,$} \\
 \centerline{$(X^* \buildrel {1 \tensor \db} \over \longrightarrow X^*
\tensor X \tensor X^* \buildrel {\ev \tensor 1} \over \longrightarrow
X^*) = 1_{X^*}\ .$}
 The monoidal category $\C$ is {\em rigid} or a {\em tensor category}
if every object of $\C$ has a dual. The full subcategory of objects in
$\C$ having duals is denoted by $\C_0$. An adjoint functor argument
shows that the dual of an object is unique up to isomorphism if it
exists.

 \subsubsection{}
 If $\omega: \B \=> \A$ is a monoidal functor and $P \in \B$ is rigid
then $\omega(P) \in \A$ is rigid with dual object $\omega(P^*)$,
evaluation $\omega(P^*) \tensor \omega(P) \iso \omega(P^* \tensor P)
\=> \omega(I_\B) \iso I_\A$, and dual basis $I_\A \iso \omega(I_\B)
\=> \omega(P \tensor P^*) \iso \omega(P) \tensor \omega(P^*)$.

 \begin{prop}
 Let $\C$ be a braided monoidal category. Then the full subcategory
$\C_0$ of rigid objects in $\C$ is a rigid braided monoidal category.
 \end{prop}

 \begin{pf}
 If the evaluation resp. the dual basis are morphisms represented by
 \bgr{40}{5}
 \put(30,0){\coeval}
 \put(0,0){\eval}
 \put(0,5){\objo{X^*}}
 \put(10,5){\objo{X}}
 \put(30,0){\obju{X}}
 \put(40,0){\obju{X^*}}
 \egr
 then the conditions are
 \bgr{100}{20}
 \put(0,10){\coeval}
 \put(70,10){\coeval}
 \put(10,5){\eval}
 \put(60,5){\eval}
 \put(0,0){\idgr{10}}
 \put(20,10){\idgr{10}}
 \put(40,0){\idgr{20}}
 \put(60,10){\idgr{10}}
 \put(80,0){\idgr{10}}
 \put(100,0){\idgr{20}}
 \put(20,20){\objo{X}}
 \put(40,20){\objo{X}}
 \put(60,20){\objo{X^*}}
 \put(100,20){\objo{X^*}}
 \put(30,8){\objo{=}}
 \put(90,8){\objo{=}}
 \put(0,0){\obju{X}}
 \put(40,0){\obju{X}}
 \put(80,0){\obju{X^*}}
 \put(100,0){\obju{X^*}}
 \egr
 If $X \in \C$ has a dual $(X^*,\ev)$ then $X^*$ has the dual
$(X,\ev\circ\sigma_{X,X^*})$ with the dual basis
 $\sigma_{X,X^*}^{-1}\circ\db$. The corresponding morphisms for $X^*$
are
 \bgr{42}{30}
 \put(0,5){\eval}
 \put(0,10){\ibraid}
 \put(0,20){\idgr{10}}
 \put(10,20){\idgr{10}}
 \put(0,30){\objo{X}}
 \put(12,30){\objo{X^*}}
 \put(30,20){\coeval}
 \put(30,10){\braid}
 \put(30,0){\idgr{10}}
 \put(40,0){\idgr{10}}
 \put(30,0){\obju{X^*}}
 \put(42,0){\obju{X}}
 \egr
 and the relations are
 \bgr{260}{60}
 \put(0,40){\coeval}
 \put(40,50){\coeval}
 \put(90,30){\coeval}
 \put(150,40){\coeval}
 \put(190,50){\coeval}
 \put(220,30){\coeval}
 \put(10,15){\eval}
 \put(40,25){\eval}
 \put(80,25){\eval}
 \put(140,15){\eval}
 \put(190,25){\eval}
 \put(230,25){\eval}
 \put(0,30){\braid}
 \put(150,30){\braid}
 \put(10,20){\ibraid}
 \put(40,30){\ibraid}
 \put(50,40){\ibraid}
 \put(140,20){\ibraid}
 \put(180,40){\ibraid}
 \put(190,30){\ibraid}
 \put(40,10){\twist{20}{20}}
 \put(80,10){\twist{20}{20}}
 \put(80,30){\twist{20}{20}}
 \put(200,10){\twist{-20}{20}}
 \put(240,10){\twist{-20}{20}}
 \put(240,30){\twist{-20}{20}}
 \put(260,0){\idgr{60}}
 \put(0,0){\idgr{30}}
 \put(20,30){\idgr{30}}
 \put(40,0){\idgr{10}}
 \put(40,40){\idgr{10}}
 \put(60,30){\idgr{10}}
 \put(60,50){\idgr{10}}
 \put(80,0){\idgr{10}}
 \put(100,50){\idgr{10}}
 \put(120,0){\idgr{60}}
 \put(140,30){\idgr{30}}
 \put(160,0){\idgr{30}}
 \put(180,50){\idgr{10}}
 \put(180,30){\idgr{10}}
 \put(200,0){\idgr{10}}
 \put(200,40){\idgr{10}}
 \put(220,50){\idgr{10}}
 \put(240,0){\idgr{10}}
 \put(20,60){\objo{X^*}}
 \put(60,60){\objo{X^*}}
 \put(100,60){\objo{X^*}}
 \put(120,60){\objo{X^*}}
 \put(140,60){\objo{X}}
 \put(180,60){\objo{X}}
 \put(220,60){\objo{X}}
 \put(260,60){\objo{X}}
 \put(30,28){\objo{=}}
 \put(70,28){\objo{=}}
 \put(110,28){\objo{=}}
 \put(170,28){\objo{=}}
 \put(210,28){\objo{=}}
 \put(250,28){\objo{=}}
 \put(0,0){\obju{X^*}}
 \put(40,0){\obju{X^*}}
 \put(80,0){\obju{X^*}}
 \put(120,0){\obju{X^*}}
 \put(160,0){\obju{X}}
 \put(200,0){\obju{X}}
 \put(240,0){\obju{X}}
 \put(260,0){\obju{X}}
 \egr
 If $X$ and $Y$ are in $\C_0$ then $X \tensor Y$ has the dual $(Y^*
\tensor X^*,\ev_Y(1_{Y^*} \tensor \ev_X \tensor 1_Y))$. The reader may
try the easy graphic and the diagrammatic proofs. Thus $\C_0$ is a
full monoidal subcategory of $\C$ which inherits the braiding and
contains the duals for every object.
 \end{pf}

 \newpage

 \subsection{Coadjoint coactions}

 \subsubsection{}
 Let $\A$ be a braided monoidal category. Let $C$ be a coalgebra in
$\A$, $H$ be a Hopf algebra in $\A$ and $z: C \=> H$ be a coalgebra
homomorphism. We define a {\em right coadjoint coaction} of $H$ on $C$
by
 $$\begin{array}{rl}
 \ad := &(1_C \tensor m_H)(1_C \tensor S \tensor 1_H)(\sigma_{H,C}
\tensor 1_H)\cr
 &\quad(z \tensor 1_C \tensor z)(1_C \tensor \Delta_C)\Delta_C: C \=>
C \tensor H.
 \end{array}$$

 \begin{prop} \label{coadcoalg}
 $C$ with the right coadjoint coaction is an $H$-co\-mo\-dule
coalgebra.
 \end{prop}

 \begin{pf}
 In graphical notation the right coadjoint coaction is
 \bgr{65}{50}
 \put(0,0){\idgr{20}}
 \put(10,0){\idgr{20}}
 \put(5,30){\idgr{20}}
 \put(0,20){\multmor{\ad}{10}}
 \put(35,5){\usebox{\rightcoadj}}
 \put(40,0){\idgr{5}}
 \put(47,45){\idgr{5}}
 \put(55,0){\idgr{5}}
 \put(5,50){\objo{C}}
 \put(47,50){\objo{C}}
 \put(25,25){\objo{=}}
 \put(0,0){\obju{C}}
 \put(10,0){\obju{H}}
 \put(40,0){\obju{C}}
 \put(55,0){\obju{H}}
 \egr
 The right coadjoint action is a counary action by
 \bgr{120}{65}
 \put(30,20){\usebox{\rightcoadj}}
 \put(80,55){\comult}
 \put(85,40){\comult}
 \put(10,10){\mor{\varepsilon}{10}}
 \put(50,5){\mor{\varepsilon}{10}}
 \put(80,45){\mor{\varepsilon}{10}}
 \put(95,30){\mor{\varepsilon}{10}}
 \put(0,30){\multmor{\ad}{10}}
 \put(0,0){\idgr{30}}
 \put(10,20){\idgr{10}}
 \put(5,40){\idgr{25}}
 \put(35,0){\idgr{20}}
 \put(42,60){\idgr{5}}
 \put(50,15){\idgr{5}}
 \put(85,60){\idgr{5}}
 \put(90,45){\idgr{10}}
 \put(85,0){\idgr{40}}
 \put(120,0){\idgr{65}}
 \put(5,65){\objo{C}}
 \put(42,65){\objo{C}}
 \put(85,65){\objo{C}}
 \put(120,65){\objo{C}}
 \put(25,30){\objo{=}}
 \put(70,30){\objo{=}}
 \put(110,30){\objo{=}}
 \put(0,0){\obju{C}}
 \put(35,0){\obju{C}}
 \put(85,0){\obju{C}}
 \put(120,0){\obju{C}}
 \egr
 The coaction is coassociative:
 \bgr{300}{100}
 \put(30,5){\usebox{\rightcoadj}}
 \put(40,55){\usebox{\rightcoadj}}
 \put(240,45){\usebox{\rightcoadj}}
 \put(85,80){\comult}
 \put(115,85){\comult}
 \put(145,80){\comult}
 \put(150,50){\comult}
 \put(170,40){\comult}
 \put(195,80){\comult}
 \put(200,40){\comult}
 \put(220,40){\comult}
 \put(255,15){\comult}
 \put(290,15){\comult}
 \put(90,85){\bcomult{15}}
 \put(100,90){\bcomult{20}}
 \put(150,85){\bcomult{15}}
 \put(200,85){\bcomult{15}}
 \put(105,35){\mult}
 \put(105,15){\mult}
 \put(150,15){\mult}
 \put(170,15){\mult}
 \put(200,25){\mult}
 \put(220,25){\mult}
 \put(85,30){\braid}
 \put(95,40){\braid}
 \put(95,20){\braid}
 \put(145,60){\braid}
 \put(150,40){\braid}
 \put(160,20){\braid}
 \put(195,60){\braid}
 \put(210,30){\braid}
 \put(65,45){\twist{-5}{10}}
 \put(105,30){\twist{5}{5}}
 \put(140,50){\twist{5}{10}}
 \put(175,45){\twist{-10}{20}}
 \put(190,50){\twist{5}{10}}
 \put(225,45){\twist{-10}{20}}
 \put(295,20){\twist{-5}{10}}
 \put(115,20){\twist{10}{10}}
 \put(85,70){\mor{z}{10}}
 \put(95,70){\mor{z}{10}}
 \put(115,65){\mor{z}{20}}
 \put(125,65){\mor{z}{20}}
 \put(145,70){\mor{z}{10}}
 \put(165,65){\mor{z}{20}}
 \put(195,70){\mor{z}{10}}
 \put(215,65){\mor{z}{20}}
 \put(85,40){\mor{S}{30}}
 \put(95,50){\mor{S}{10}}
 \put(150,30){\mor{S}{10}}
 \put(160,30){\mor{S}{10}}
 \put(205,45){\mor{S}{15}}
 \put(0,30){\multmor{\ad}{10}}
 \put(10,60){\multmor{\ad}{10}}
 \put(280,50){\multmor{\ad}{10}}
 \put(0,0){\idgr{30}}
 \put(10,0){\idgr{30}}
 \put(10,40){\idgr{20}}
 \put(20,0){\idgr{60}}
 \put(15,70){\idgr{30}}
 \put(35,0){\idgr{5}}
 \put(50,0){\idgr{5}}
 \put(45,45){\idgr{10}}
 \put(65,0){\idgr{45}}
 \put(52,95){\idgr{5}}
 \put(85,0){\idgr{30}}
 \put(95,0){\idgr{20}}
 \put(95,60){\idgr{10}}
 \put(105,50){\idgr{35}}
 \put(115,40){\idgr{25}}
 \put(110,95){\idgr{5}}
 \put(110,0){\idgr{15}}
 \put(125,30){\idgr{35}}
 \put(140,0){\idgr{50}}
 \put(155,0){\idgr{15}}
 \put(150,20){\idgr{10}}
 \put(155,70){\idgr{10}}
 \put(155,55){\idgr{5}}
 \put(157,90){\idgr{10}}
 \put(170,30){\idgr{10}}
 \put(180,20){\idgr{20}}
 \put(190,0){\idgr{50}}
 \put(200,30){\idgr{10}}
 \put(205,0){\idgr{25}}
 \put(205,70){\idgr{10}}
 \put(207,90){\idgr{10}}
 \put(225,0){\idgr{25}}
 \put(230,30){\idgr{10}}
 \put(245,0){\idgr{45}}
 \put(255,0){\idgr{15}}
 \put(252,85){\idgr{15}}
 \put(175,0){\idgr{15}}
 \put(260,20){\idgr{25}}
 \put(280,0){\idgr{50}}
 \put(285,60){\idgr{40}}
 \put(290,30){\idgr{20}}
 \put(290,0){\idgr{15}}
 \put(300,0){\idgr{15}}
 \put(265,0){\idgr{15}}
 \put(15,100){\objo{C}}
 \put(52,100){\objo{C}}
 \put(110,100){\objo{C}}
 \put(157,100){\objo{C}}
 \put(207,100){\objo{C}}
 \put(252,100){\objo{C}}
 \put(285,100){\objo{C}}
 \put(30,50){\objo{=}}
 \put(72,50){\objo{=}}
 \put(132,50){\objo{=}}
 \put(182,50){\objo{=}}
 \put(232,50){\objo{=}}
 \put(272,50){\objo{=}}
 \put(0,0){\obju{C}}
 \put(10,0){\obju{H}}
 \put(20,0){\obju{H}}
 \put(35,0){\obju{C}}
 \put(50,0){\obju{H}}
 \put(65,0){\obju{H}}
 \put(85,0){\obju{C}}
 \put(95,0){\obju{H}}
 \put(110,0){\obju{H}}
 \put(140,0){\obju{C}}
 \put(155,0){\obju{H}}
 \put(175,0){\obju{H}}
 \put(190,0){\obju{C}}
 \put(205,0){\obju{H}}
 \put(225,0){\obju{H}}
 \put(245,0){\obju{C}}
 \put(255,0){\obju{H}}
 \put(265,0){\obju{H}}
 \put(280,0){\obju{C}}
 \put(290,0){\obju{H}}
 \put(300,0){\obju{H}}
 \egr
 The comultiplication is an $H$-comodule morphism by
 \bgr{280}{85}
 \put(50,30){\usebox{\rightcoadj}}
 \put(85,30){\usebox{\rightcoadj}}
 \put(210,30){\usebox{\rightcoadj}}
 \put(135,65){\comult}
 \put(175,70){\comult}
 \put(155,45){\comult}
 \put(210,25){\comult}
 \put(260,20){\comult}
 \put(5,50){\bcomult{20}}
 \put(62,70){\bcomult{35}}
 \put(140,70){\bcomult{20}}
 \put(150,75){\bcomult{30}}
 \put(20,25){\mult}
 \put(90,5){\mult}
 \put(160,5){\mult}
 \put(170,15){\mult}
 \put(155,30){\mult}
 \put(10,30){\braid}
 \put(80,10){\braid}
 \put(150,10){\braid}
 \put(160,20){\braid}
 \put(135,45){\braid}
 \put(80,20){\twist{-10}{10}}
 \put(100,10){\twist{5}{20}}
 \put(150,20){\twist{-5}{15}}
 \put(170,10){\twist{5}{5}}
 \put(180,20){\twist{5}{10}}
 \put(265,25){\twist{5}{15}}
 \put(170,30){\twist{5}{10}}
 \put(135,55){\mor{z}{10}}
 \put(160,50){\mor{z}{20}}
 \put(185,50){\mor{z}{20}}
 \put(145,35){\mor{S}{10}}
 \put(165,35){\mor{S}{10}}
 \put(0,40){\multmor{\ad}{10}}
 \put(20,40){\multmor{\ad}{10}}
 \put(270,40){\multmor{\ad}{10}}
 \put(0,0){\idgr{40}}
 \put(10,0){\idgr{30}}
 \put(25,0){\idgr{25}}
 \put(15,55){\idgr{30}}
 \put(30,30){\idgr{10}}
 \put(55,0){\idgr{30}}
 \put(80,0){\idgr{10}}
 \put(95,0){\idgr{5}}
 \put(80,75){\idgr{10}}
 \put(90,20){\idgr{10}}
 \put(135,0){\idgr{45}}
 \put(150,0){\idgr{10}}
 \put(145,55){\idgr{10}}
 \put(155,35){\idgr{10}}
 \put(165,0){\idgr{5}}
 \put(165,80){\idgr{5}}
 \put(175,40){\idgr{30}}
 \put(185,30){\idgr{20}}
 \put(210,0){\idgr{25}}
 \put(220,0){\idgr{25}}
 \put(222,70){\idgr{15}}
 \put(230,0){\idgr{30}}
 \put(260,0){\idgr{20}}
 \put(270,0){\idgr{20}}
 \put(280,0){\idgr{40}}
 \put(275,50){\idgr{35}}
 \put(15,85){\objo{C}}
 \put(80,85){\objo{C}}
 \put(165,85){\objo{C}}
 \put(222,85){\objo{C}}
 \put(275,85){\objo{C}}
 \put(40,45){\objo{=}}
 \put(120,45){\objo{=}}
 \put(200,40){\objo{=}}
 \put(250,40){\objo{=}}
 \put(0,0){\obju{C}}
 \put(10,0){\obju{C}}
 \put(25,0){\obju{H}}
 \put(55,0){\obju{C}}
 \put(80,0){\obju{C}}
 \put(95,0){\obju{H}}
 \put(135,0){\obju{C}}
 \put(150,0){\obju{C}}
 \put(165,0){\obju{H}}
 \put(210,0){\obju{C}}
 \put(220,0){\obju{C}}
 \put(230,0){\obju{H}}
 \put(260,0){\obju{C}}
 \put(270,0){\obju{C}}
 \put(280,0){\obju{H}}
 \egr
 and preserves the counit of $C$
 \bgr{120}{60}
 \put(30,15){\usebox{\rightcoadj}}
 \put(85,35){\comult}
 \put(85,20){\mult}
 \put(0,15){\mor{\varepsilon}{10}}
 \put(35,5){\mor{\varepsilon}{10}}
 \put(120,30){\mor{\varepsilon}{10}}
 \put(90,40){\mor{z}{20}}
 \put(85,25){\mor{S}{10}}
 \put(120,15){\mor{\mu}{10}}
 \put(0,30){\multmor{\ad}{10}}
 \put(0,25){\idgr{5}}
 \put(5,40){\idgr{20}}
 \put(10,0){\idgr{30}}
 \put(42,55){\idgr{5}}
 \put(50,0){\idgr{15}}
 \put(90,0){\idgr{20}}
 \put(95,25){\idgr{10}}
 \put(120,0){\idgr{15}}
 \put(120,40){\idgr{20}}
 \put(5,60){\objo{C}}
 \put(42,60){\objo{C}}
 \put(90,60){\objo{C}}
 \put(120,60){\objo{C}}
 \put(25,30){\objo{=}}
 \put(70,30){\objo{=}}
 \put(105,30){\objo{=}}
 \put(10,0){\obju{H}}
 \put(50,0){\obju{H}}
 \put(90,0){\obju{H}}
 \put(120,0){\obju{H}}
 \egr
 \end{pf}

 Now we want to slightly generalize the notion of a right coadjoint
coaction to the case where $H$ is only a bialgebra.
 \begin{lma} \label{rcoadjact}
 Let $H$ be a Hopf algebra, $C$ be a coalgebra, and $z:C \=> H$ be a
coalgebra morphism. A right coaction $\ad: C \=> C \tensor H$ is the
right coadjoint coaction iff
 \bgr{50}{40}
 \put(0,5){\usebox{\halfadj}}
 \put(40,25){\comult}
 \put(50,15){\mor{z}{10}}
 \put(0,0){\idgr{5}}
 \put(15,0){\idgr{5}}
 \put(40,0){\idgr{25}}
 \put(50,0){\idgr{15}}
 \put(8,35){\idgr{5}}
 \put(45,30){\idgr{10}}
 \put(8,40){\objo{C}}
 \put(45,40){\objo{C}}
 \put(32,20){\objo{=}}
 \put(0,0){\obju{C}}
 \put(15,0){\obju{H}}
 \put(40,0){\obju{C}}
 \put(50,0){\obju{H}}
 \egr
 \end{lma}

 \begin{pf}
 If $\ad$ is the right coadjoint coaction then the equation of the
lemma holds by

 \bgr{150}{70}
 \put(0,20){\usebox{\halfadj}}
 \put(45,20){\usebox{\rightcoadj}}
 \put(90,40){\comult}
 \put(110,55){\comult}
 \put(140,40){\comult}
 \put(95,60){\bcomult{20}}
 \put(40,60){\bcomult{20}}
 \put(90,25){\mult}
 \put(50,5){\bmult{15}}
 \put(105,10){\bmult{15}}
 \put(40,10){\braid}
 \put(95,15){\braid}
 \put(105,25){\twist{5}{10}}
 \put(40,20){\mor{z}{40}}
 \put(95,45){\mor{z}{15}}
 \put(100,30){\mor{S}{10}}
 \put(120,15){\mor{z}{40}}
 \put(150,0){\mor{z}{40}}
 \put(0,0){\idgr{20}}
 \put(7,50){\idgr{20}}
 \put(15,0){\idgr{20}}
 \put(40,0){\idgr{10}}
 \put(60,0){\idgr{5}}
 \put(50,65){\idgr{5}}
 \put(65,10){\idgr{10}}
 \put(95,0){\idgr{15}}
 \put(90,30){\idgr{10}}
 \put(105,65){\idgr{5}}
 \put(110,0){\idgr{10}}
 \put(110,35){\idgr{20}}
 \put(140,0){\idgr{40}}
 \put(145,45){\idgr{25}}
 \put(7,70){\objo{C}}
 \put(50,70){\objo{C}}
 \put(105,70){\objo{C}}
 \put(145,70){\objo{C}}
 \put(30,35){\objo{=}}
 \put(83,35){\objo{=}}
 \put(133,35){\objo{=}}
 \put(0,0){\obju{C}}
 \put(15,0){\obju{H}}
 \put(40,0){\obju{C}}
 \put(60,0){\obju{H}}
 \put(95,0){\obju{C}}
 \put(110,0){\obju{H}}
 \put(140,0){\obju{C}}
 \put(150,0){\obju{H}}
 \egr

 Conversely if this equality holds then the right coaction $\ad: C \=>
C \tensor H$ is the right coadjoint coaction since
 \bgr{170}{70}
 \put(0,10){\usebox{\rightcoadj}}
 \put(60,30){\usebox{\halfadj}}
 \put(105,40){\comult}
 \put(50,60){\bcomult{18}}
 \put(110,60){\bcomult{20}}
 \put(105,25){\mult}
 \put(120,10){\mult}
 \put(60,5){\bmult{15}}
 \put(50,20){\braid}
 \put(110,15){\braid}
 \put(120,25){\twist{5}{10}}
 \put(130,15){\twist{5}{10}}
 \put(50,30){\mor{z}{20}}
 \put(60,10){\mor{S}{10}}
 \put(110,50){\mor{z}{10}}
 \put(115,30){\mor{S}{10}}
 \put(125,50){\multmor{\ad}{10}}
 \put(160,30){\multmor{\ad}{10}}
 \put(5,0){\idgr{10}}
 \put(20,0){\idgr{10}}
 \put(12,50){\idgr{20}}
 \put(50,0){\idgr{20}}
 \put(50,50){\idgr{10}}
 \put(60,65){\idgr{5}}
 \put(65,0){\idgr{5}}
 \put(75,10){\idgr{20}}
 \put(105,30){\idgr{10}}
 \put(110,0){\idgr{15}}
 \put(110,45){\idgr{5}}
 \put(120,65){\idgr{5}}
 \put(125,0){\idgr{10}}
 \put(125,35){\idgr{15}}
 \put(135,25){\idgr{25}}
 \put(160,0){\idgr{30}}
 \put(170,0){\idgr{30}}
 \put(165,40){\idgr{30}}
 \put(12,70){\objo{C}}
 \put(60,70){\objo{C}}
 \put(120,70){\objo{C}}
 \put(165,70){\objo{C}}
 \put(38,33){\objo{=}}
 \put(92,33){\objo{=}}
 \put(145,33){\objo{=}}
 \put(5,0){\obju{C}}
 \put(20,0){\obju{H}}
 \put(50,0){\obju{C}}
 \put(65,0){\obju{H}}
 \put(110,0){\obju{C}}
 \put(125,0){\obju{H}}
 \put(160,0){\obju{C}}
 \put(170,0){\obju{H}}
 \egr
 \end{pf}

 \subsubsection{} \label{coadjcoact}
 We say that a coaction $\ad: C \=> C \tensor B$ for a given $z: C \=>
B$, $B$ a bialgebra, is a {\em right coadjoint coaction} if the
equation in Lemma \ref{rcoadjact} holds.

 More generally if $z: C \=> B$ is $\star$-invertible then a coadjoint
coaction can be constructed in the same way as above. We don't know if
there are more general conditions for $z: C \=> B$ such that a right
coadjoint coaction exists nor whether it is unique then.

 \subsubsection{}
 In the dual situation let $f:H \=> A$ be an algebra homomorphism with
a Hopf algebra $H$. The right adjoint action $ah = \sum f(S(h_1))
\cdot a \cdot f(h_2)$ is characterized by the equation $\sum f(h_1)
\cdot (ah_2) = a \cdot f(h)$.

 \subsection{$\C_0$-generated coalgebras}

 We still need another somewhat more general setup. Let $\C$ be a
monoidal category, $\C_0$ be a full monoidal subcategory of $\C$. In
this situation we consider a special type of coalgebra in $\C$.

 \begin{defn}
 Let $C \in \C$ be a coalgebra satisfying the following conditions:
 \begin{enumerate}
 \item $C$ is a colimit in $\C$ of a diagram of objects $C_i$ in
$\C_0$.
 \item All morphisms $X \tensor \iota_i \tensor M:X \tensor C_i
\tensor M \=> X \tensor C \tensor M$ are monomorphisms in $\C$ where
$X \in C_0$, $M \in \C$ and the $\iota_i: C_i \=> C$ are the
injections of the colimit diagram.
 \item Every $C_i$ is a subcoalgebra of $C$ via $\iota_i: C_i \=> C$.
 \item If $(P,\delta_P: P \=> P \tensor C)$ is a comodule over $C$ and
$P \in \C_0$, then there exists a $C_i$ in the diagram for $C$ and a
morphism $\delta_{P,i}:P \=> P \tensor C_i$ such that
 $$\bfig
 \putmorphism(0, 400)(1, 0)[P`P \tensor C_i`\sst \delta_{P,i}]{400}1a
 \putmorphism(0, 400)(1, -1)[`P \tensor C`\sst \delta_P]{400}1l
 \putmorphism(400, 400)(0, -1)[``\sst 1 \tensor \iota_i]{400}1r
 \efig$$
 commutes.
 \end{enumerate}
 Then $C$ is called a $\C_0${\em -generated coalgebra}.
 \end{defn}

 \subsubsection{}
 If $\C_0 = \C$ then the conditions in the previous definition are
trivially satisfied. If $\C = \Vek$ and $C_0 = \vek$, the category of
finite-dimensional vector spaces, then every coalgebra in $\C$ is a
$\C_0$-generated coalgebra by the fundamental theorem for coalgebras
(\cite{SW} Thm. 2.2.1) and its generalization to the fundamental
theorem for comodules.

 \subsubsection{}
 We denote by $\C_0^C$ the category of $C$-comodules in $\C_0$. Then
$\C_0^C$ is a $\C_0$-category and the forgetful functor $\omega:
\C_0^C \=> \C_0$ is a $\C_0$-functor.

 \subsubsection{}
 It is an easy exercise to show for a $\C_0$-generated coalgebra, that
the $(P,\delta_{P,i}: P \=> P \tensor C_i)$ are comodules.

 \section{Reconstruction properties}

 For the rest of this paper let the control category $\C$ be a braided
monoidal category and the base category $\A$ be a $\C$-monoidal
category.

 \subsection{Reconstruction of coalgebras}

 \begin{defn}
 We define the category $\overline\J(\C)$ of all $\C$-categories
``over'' $\A$ as follows. The objects are pairs $(\B,\omega)$
consisting of a $\C$-category $\B$ and of a $\C$-functor $\omega: \B
\=> \A$. A morphism $[\chi,\zeta]: (\B,\omega) \=> (\B',\omega')$ is
an equivalence class of pairs $(\chi,\zeta)$ with $\chi: \B \=> \B'$ a
$\C$-functor and $\zeta: \omega \=> \omega'\chi$ a $\C$-isomorphism.
Two such pairs $(\chi,\zeta)$ and $(\chi',\zeta')$ are equivalent if
there is a $\C$-isomorphism $\varphi: \chi \=> \chi'$ with $\zeta' =
\omega'\varphi \circ \zeta$. Composition is given by
$[\chi',\zeta']\circ[\chi,\zeta] = [\chi'\chi, \zeta'\chi \circ
\zeta]$.

 Let $\A(\C)$ be a full subcategory of $\overline\J(\C)$.
 \end{defn}

 \subsubsection{} \label{coalgcat}
 Theorem \ref{comodcats} defines a functor $\A^\x : \A\coalg \=>
\overline\J(\C)$ by $\A^\x (C) := (\A^C,\omega)$ where $\omega$ is the
forgetful functor. Furthermore $\A^\x (z) := [\A^z,\id]$.

 \subsubsection{}
 If $\A_0$ is a full $\C$-monoidal subcategory of $\A$, then we define
the full subcategory $\J_0(\C)$ of $\overline\J(\C)$ to consist of
those $\C$-categories $\B$ over $\A$ whose forgetful functor $\omega:
\B \=> \A$ factors through $\A_0$.

 \subsubsection{}
 In this case we obtain a functor $\A_0^\x : \A\coalg \=> \J_0(\C)$ by
$\A_0^\x (C) := (\A_0^C,\omega)$ where $\A_0^C$ denotes the full
subcategory of $\A^C$ of those $C$-comodules whose underlying object
is in $\A_0$.

 \bigskip
 Now we address the question which properties of an algebra $A$ or a
coalgebra $C$ in a monoidal category $\A$ can be recovered from the
category of its modules $\A_A$ resp. comodules $\A^C$. Since
reconstruction of coalgebras is somewhat simpler (see Theorem
\ref{finiterec}) we will perform the reconstruction of a coalgebra
$C$ from $\A^C$ explicitly and derive the reconstruction of an algebra
by duality. As we remarked in the introduction $C$ is not uniquely
determined by $\A^C$. But if we use additional information about the
forgetful $\C$-functor $\omega: \A^C \=> \A$ we can reconstruct $C$ up
to isomorphism with its full structure.

 In many cases we can actually ``reconstruct'' a coalgebra $C$ from a
fairly arbitrary $\C$-functor $\omega: \B \=> \A$. We will postpone
the discussion of how to obtain the {\em object} $C \in \A$ from a
functor $\omega: \B \=> \A$ to the next section. In this section we
will give the general definition and discuss the {\em structure} of
such a reconstructed object $C$.

 A different point of view is how to find a left adjoint functor to
the functor $\A^\x : \A\coalg \=> \J(\C)$. If such a left adjoint
functor does not exist ``globally'', it might still exist ``locally'',
i.e. a certain functor is representable.

 \begin{defn}
 Let $\B$ be a $\C$-category and $\omega: \B \=> \A$ be a
 $\C$-functor. Then the sets $\Nat_\C(\omega,\omega \tensor M)$ depend
functorially on $M \in \A$, i.e.~we have a functor
 $$\Nat_\C(\omega,\omega \tensor \X ): \A \=> \Set.$$
 If this functor is representable then the representing object will be
denoted by $\coend_\C(\omega)$. It is unique up to isomorphism. (In
the dual situation a representing object for $\Nat_\C(\omega \tensor
\X ,\omega)$ will be denoted by $\rend_\C(\omega)$.) So we have
 $$\Nat_\C(\omega, \omega \tensor M) \iso \A(\coend_\C(\omega),M).$$
 The universal arrow for this functor
 $$\delta: \omega \=> \omega \tensor \coend_\C(\omega)$$
 is a $\C$-morphism, the image of the identity in
$\A(\coend_\C(\omega), \coend_\C(\omega))$. It solves the following
universal problem
 \begin{itemize}
 \item for every  $M \in \A$ and every $\C$-morphism $\varphi: \omega
\=> \omega \tensor M$ there is a unique morphism $f: C \=> M$ such
that
 $$\bfig
 \putmorphism(0, 400)(1, 0)[\omega`\omega \tensor C`\sst
\delta]{400}1a
 \putmorphism(400, 400)(0, -1)[`\omega \tensor M`\sst 1 \tensor
f]{400}1r
 \putmorphism(0, 400)(1, -1)[``\sst \varphi]{400}1l
 \efig$$ commutes.
 \end{itemize}
 This universal property is in fact equivalent to the representability
of $\Nat_\C(\omega, \omega \tensor \X )$ and induces a universal
factorization of $\omega$ through the category of comodules $\A^C$.
 \end{defn}

 The study of $\C$-functors as conducted here has many properties in
common with similar results for general functors. In fact general
categories, functors and natural transformations may also be
considered as $\C$-categories, $\C$-functors, resp. $\C$-morphisms for
the monoidal category $\C$ with one object $I$ and one morphism
$\id_I$.

 Many of the following propositions are well known for the case of a
monoidal category $\C = \{I\}$ and can be proved by standard universal
abstract nonsense. So we only sketch the idea of the proofs. There
are, however, subtle difficulties and restrictions with respect to
braidings that do not occur in the case of a symmetric control
category $\C$.

 \begin{prop}\label{fuprcoalgebra}
 If \ $\Nat_\C(\omega, \omega \tensor \X )$ is representable, then the
representing object $C = \coend_\C(\omega)$ is a coalgebra in $\A$.
This coalgebra is uniquely determined up to isomorphisms of
coalgebras.

 Furthermore every object $\omega(P) \in \A$ with $P \in \B$ is a
 $C$-comodule via $\delta: \omega(P) \=> \omega(P) \tensor C$ and
every morphism $\omega(f)$ is a morphism of $C$-comodules.

 Every object $\omega(X \tensor P) \in \A$ with $X \in \C$ and $P \in
\B$ is isomorphic as a $C$-comodule to $X \tensor \omega(P)$ with the
structure induced by $\omega(P)$.
 \end{prop}

 \begin{pf}
 The comultiplication $\Delta$ and counit $\varepsilon$ are uniquely
defined by $(1_\omega \tensor \Delta) \delta = (\delta \tensor
1_C)\delta$ and $(1_\omega \tensor \varepsilon)\delta =
 \rho^{-1}_\omega$. The last claim follows since $\delta$ is a
 $\C$-morphism.
 \end{pf}

 We will encounter situations of comodules $(P,\vartheta: P \=> P
\tensor C)$ in $\A$ where we want to know if this comodule comes about
as in the previous Proposition. So we define

 \begin{defn}
 Let \ $\Nat_\C(\omega, \omega \tensor \X )$ be representable by $C
\in \A$. Then a comodule $(P, \vartheta: P \=> P \tensor C)$ in $\A$
can be {\em lifted along} $\omega$ if there is an object $Q \in \B$
and a comodule isomorphism $(\omega(Q),\delta) \iso (P, \vartheta)$.
In this case the comodule $(P, \vartheta)$ is called {\em liftable
along} $\omega$ and $Q \in B$ a {\em lifting}.
 \end{defn}

 \subsection{Reconstruction of bialgebras}

 Assume now, that the base category $\A$ is a $\C$-braided
 $\C$-mo\-noid\-al category. (It will be clear from the context which
tensor product is being used, so we simply use $\tensor$ for the
tensor product in $\A$.)

 Consider a $\C$-functor $\omega: \B \=> \A$. Then the bifunctors
$\omega \tensor \omega = \omega^2: \B \times \B \=> \A$ and  $\omega^2
\tensor M: \B \times \B \=> \A$ are $\C$-bifunctors as can be easily
checked. The sets $\Nat_\C(\omega^2,\omega^2 \tensor M)$ of
 $\C$-bimorphisms depend functorially on $M$, i.e.~we have a functor
 $$\Nat_\C(\omega^2,\omega^2 \tensor \X ): \A \=> \Set.$$
 Let $\Nat_\C(\omega, \omega \tensor \X )$ be representable with
universal $\C$-morphism $\delta: \omega \=> \omega \tensor C$. In
general the morphism
 $$\delta_2 := (1_\omega \tensor \sigma_{\coend_\C(\omega),\omega}
\tensor 1_{\coend_\C(\omega)})(\delta \tensor \delta): \omega^2 \=>
\omega^2 \tensor \coend_\C(\omega)^2$$
 will not be a $\C$-bimorphism. This is, however, the case if $C =
\coend_\C(\omega)^2$ is $\C$-central (see Definition \ref{central}).
Similarly $\delta^{n}$ is a $\C$-multimorphism if $C$ is $\C$-central.

 \begin{defn}
 If the functor $\Nat_\C(\omega, \omega \tensor \X )$ is representable
with universal $\C$-morphism $\delta: \omega \=> \omega \tensor C$, if
$C$ is $\C$-central and if $\Nat_\C(\omega^2,\omega^2 \tensor \X )$ is
also representable with the special universal $\C$-bimorphism
 $$\delta_2 := (1_\omega \tensor \sigma_{C,\omega} \tensor
1_{C})(\delta \tensor \delta): \omega \tensor \omega \=> \omega
\tensor \omega \tensor C \tensor C$$
 then we say that $\Nat_\C(\omega,\omega \tensor \X )$ is {\em
birepresentable}.

 In a similar way we proceed for the multifunctor $\omega \tensor
\ldots \tensor \omega = \omega^n$. If $C$ is $\C$-central and the
functor $\Nat_\C(\omega^n,\omega^n \tensor \X )$ is representable with
the universal morphism
 $$\delta^{(n)} := \tau (\delta \tensor \ldots \tensor \delta):
\omega^n \=> \omega^n \tensor \coend_\C(\omega)^n$$
 with the obvious choice of $\tau \in B_{2n}$, the Artin braid group,
then we say that $\Nat_\C(\omega,\omega \tensor \X )$ is
 $n${\em -representable}. If this holds for all $n \in {\Bbb N}$ we
say that the functor $\Nat_\C(\omega,\omega \tensor \X )$ is {\em
multirepresentable} (fully representable in \cite{MAJ2}).
 \end{defn}

 \begin{prop} \label{fuprbialgebra}
 Let $\A$ be $\C$-braided $\C$-monoidal, $\B$ be $\C$-monoidal and
$\omega: \B \=> \A$ be a $\C$-monoidal functor. If $\Nat_\C(\omega,
\omega \tensor \X )$ is multirepresentable, then $B :=
\coend_\C(\omega)$ is a bialgebra in $\A$. This bialgebra is uniquely
determined up to isomorphisms of bialgebras.

 If in addition $\B$ is $\C$-braided ($\omega$ will usually not
preserve the braiding), then $\coend_\C(\omega)$ is coquasitriangular
in $\A$.

 If $\omega$ factors through the full subcategory $\A_0$ of rigid
objects in $\A$ then $\coend_\C(\omega)$ is a Hopf algebra in $\A$.

 Furthermore for any objects $P,Q \in \B$ the $B$-comodule structure
on $\omega'(P) \tensor \omega'(Q)$ is defined by the multiplication on
$B$.
 \end{prop}

 \begin{pf}
 Similar to \cite{MAJ4} Theorem 3.2. resp. 3.11. We check only that
the relevant morphisms that are factored through the universal
morphisms are $\C$-morphisms.

 The multiplication of $B$ is defined by the $\C$-bimorphism
$\delta'_{P \tensor Q}: \omega(P) \tensor \omega(Q) \iso \omega(P
\tensor Q) \=> \omega(P \tensor Q) \tensor B$ as the uniquely
determined morphism $\widetilde m_B: B \tensor B \=> B$ such that
 $$(1_{\omega(P)} \tensor 1_{\omega(Q)} \tensor \widetilde
m)(1_{\omega(P)} \tensor \sigma_{B,Q} \tensor 1_B)(\delta_P \tensor
\delta_Q) = \delta'_{P \tensor Q}.$$

 The morphism $\delta'_{P \tensor Q \tensor R}: \omega(P) \tensor
\omega(Q) \tensor \omega(R) \=> \omega(P \tensor Q \tensor R) \tensor
B$ responsible for associativity is a $\C$-trimorphism.

 The coquasitriangular structure $r: B \tensor B \=> I$ is defined by
the $\C$-bimorphism $\sigma_\A^{-1} (\omega(Q),\omega(P)) \omega(
\sigma_\B(P,Q)): \omega(P) \tensor \omega(Q) \=> \omega(P) \tensor
\omega(Q) \tensor I$ and the braid equation
 \bgr{60}{40}
 \put(40,15){\braid}
 \put(0,10){\ibraid}
 \put(0,20){\brmor}
 \put(30,25){\multmor{\delta}{10}}
 \put(50,25){\multmor{\delta}{10}}
 \put(50,5){\multmor{r}{10}}
 \put(0,0){\idgr{10}}
 \put(0,30){\idgr{10}}
 \put(10,0){\idgr{10}}
 \put(10,30){\idgr{10}}
 \put(30,0){\idgr{25}}
 \put(35,35){\idgr{5}}
 \put(40,0){\idgr{15}}
 \put(55,35){\idgr{5}}
 \put(60,15){\idgr{10}}
 \put(20,20){\objo{=}}
 \put(0,40){\objo{P}}
 \put(10,40){\objo{Q}}
 \put(35,40){\objo{P}}
 \put(55,40){\objo{Q}}
 \put(0,0){\obju{P}}
 \put(10,0){\obju{Q}}
 \put(30,0){\obju{P}}
 \put(40,0){\obju{Q}}
 \egr
 where the braid marked with a circle represents the braiding of the
category $\B$. This diagram represents the equation
 $$\sigma_\A^{-1}(\omega(Q),\omega(P))\omega(\sigma_\B(P,Q)) =
 (1_{\omega(P)} \tensor 1_{\omega(Q)} \tensor r)(1_{\omega(P)} \tensor
\sigma_\A(B,\omega(Q)) \tensor 1_B)(\delta_P \tensor \delta_Q).$$
 Observe that the braiding of $\A^B$ for a braided bialgebra $B$ is
described by the equation
 \bgr{110}{65}
 \put(30,10){\braid}
 \put(40,30){\braid}
 \put(85,30){\braid}
 \put(0,25){\brmor}
 \put(80,20){\twist{5}{10}}
 \put(95,40){\twist{5}{10}}
 \put(30,40){\multmor{\delta}{10}}
 \put(50,40){\multmor{\delta}{10}}
 \put(90,20){\multmor{\delta}{10}}
 \put(100,50){\multmor{\delta}{10}}
 \put(50,20){\multmor{r}{10}}
 \put(100,5){\multmor{r}{10}}
 \put(30,0){\idgr{10}}
 \put(30,20){\idgr{20}}
 \put(35,50){\idgr{15}}
 \put(40,0){\idgr{10}}
 \put(40,20){\idgr{10}}
 \put(55,50){\idgr{15}}
 \put(60,30){\idgr{10}}
 \put(80,0){\idgr{20}}
 \put(85,40){\idgr{25}}
 \put(90,0){\idgr{20}}
 \put(100,15){\idgr{5}}
 \put(105,60){\idgr{5}}
 \put(110,15){\idgr{35}}
 \put(0,0){\idgr{25}}
 \put(0,35){\idgr{30}}
 \put(10,0){\idgr{25}}
 \put(10,35){\idgr{30}}
 \put(35,65){\objo{P}}
 \put(55,65){\objo{Q}}
 \put(85,65){\objo{P}}
 \put(105,65){\objo{Q}}
 \put(0,65){\objo{P}}
 \put(10,65){\objo{Q}}
 \put(70,30){\objo{=}}
 \put(20,30){\objo{=}}
 \put(30,0){\obju{P}}
 \put(40,0){\obju{Q}}
 \put(80,0){\obju{P}}
 \put(90,0){\obju{Q}}
 \put(0,0){\obju{P}}
 \put(10,0){\obju{Q}}
 \egr

 Finally the antipode is defined by the morphism
 $(\omega(\ev_P) \tensor \sigma_{B,\omega(P)})
 (1_{\omega(P)} \tensor \delta_{\omega(P)^*} \tensor 1_{\omega(P)})
 (1_{\omega(P)} \tensor \omega(\db_P)): \omega(P) \=> \omega(P)
\tensor B$. To show that this is a $\C$-morphism, we first show the
following claim. If $P \in \B$ defines a trivial $B$-comodule then
$\omega(P)^*$ is also trivial. This follows from
 \bgr{260}{40}
 \put(110,5){\mult}
 \put(170,5){\mult}
 \put(80,15){\eval}
 \put(140,15){\eval}
 \put(200,15){\eval}
 \put(30,15){\eval}
 \put(210,30){\coeval}
 \put(40,30){\coeval}
 \put(90,30){\bcoeval{20}}
 \put(150,30){\bcoeval{20}}
 \put(100,10){\braid}
 \put(160,10){\braid}
 \put(160,20){\mor{u}{10}}
 \put(60,30){\mor{u}{10}}
 \put(10,30){\mor{u}{10}}
 \put(90,20){\multmor{\delta}{10}}
 \put(110,20){\multmor{\delta}{10}}
 \put(170,20){\multmor{\delta}{10}}
 \put(220,20){\multmor{\delta}{10}}
 \put(250,20){\multmor{\delta}{10}}
 \put(0,0){\idgr{40}}
 \put(10,0){\idgr{30}}
 \put(30,20){\idgr{20}}
 \put(40,20){\idgr{10}}
 \put(50,0){\idgr{30}}
 \put(60,0){\idgr{30}}
 \put(80,20){\idgr{20}}
 \put(100,0){\idgr{10}}
 \put(115,0){\idgr{5}}
 \put(120,10){\idgr{10}}
 \put(140,20){\idgr{20}}
 \put(150,20){\idgr{10}}
 \put(160,0){\idgr{10}}
 \put(175,0){\idgr{5}}
 \put(180,10){\idgr{10}}
 \put(200,20){\idgr{20}}
 \put(210,20){\idgr{10}}
 \put(220,0){\idgr{20}}
 \put(230,0){\idgr{20}}
 \put(250,0){\idgr{20}}
 \put(250,30){\idgr{10}}
 \put(260,0){\idgr{20}}
 \put(0,40){\objo{P^*}}
 \put(30,40){\objo{P^*}}
 \put(80,40){\objo{P^*}}
 \put(140,40){\objo{P^*}}
 \put(200,40){\objo{P^*}}
 \put(250,40){\objo{P^*}}
 \put(20,20){\objo{=}}
 \put(70,20){\objo{=}}
 \put(130,20){\objo{=}}
 \put(190,20){\objo{=}}
 \put(240,20){\objo{=}}
 \put(0,0){\obju{P^*}}
 \put(10,0){\obju{B}}
 \put(50,0){\obju{P^*}}
 \put(60,0){\obju{B}}
 \put(100,0){\obju{P^*}}
 \put(115,0){\obju{B}}
 \put(160,0){\obju{P^*}}
 \put(175,0){\obju{B}}
 \put(220,0){\obju{P^*}}
 \put(230,0){\obju{B}}
 \put(250,0){\obju{P^*}}
 \put(260,0){\obju{B}}
 \egr
 Observe that we have $X \tensor P \iso X \tensor (I \tensorhat P)
\iso (X \tensor I) \tensorhat P = \widetilde X \tensorhat P$ in $\B$
for $\widetilde X := X \tensor I$ which gives trivial $B$-comodules
$\omega(\widetilde X) \iso X \tensor \omega(I)$ and $\omega(\widetilde
X)^*$. So the diagram (where $X$ denotes $\omega(\widetilde X)$ and
$P$ denotes $\omega(P)$)
 \bgr{340}{70}
 \put(110,35){\mult}
 \put(190,35){\mult}
 \put(10,45){\eval}
 \put(80,45){\eval}
 \put(160,45){\eval}
 \put(240,45){\eval}
 \put(310,45){\eval}
 \put(0,35){\beval{30}}
 \put(70,35){\beval{30}}
 \put(150,35){\beval{30}}
 \put(230,35){\beval{30}}
 \put(270,50){\coeval}
 \put(90,65){\bcoeval{50}}
 \put(170,65){\bcoeval{50}}
 \put(20,65){\bcoeval{40}}
 \put(250,60){\bcoeval{40}}
 \put(30,60){\bcoeval{20}}
 \put(110,60){\bcoeval{20}}
 \put(190,60){\bcoeval{20}}
 \put(320,60){\bcoeval{20}}
 \put(40,40){\braid}
 \put(50,30){\braid}
 \put(100,40){\braid}
 \put(120,20){\braid}
 \put(130,10){\braid}
 \put(180,40){\braid}
 \put(200,20){\braid}
 \put(210,10){\braid}
 \put(260,40){\braid}
 \put(270,30){\braid}
 \put(280,20){\braid}
 \put(330,40){\braid}
 \put(120,30){\twist{-5}{5}}
 \put(200,30){\twist{-5}{5}}
 \put(200,50){\mor{u}{10}}
 \put(90,50){\multmor{\delta}{10}}
 \put(110,50){\multmor{\delta}{10}}
 \put(170,50){\multmor{\delta}{10}}
 \put(250,50){\multmor{\delta}{10}}
 \put(320,50){\multmor{\delta}{10}}
 \put(20,50){\multmor{\delta}{20}}
 \put(0,40){\idgr{30}}
 \put(10,50){\idgr{20}}
 \put(20,60){\idgr{5}}
 \put(30,40){\idgr{10}}
 \put(40,0){\idgr{40}}
 \put(50,0){\idgr{30}}
 \put(50,50){\idgr{10}}
 \put(60,0){\idgr{30}}
 \put(60,40){\idgr{25}}
 \put(70,40){\idgr{30}}
 \put(80,50){\idgr{20}}
 \put(90,60){\idgr{5}}
 \put(120,0){\idgr{20}}
 \put(120,40){\idgr{10}}
 \put(130,0){\idgr{10}}
 \put(130,30){\idgr{30}}
 \put(140,0){\idgr{10}}
 \put(140,20){\idgr{45}}
 \put(150,40){\idgr{30}}
 \put(160,50){\idgr{20}}
 \put(170,60){\idgr{5}}
 \put(190,50){\idgr{10}}
 \put(200,40){\idgr{10}}
 \put(200,0){\idgr{20}}
 \put(210,0){\idgr{10}}
 \put(210,30){\idgr{30}}
 \put(220,0){\idgr{10}}
 \put(220,20){\idgr{45}}
 \put(230,40){\idgr{30}}
 \put(240,50){\idgr{20}}
 \put(270,0){\idgr{30}}
 \put(280,0){\idgr{20}}
 \put(280,40){\idgr{10}}
 \put(290,0){\idgr{20}}
 \put(290,30){\idgr{30}}
 \put(300,0){\idgr{70}}
 \put(310,50){\idgr{20}}
 \put(330,0){\idgr{40}}
 \put(340,0){\idgr{40}}
 \put(340,50){\idgr{10}}
 \put(0,70){\objo{X}}
 \put(10,70){\objo{P}}
 \put(70,70){\objo{X}}
 \put(80,70){\objo{P}}
 \put(150,70){\objo{X}}
 \put(160,70){\objo{P}}
 \put(230,70){\objo{X}}
 \put(240,70){\objo{P}}
 \put(300,70){\objo{X}}
 \put(310,70){\objo{P}}
 \put(65,30){\objo{=}}
 \put(145,30){\objo{=}}
 \put(225,30){\objo{=}}
 \put(295,30){\objo{=}}
 \put(40,0){\obju{X}}
 \put(50,0){\obju{P}}
 \put(60,0){\obju{B}}
 \put(120,0){\obju{X}}
 \put(130,0){\obju{P}}
 \put(140,0){\obju{B}}
 \put(200,0){\obju{X}}
 \put(210,0){\obju{P}}
 \put(220,0){\obju{B}}
 \put(270,0){\obju{X}}
 \put(280,0){\obju{P}}
 \put(290,0){\obju{B}}
 \put(300,0){\obju{X}}
 \put(330,0){\obju{P}}
 \put(340,0){\obju{B}}
 \egr
 shows that the morphism  $(\omega(\ev_P) \tensor
\sigma_{B,\omega(P)})
 (1_{\omega(P)} \tensor \delta_{\omega(P)^*} \tensor 1_{\omega(P)})
 (1_{\omega(P)} \tensor \omega(\db_P)): \omega(P) \=> \omega(P)
\tensor B$ is a $\C$-morphism.
 \end{pf}

 Further interesting properties of $\A$ resp. $\omega$ for
reconstruction may be found in \cite{DR2, KT, YE}

 \subsection{Reconstruction of morphisms}

 \subsubsection{}\label{coalgmor}
 Let $(\B,\omega)$ and $(\B',\omega')$ be objects in $\J(\C)$ and let
$[\chi,\zeta]: (\B,\omega) \=> (\B',\omega')$ be a morphism in
$\J(\C)$. Then $\omega: \B \=> \A$, $\omega': \B' \=> \A$, and $\chi:
\B \=> \B'$ are $\C$-functors, and $\zeta: \omega \iso \omega'\chi$ is
a $\C$-isomorphism for the diagram:
 $$\bfig
 \putmorphism(0, 400)(1, 0)[\B`\B'`\sst \chi]{400}1a
 \putmorphism(0, 400)(2, -3)[`\A.`\sst \omega]{200}1l
 \putmorphism(400, 400)(-2, -3)[``\sst \omega']{200}1r
 \efig$$
 Let $\delta: \omega \=> \omega \tensor C$ and $\partial: \omega' \=>
\omega' \tensor C'$ be universal $\C$-morphisms. Since $\zeta: \omega
\=> \omega'\chi$ is a $\C$-isomorphism there is a unique morphism
$\coend_\C([\chi,\zeta]) = z: C \=> C'$ such that
 $$\bfig
 \putmorphism(0, 400)(1, 0)[\omega`\omega \tensor C`\sst
\delta]{600}1a
 \putmorphism(0, 0)(1, 0)[\omega'\chi`\omega'\chi \tensor C'`\sst
\partial\chi]{600}1a
 \putmorphism(0, 400)(0, -1)[``\sst \zeta]{400}1l
 \putmorphism(600, 400)(0, -1)[``\sst \zeta \tensor z]{400}1r
 \efig$$
 commutes. If $(\chi,\zeta)$ and $(\chi',\zeta')$ are equivalent
(representatives of $[\chi,\zeta]$) by $\varphi: \chi \=> \chi'$ with
$\zeta' = \omega'\varphi \circ \zeta$ then the diagram
 \newpage
 $$\bfig
 \putmorphism(0, 800)(1, 0)[\omega`\omega \tensor C`\sst
\delta]{1000}1a
 \putmorphism(200, 400)(1, 0)[\omega'\chi`\omega'\chi \tensor C'`\sst
\partial\chi]{600}1a
 \putmorphism(0, 0)(1, 0)[\omega'\chi'`\omega'\chi' \tensor C'`\sst
\partial\chi']{1000}1a
 \putmorphism(0, 800)(1, -2)[``\sst \zeta]{200}1r
 \putmorphism(0, 800)(0, -1)[``\sst \zeta']{800}1l
 \putmorphism(200, 400)(-1, -2)[``\sst \omega'\varphi]{200}1r
 \putmorphism(1000, 800)(-1, -2)[``\sst \zeta \tensor z]{200}1l
 \putmorphism(1050, 800)(0, -1)[``\sst \zeta' \tensor z']{800}1r
 \putmorphism(800, 400)(1, -2)[``\sst \omega'\varphi \tensor 1]{200}1l
 \efig$$
 commutes. By the uniqueness of the induced morphism from $C$ to $C'$
we get $z = z'$, hence $z$ is uniquely defined by the class
$[\chi,\zeta]$.

 It is easy to see that $z$ is a coalgebra morphism.

 Observe that $\omega(P) \buildrel \delta \over \longrightarrow
\omega(P) \tensor C \buildrel 1 \tensor z \over \longrightarrow
\omega(P) \tensor C'$ defines a $C'$-coaction on $\omega(P)$ for every
$P \in \B$.

 \subsubsection{}\label{zrightcoadj}
 Let $\A$ be a $\C$-braided $\C$-monoidal category. Let $[\chi,\zeta]:
(\B,\omega) \=> (\B',\omega')$ and $\delta$ and $\partial$ be as
before. Furthermore let $\B'$ be a $\C$-monoidal category and
$\omega': \B' \=> \A$ be a $\C$-monoidal functor. Assume that
$\Nat_\C(\omega', \omega' \tensor \X )$ is multirepresentable with the
universal morphism $\partial: \omega' \=> \omega' \tensor B$. By
Proposition \ref{fuprbialgebra} $B$ is a bialgebra.

 We call $\Nat_\C(\omega,\omega \tensor \X ): \B \=> \Set$
 $\omega'${\em -representable} if there is an object $\bar C \in \B'$
and a $\C$-morphism $d: \chi \=> \chi \tensor \bar C$ such that the
induced morphism $y: C \=> \omega'\bar C$ in the commutative diagram
  $$\bfig
 \putmorphism(0, 300)(1, 0)[\omega`\omega \tensor C`\sst
\delta]{600}1a
 \putmorphism(600, 300)(1, 0)[\phantom{\omega \tensor C}`\omega
\tensor \omega' \bar C`\sst 1 \tensor y]{800}1a
 \putmorphism(0, 0)(1, 0)[\omega'\chi`\omega'(\chi \tensor \bar
C)`\sst \omega' d]{600}1a
 \putmorphism(600, 0)(1, 0)[\phantom{\omega'(\chi \tensor \bar
C)}`\omega'\chi \tensor \omega' \bar C`\sst \upsilon]{800}1a
 \putmorphism(0, 300)(0, -1)[``\sst \zeta]{300}1l
 \putmorphism(1400, 300)(0, -1)[``\sst \zeta \tensor 1]{300}1r
 \efig$$
 is an isomorphism. Equivalently the morphism
 $$\omega \buildrel \zeta \over \longrightarrow \omega'\chi \buildrel
\omega' d \over \longrightarrow \omega'(\chi \tensor \bar C) \buildrel
\upsilon \over \longrightarrow \omega'\chi \tensor \omega' \bar C
\buildrel \zeta^{-1} \tensor 1 \over \longrightarrow \omega \tensor
\omega' \bar C$$
 is a universal $\C$-morphism.

 \subsubsection{}
 Observe that every morphism $f$ in $\B$ induces a $C$-comodule
morphism $\omega(f)$ in $\A$, and that every morphism $g$ in $\B'$
induces a $B$-comodule morphism $\omega'(g)$ in $\A$. In particular
$\omega'(d) \iso \delta$ is a $B$-comodule morphism in $\A$.
Furthermore $\zeta$ is a $B$-comodule isomorphism with the
 $B$-comodule structure on $\omega$ as defined in \ref{coalgmor}.

 \bigskip
 Let $\ad: C \=> C \tensor B$ denote the coaction
 $$C
 \buildrel y \over \longrightarrow \omega'(\bar C)
 \buildrel \partial(\bar C) \over \longrightarrow \omega'(\bar C)
\tensor B
 \buildrel {y^{-1} \tensor B} \over \longrightarrow C \tensor B.$$

 \begin{prop} \label{coadjact}
 Under the conditions of \ref{zrightcoadj} $[\chi,\zeta]$ induces a
coalgebra morphism $z: C \=> B$ and the coaction of $B$ on $C$ is a
right coadjoint coaction.
 \end{prop}

 \begin{pf}
 The induced coalgebra morphism was defined in \ref{coalgmor}. For
every $P \in \B$ and $P' := \omega(P)$ we get
 \bgr{160}{65}
 \put(140,5){\usebox{\halfadj}}
 \put(10,20){\comult}
 \put(100,5){\mult}
 \put(90,10){\braid}
 \put(105,30){\twist{-10}{20}}
 \put(20,0){\mor{z}{20}}
 \put(60,0){\mor{z}{40}}
 \put(90,20){\mor{z}{15}}
 \put(0,40){\multmor{\delta}{15}}
 \put(40,10){\multmor{\delta}{10}}
 \put(45,40){\multmor{\delta}{15}}
 \put(80,35){\multmor{\delta}{10}}
 \put(85,50){\multmor{\delta}{10}}
 \put(130,40){\multmor{\delta}{15}}
 \put(100,20){\multmor{\ad}{10}}
 \put(0,0){\idgr{40}}
 \put(10,0){\idgr{20}}
 \put(8,50){\idgr{15}}
 \put(15,25){\idgr{15}}
 \put(40,0){\idgr{10}}
 \put(50,0){\idgr{10}}
 \put(45,20){\idgr{20}}
 \put(53,50){\idgr{15}}
 \put(80,0){\idgr{35}}
 \put(85,45){\idgr{5}}
 \put(90,0){\idgr{10}}
 \put(90,60){\idgr{5}}
 \put(105,0){\idgr{5}}
 \put(110,10){\idgr{10}}
 \put(130,0){\idgr{40}}
 \put(135,50){\idgr{15}}
 \put(140,0){\idgr{5}}
 \put(155,0){\idgr{5}}
 \put(145,35){\idgr{5}}
 \put(8,65){\objo{P'}}
 \put(53,65){\objo{P'}}
 \put(90,65){\objo{P'}}
 \put(135,65){\objo{P'}}
 \put(30,30){\objo{=}}
 \put(70,30){\objo{=}}
 \put(120,30){\objo{=}}
 \put(0,0){\obju{P'}}
 \put(10,0){\obju{C}}
 \put(20,0){\obju{B}}
 \put(40,0){\obju{P'}}
 \put(50,0){\obju{C}}
 \put(60,0){\obju{B}}
 \put(80,0){\obju{P'}}
 \put(90,0){\obju{C}}
 \put(105,0){\obju{B}}
 \put(130,0){\obju{P'}}
 \put(140,0){\obju{C}}
 \put(155,0){\obju{B}}
 \egr
 where the first and last equalities arise from the coassociativity of
the cooperation $\delta$ and the middle equality is the fact that (the
left lower resp. right upper) $\delta$ is a $B$-comodule morphism, the
coaction of $B$ on $P'$ as in \ref{coalgmor} and on $P' \tensor C$ via
the multiplication of $B$. Since the natural transformation $\omega'
\chi \buildrel \delta \over \longrightarrow \omega' \chi \tensor C \=>
\omega' \chi \tensor C \tensor B$ given by the above graphic diagram
induces precisely one morphism $C \=> C \tensor B$ we get
(\ref{coadjcoact}) that $\ad : C \=> C \tensor B$ is a right coadjoint
coaction.

 Furthermore we know that $C \iso \omega'(\bar C)$ is a $B$-comodule.
 \end{pf}

 The previous Proposition says in particular that $C$ with the given
(coadjoint) $B$-comodule structure can be lifted along $\omega': \B'
\=> \A$.

 \begin{cor} \label{adjcomodalg}
 Under the conditions of Proposition \ref{coadjact} if $B$ is a Hopf
algebra in $\A$ then $C$ is a $B$-comodule coalgebra under the right
coadjoint coaction defined by $z: C \=> B$.
 \end{cor}

 \begin{pf}
 By Lemma \ref{rcoadjact} $\ad: C \=> C \tensor B$ is the uniquely
defined coadjoint coaction which by Proposition \ref{coadcoalg} makes
$C$ a $B$-comodule coalgebra.
 \end{pf}

 \subsubsection{}
 If $\coend_\C(\omega)$ exists for all objects $(\B,\omega)$ in
$\J(\C)$ then we have a functor $\coend_C: \J(\C) \=> \A\coalg$. If
furthermore all full comodule categories $\A^C$ together with the
forgetful functor $\omega: \A^C \=> \A$ are objects in $\J(\C)$ then
$\coend_\C: \J(\C) \=> \A\coalg$ is left adjoint to $\A^\x : \A\coalg
\=> \J(\C)$ (from \ref{coalgcat}).

 \subsection{Applications}

 \subsubsection{}
 We specialize $\C$ to the case of a one-element category. We call
this case {\em full reconstruction}. Let $\omega: \B \=> \A$ be a
functor. Let $\delta: \omega \=> \omega \tensor C$ be a universal
morphism. Then the propositions of this section specialize to:

 If $\Nat(\omega, \omega \tensor \X )$ is representable, then the
representing object $C = \coend(\omega)$ is a coalgebra in $\A$. This
coalgebra is uniquely determined up to isomorphisms of coalgebras.

 Furthermore every object $\omega(P) \in \A$ with $P \in \B$ is a
 $C$-comodule via $\delta: \omega(P) \=> \omega(P) \tensor C$ and
every morphism $\omega(f)$ is a $C$-comodule morphism.

 Let $\A$ be braided monoidal, $\B$ be monoidal and $\omega: \B \=>
\A$ be a monoidal functor. If $\Nat(\omega, \omega \tensor \X )$ is
multirepresentable, then $B = \coend(\omega)$ is a bialgebra in $\A$.
This bialgebra is uniquely determined up to isomorphisms of
bialgebras.

 If in addition $\B$ is braided, then $\coend(\omega)$ is
coquasitriangular in $\A$.

 If $\B$ is rigid then $\coend(\omega)$ is a Hopf algebra in $\A$.

 Furthermore for any objects $P,Q \in \B$ the $B$-comodule structure
on $\omega(P) \tensor \omega(Q)$ is defined by the comultiplication on
$\coend(\omega)$.

 \subsubsection{}
 We specialize $\C = \A_0$ with $\A$ a braided monoidal category and
$\A_0$ a full (braided) monoidal subcategory. We will call this case
{\em restricted reconstruction}. Let $\omega: \B \=> \A_0$ be an
$\A_0$-functor. Let $\delta: \omega \=> \omega \tensor C$ be a
universal $\A_0$-morphisms. Then the propositions of this section
specialize to:

  If $\Nat_{\A_0}(\omega, \omega \tensor \X )$ is representable, then
the representing object $C = \coend_{\A_0}(\omega)$ is a coalgebra in
$\A$. This coalgebra is uniquely determined up to isomorphisms of
coalgebras.

 Furthermore every object $\omega(P) \in \A_0$ with $P \in \B$ is a
$C$-comodule via $\delta: \omega(P) \=> \omega(P) \tensor C$ and every
morphism $\omega(f)$ is a $C$-comodule morphism.

 Let $\B$ be $\A_0$-monoidal and $\omega: \B \=> \A_0$ be an
 $\A_0$-monoidal functor. If $\Nat_{\A_0}(\omega,\break \omega \tensor
\X )$ is multirepresentable, then $B = \coend_{\A_0}(\omega)$ is a
bialgebra in $\A$. This bialgebra is uniquely determined up to
isomorphisms of bialgebras.

 If in addition $\B$ is $\A_0$-braided, then $\coend_{\A_0}(\omega)$
is coquasitriangular in $\A$.

 If $\B$ is rigid then $\coend_{\A_0}(\omega)$ is a Hopf algebra in
$\A$.

 Furthermore for any objects $P,Q \in \B$ the $B$-comodule structure
on $\omega(P) \tensor \omega(Q)$ is defined by the multiplication on
$\coend_{\A_0}(\omega)$.

 \section{Existence theorems in reconstruction theory}

 \subsection{Restricted reconstruction}

 In this section we will study reconstruction of a given coalgebra $C$
with help of the functor $\Nat_{\C_0}(\omega,\omega \tensor \X )$. We
call this {\em restricted reconstruction}.

 For this purpose let $\C$ be a braided monoidal category and $\C_0$
be a full (braided) monoidal subcategory of $\C$. For a Hopf algebra
$H$ in $\C$, a coalgebra $C$ in $\C$ and a coalgebra morphism $z: C
\=> H$ we know that $C$ is a coalgebra in $\C^H$ with respect to the
right coadjoint coaction $\ad: C \=> C \tensor H$ by Proposition
\ref{coadcoalg}. If $C$ is a $\C_0$-generated coalgebra then the
subcoalgebras $C_i$ are also in $\C^H$ (actually in $\C_0^H$) by the
coadjoint action. We consider the underlying functor $\omega: \C_0^C
\=> \C^H$.

 \begin{thm} \label{reccoalg}
 Let $C$ be a $\C_0$-generated coalgebra in $\C$ and $H$ be a Hopf
algebra in $\C$. Let $z: C \=> H$ be a coalgebra morphism. Let $\omega
:= \C_0^z: \C_0^C \=> \C_0^H \subseteq \C^H$ be the functor induced by
$z$. Then
 $$\Nat_{\C_0}(\omega, \omega \tensor \X ) : \C^H \=> \Set$$
 is representable by the coalgebra $C = \coend_{\C_0}(\omega)$ in
$\C^H$ with the canonical morphism $\delta: \omega \=> \omega \tensor
C$.
 \end{thm}

 \begin{pf}
 We define maps
 $$\Sigma: \C^H(C,M) \=> \Nat_{\C_0}(\omega,\omega \tensor M)$$
 and
 $$\Pi: \Nat_{\C_0}(\omega,\omega \tensor M) \=> \C^H(C,M).$$
 The first map is defined by $\Sigma(f)(P,\delta_P) := (1_P \tensor
f)\delta_P: P \=> P \tensor C \=> P \tensor M$. Then
$\Sigma(f)(P,\delta_P): P \=> P \tensor M$ is an $H$-comodule morphism
since the following diagram commutes
 $$\bfig
 \putmorphism(0, 1200)(1, 0)[P`P \tensor C`\sst \delta_P]{800}1a
 \putmorphism(800, 1200)(1, 0)[\phantom{P \tensor C}`P \tensor M`\sst
1 \tensor f]{1300}1a
 \putmorphism(800, 900)(1, 0)[P \tensor C \tensor C \tensor H`P
\tensor C \tensor M \tensor H`\sst 1 \tensor 1 \tensor f \tensor
1]{1300}1a
 \putmorphism(800, 600)(1, 0)[P \tensor H \tensor C \tensor H`P
\tensor H \tensor M \tensor H`\sst 1 \tensor 1 \tensor f \tensor
1]{1300}1a
 \putmorphism(800, 300)(1, 0)[P \tensor C \tensor H \tensor H`P
\tensor M \tensor H \tensor H`\sst 1 \tensor f \tensor 1 \tensor
1]{1300}1a
 \putmorphism(0, 0)(1, 0)[P \tensor H`P \tensor C \tensor H`\sst
\delta_P \tensor 1]{800}1a
 \putmorphism(800, 0)(1, 0)[\phantom{P \tensor C \tensor H}`P \tensor
M \tensor H`\sst 1 \tensor f \tensor 1]{1300}1a
 \putmorphism(0, 1200)(0, -1)[`P \tensor C`\sst \delta_P]{600}1l
 \putmorphism(0, 600)(0, -1)[``\sst 1 \tensor z]{600}1l
 \putmorphism(800, 1200)(0, -1)[``\sst \delta_P \tensor \ad]{300}1l
 \putmorphism(800, 900)(0, -1)[``\sst 1 \tensor z \tensor 1 \tensor
1]{300}1l
 \putmorphism(800, 600)(0, -1)[``\sst 1 \tensor \sigma \tensor
1]{300}1l
 \putmorphism(800, 300)(0, -1)[``\sst 1 \tensor 1 \tensor m_H]{300}1l
 \putmorphism(2100, 1200)(0, -1)[``\sst \delta_P \tensor
\delta_M]{300}1r
 \putmorphism(2100, 900)(0, -1)[``\sst 1 \tensor z \tensor 1 \tensor
1]{300}1r
 \putmorphism(2100, 600)(0, -1)[``\sst 1 \tensor \sigma \tensor
1]{300}1r
 \putmorphism(2100, 300)(0, -1)[``\sst 1 \tensor 1 \tensor m_H]{300}1r
 \efig$$
 where the left hand side commutes by
 \bgr{70}{60}
 \put(50,30){\multmor{\delta}{10}}
 \put(0,30){\multmor{\delta}{10}}
 \put(5,45){\multmor{\delta}{20}}
 \put(55,45){\multmor{\delta}{15}}
 \put(20,5){\mult}
 \put(10,10){\braid}
 \put(10,20){\mor{z}{10}}
 \put(70,15){\mor{z}{30}}
 \put(20,20){\multmor{\ad}{10}}
 \put(5,40){\idgr{5}}
 \put(55,40){\idgr{5}}
 \put(0,0){\idgr{30}}
 \put(10,0){\idgr{10}}
 \put(15,55){\idgr{5}}
 \put(25,0){\idgr{5}}
 \put(25,30){\idgr{15}}
 \put(30,10){\idgr{10}}
 \put(50,0){\idgr{30}}
 \put(60,0){\idgr{30}}
 \put(62,55){\idgr{5}}
 \put(70,0){\idgr{15}}
 \put(15,60){\objo{P}}
 \put(62,60){\objo{P}}
 \put(40,25){\objo{=}}
 \put(0,0){\obju{P}}
 \put(10,0){\obju{C}}
 \put(25,0){\obju{H}}
 \put(50,0){\obju{P}}
 \put(60,0){\obju{C}}
 \put(70,0){\obju{H}}
 \egr
 which follows from Lemma \ref{rcoadjact}. Hence $\Sigma(f): \omega
\=> \omega \tensor M$ is a natural transformation. Furthermore we have
$\Sigma(f)(X \tensor (P,\delta_P)) = (1_{X \tensor P} \tensor
f)\delta_{X \tensor P} = (1_X \tensor 1_P \tensor f)(1_X \tensor
\delta_P) = 1_X \tensor \Sigma(f)(P,\delta_P)$, so $\Sigma(f)$ is a
$\C_0$-morphism.

 To define the map $\Pi$ let $\varphi: \omega \=> \omega \tensor M$ be
given. Define $\lim \varphi(C_i) = \varphi(C): C \=> \C \tensor M$ as
the uniquely determined morphism so that
 $$\bfig
 \putmorphism(0, 400)(1, 0)[C_i`C_i \tensor M`\sst
\varphi(C_i)]{600}1a
 \putmorphism(0, 0)(1, 0)[C`C \tensor M`\sst \varphi(C)]{600}1a
 \putmorphism(0, 400)(0, -1)[``\sst \iota_i]{400}1l
 \putmorphism(600, 400)(0, -1)[``\sst \iota_i \tensor 1_M]{400}1r
 \efig$$
 commutes and let
 $$\Pi(\varphi) := (\epsilon \tensor 1_M)\varphi(C): C \=> C \tensor M
\=> M.$$

 Since $C_i$ is a subcoalgebra of $C$ it is an $H$-subcomodule by the
coadjoint coaction induced by the coalgebra morphism $z\iota_i: C_i
\=> H$. Hence, $C$ is a colimit of $H$-comodule coalgebras and
$\varphi(C): C \=> C \tensor M$ is an $H$-comodule morphism.
Consequently $\Pi(\varphi)$ is in $\C^H$.

 We have $\Pi\Sigma(f) = (\epsilon \tensor 1_M)\lim((1_{C_i} \tensor
f)\delta_i) = (\epsilon \tensor 1_M)(1_C \tensor f)\Delta = f(\epsilon
\tensor 1_C)\Delta_C = f$.

 Now observe that $\delta_P: P \=> P \tensor C$ is a $C$-comodule
morphism with the $C$-structure on $P \tensor C$ coming from the one
of $C$. Thus we get $(1 \tensor \delta_j)\delta_{P,j} = (\delta_{P,j}
\tensor 1_C)\delta_P$ for the morphism $\delta_{P,j}: P \=> P \tensor
C_j$ which exists by the assumptions about a $\C_0$-generated
coalgebra. So $\delta_{P,j}$ is a $C$-comodule morphism as well, hence
$\varphi(P \tensor C_j,1_P \tensor \delta_j)\delta_{P,j} =
(\delta_{P,j} \tensor 1_M)\varphi(P,\delta_P)$. From this we get

 $$\begin{array}{rl}
 \Sigma\Pi(\varphi)(P,\delta_P) &= (1_P \tensor (\epsilon \tensor 1_M)
\varphi(C))\delta_P\\
 &= (1_P \tensor \epsilon \tensor 1_M)(1_P \tensor
\lim\varphi(C_i))(1_P \tensor \iota_j)\delta_{P,j}\\
 &= (1_P \tensor \epsilon \tensor 1_M)(1_P \tensor \iota_j \tensor
1_M)(1_P \tensor \varphi(C_j))\delta_{P,j}\\
 &= (1_P \tensor \epsilon \tensor 1_M)(1_P \tensor \iota_j \tensor
1_M)\varphi(P \tensor C_j)\delta_{P,j}\\
 &= (1_P \tensor \epsilon \tensor 1_M)(1_P \tensor \iota_j \tensor
1_M)(\delta_{P,j} \tensor 1_M)\varphi(P)\\
 &= \varphi(P,\delta_P).
 \end{array}$$
 So $\Sigma$ and $\Pi$ are inverses of each other. The claim about the
coalgebra structure is clear from the uniqueness of the reconstructed
coalgebra.
 \end{pf}

 \subsubsection{}
 Observe that in general the set of {\em all} natural transformations
-- not just the $\C$-morphisms -- from $\omega$ to $\omega \tensor M$
is too large for the reconstruction of $C$ as we have seen in
\ref{Cexamples}. By Proposition \ref{nattransisC}, however, this
difference does not occur if the base category $\C$ is the category of
vector spaces $\Vek$ over $\kl$. This explains the usual {\em full
reconstruction} with the functor $\Nat(\omega,\omega \tensor \X )$
like in \cite{DEL, UL1, MAJ4}.

 Examples of natural transformations which are not $\C$-morphisms can
be derived from \ref{Cexamples}.

 Now assume that $\C$ is a cocomplete braided monoidal category and
$\C_0$ is a (braided) full monoidal subcategory of $\C$. Furthermore
assume that the tensor product in $\C$ preserves arbitrary colimits in
both variables.

 \begin{thm} \label{recbialg}
 Let $B$ be a $\C_0$-generated coalgebra in $\C$ and a $\C_0$-central
bialgebra, let $H$ be a braided $\C_0$-central Hopf algebra in $\C$.
Let $z: B \=> H$ be a bialgebra morphism. Let $\omega := \C_0^z:
\C_0^B \=> \C_0^H \subseteq \C^H$ be the functor induced by $z$. Then
$\Nat_{\C_0}(\omega, \omega \tensor \X ) : \C^H \=> \Set$ is
multirepresentable by the bialgebra $\coend_{\C_0}(\omega)$ in $\C^H$
which is equal to $B$ as an $H$-comodule coalgebra under the coadjoint
coaction, but carries a different multiplicative structure $\widetilde
m_B: B \tensor B \=> B$, the transmuted multiplication.
 \end{thm}

 \begin{pf}
 We extend the proof of \ref{reccoalg} to $\omega^2$. Define maps
 $$\Sigma: \C^H(B \tensor B,M) \=> \Nat_{\C_0}(\omega \tensor
\omega,\omega \tensor \omega \tensor M)$$
 and
 $$\Pi: \Nat_{\C_0}(\omega \tensor \omega,\omega \tensor \omega
\tensor M) \=> \C^H(B \tensor B,M).$$
 The first map is given by $\Sigma(f)((P,\delta_P),(Q,\delta_Q)) :=
(1_{P \tensor Q} \tensor f)(1_P \tensor \sigma_{B,Q} \tensor
1_B)(\delta_P \tensor \delta_Q): P \tensor Q \=> P \tensor B \tensor Q
\tensor B \=> P \tensor Q \tensor B \tensor B \=> P \tensor Q \tensor
M$. Then $\Sigma(f) ((P,\delta_P), (Q,\delta_Q)): P \tensor Q \=> P
\tensor Q \tensor M$ is an $H$-comodule morphism by a similar proof as
in \ref{reccoalg} replacing $P$ by $P \tensor Q$. Hence $\Sigma(f):
\omega \tensor \omega \=> \omega \tensor \omega \tensor M$ is a
natural transformation. To show that $\Sigma(f)$ is a $\C_0$-morphism
we observe that $\xi: \omega(X \tensor P) \=> X \tensor \omega(P)$ is
the identity. So we have
 $$\begin{array}{l}
 \Sigma(f)(X \tensor (P,\delta_P),(Q,\delta_Q)) \\
 \hskip1cm = (1_X \tensor 1_P \tensor 1_Q \tensor f)(1_X \tensor 1_P
\tensor \sigma_{B,Q} \tensor 1_B)(1_X \tensor \delta_P \tensor
\delta_Q)\\
 \hskip1cm = 1_X \tensor (1_P \tensor 1_Q \tensor f)(1_P \tensor
\sigma_{B,Q} \tensor 1_B)(\delta_P \tensor \delta_Q)\\
 \hskip1cm = 1_X \tensor \Sigma(f)((P,\delta_P),(Q,\delta_Q)),
 \end{array}$$
 $$\begin{array}{l}
 (\sigma_{P,Y} \tensor 1_{Q \tensor M})\Sigma(f)((P,\delta_P), Y
\tensor (Q, \delta_Q))\\
 \hskip1cm = (\sigma_{P,Y} \tensor 1_{Q \tensor M})(1_{P \tensor Y
\tensor Q} \tensor f)(1_P \tensor \sigma_{B,Y \tensor Q} \tensor
1_B)(\delta_P \tensor \delta_{Y \tensor Q}) \\
 \hskip1cm = (1_{Y \tensor P \tensor Q} \tensor f)(1_{Y \tensor P}
\tensor \sigma_{B,Q} \tensor 1_B) (1_Y \tensor \delta_P \tensor
\delta_Q)(\sigma_{P,Y} \tensor 1_Q)\\
 \hskip1cm = (1_Y \tensor
\Sigma(f)((P,\delta_P),(Q,\delta_Q)))(\sigma_{P,Y} \tensor 1_Q)
 \end{array}$$
 or as a braid diagram
 \bgr{100}{55}
 \put(0,40){\multmor{\delta}{10}}
 \put(30,40){\multmor{\delta}{10}}
 \put(70,25){\multmor{\delta}{10}}
 \put(90,25){\multmor{\delta}{10}}
 \put(0,0){\braid}
 \put(10,30){\braid}
 \put(20,20){\braid}
 \put(60,45){\braid}
 \put(80,15){\braid}
 \put(75,35){\twist{-5}{10}}
 \put(30,10){\multmor{f}{10}}
 \put(90,5){\multmor{f}{10}}
 \put(0,10){\idgr{30}}
 \put(10,10){\idgr{20}}
 \put(5,50){\idgr{5}}
 \put(20,0){\idgr{20}}
 \put(20,40){\idgr{15}}
 \put(30,30){\idgr{10}}
 \put(35,0){\idgr{10}}
 \put(35,50){\idgr{5}}
 \put(40,20){\idgr{20}}
 \put(60,0){\idgr{45}}
 \put(70,0){\idgr{25}}
 \put(80,0){\idgr{15}}
 \put(95,0){\idgr{5}}
 \put(95,35){\idgr{20}}
 \put(100,15){\idgr{10}}
 \put(5,55){\objo{P}}
 \put(20,55){\objo{Y}}
 \put(35,55){\objo{Q}}
 \put(60,55){\objo{P}}
 \put(70,55){\objo{Y}}
 \put(95,55){\objo{Q}}
 \put(50,22){\objo{=}}
 \put(0,0){\obju{Y}}
 \put(10,0){\obju{P}}
 \put(20,0){\obju{Q}}
 \put(35,0){\obju{M}}
 \put(60,0){\obju{Y}}
 \put(70,0){\obju{P}}
 \put(80,0){\obju{Q}}
 \put(95,0){\obju{M}}
 \egr
 and
 $$\begin{array}{l}
 \Sigma(f)((P,\delta_P), Y \tensor (Q, \delta_Q))(\sigma_{Y,P} \tensor
1_Q)\\
 \hskip1cm = (1_{P \tensor Y \tensor Q} \tensor f)(1_P \tensor
\sigma_{B,Y \tensor Q} \tensor 1_B)(\delta_P \tensor \delta_{Y \tensor
Q})(\sigma_{Y,P} \tensor 1_Q)\\
 \hskip1cm = (1_{P \tensor Y \tensor Q} \tensor f)(1_P \tensor
\sigma_{B,Y \tensor Q} \tensor 1_B)(\sigma_{Y,P \tensor B} \tensor
1_{Q \tensor B})(1_Y \tensor \delta_P \tensor \delta_Q)\\
 \hskip1cm = (1_{P \tensor Y \tensor Q} \tensor f)(1_{P \tensor Y}
\tensor \sigma_{B,Q} \tensor 1_B)(1_P \tensor
(\sigma_{B,Y}\sigma_{Y,B} \tensor 1_{Q \tensor B}))\\
 \hskip2cm(\sigma_{Y,P} \tensor 1_{B \tensor Q \tensor B})(1_Y \tensor
\delta_P \tensor \delta_Q)
 \end{array}$$
 \newpage
 $$\begin{array}{l}
 \hskip1cm = (1_{P \tensor Y \tensor Q} \tensor f)(1_{P \tensor Y}
\tensor \sigma_{B,Q} \tensor 1_B)(\sigma_{Y,P} \tensor 1_{B \tensor Q
\tensor B})(1_Y \tensor \delta_P \tensor \delta_Q)\\
 \hskip1cm = (1_{P \tensor Y \tensor Q} \tensor f)(\sigma_{Y,P}
\tensor \sigma_{B,Q} \tensor 1_B)(1_Y \tensor \delta_P \tensor
\delta_Q)\\
 \hskip1cm = (\sigma_{Y,P} \tensor 1_{Q \tensor M})(1_{Y \tensor P
\tensor Q} \tensor f)(1_{Y \tensor P} \sigma_{B,Q} \tensor 1_B)(1_Y
\tensor \delta_P \tensor \delta_Q)\\
 \hskip1cm = (\sigma_{Y,P} \tensor 1_{Q \tensor M})(1_Y \tensor
\Sigma(f)((P,\delta_P), (Q,\delta_Q))
 \end{array}$$
 or as a braid diagram
 \bgr{160}{70}
 \put(30,5){\multmor{f}{10}}
 \put(35,0){\idgr{5}}
 \put(90,5){\multmor{f}{10}}
 \put(95,0){\idgr{5}}
 \put(150,5){\multmor{f}{10}}
 \put(155,0){\idgr{5}}

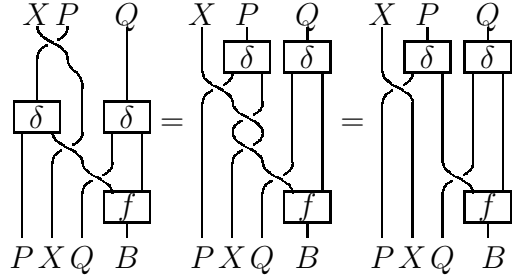
 \put(10,25){\braid}
 \put(20,15){\braid}
 \put(5,60){\braid}
 \put(60,45){\braid}
 \put(70,25){\braid}
 \put(70,35){\braid}
 \put(80,15){\braid}
 \put(120,45){\braid}
 \put(140,15){\braid}
 \put(20,50){\twist{-5}{10}}
 \put(0,35){\multmor{\delta}{10}}
 \put(30,35){\multmor{\delta}{10}}
 \put(70,55){\multmor{\delta}{10}}
 \put(90,55){\multmor{\delta}{10}}
 \put(130,55){\multmor{\delta}{10}}
 \put(150,55){\multmor{\delta}{10}}
 \put(0,0){\idgr{35}}
 \put(5,45){\idgr{15}}
 \put(10,0){\idgr{25}}
 \put(20,0){\idgr{15}}
 \put(20,35){\idgr{15}}
 \put(30,25){\idgr{10}}
 \put(35,45){\idgr{25}}
 \put(60,0){\idgr{45}}
 \put(60,55){\idgr{15}}
 \put(70,0){\idgr{25}}
 \put(75,65){\idgr{5}}
 \put(80,0){\idgr{15}}
 \put(80,45){\idgr{10}}
 \put(90,25){\idgr{30}}
 \put(95,65){\idgr{5}}
 \put(100,15){\idgr{40}}
 \put(120,0){\idgr{45}}
 \put(120,55){\idgr{15}}
 \put(130,0){\idgr{45}}
 \put(135,65){\idgr{5}}
 \put(140,0){\idgr{15}}
 \put(140,25){\idgr{30}}
 \put(150,25){\idgr{30}}
 \put(155,65){\idgr{5}}
 \put(160,15){\idgr{40}}
 \put(40,15){\idgr{20}}
 \put(5,70){\objo{X}}
 \put(15,70){\objo{P}}
 \put(35,70){\objo{Q}}
 \put(60,70){\objo{X}}
 \put(75,70){\objo{P}}
 \put(95,70){\objo{Q}}
 \put(120,70){\objo{X}}
 \put(135,70){\objo{P}}
 \put(155,70){\objo{Q}}
 \put(50,35){\objo{=}}
 \put(110,35){\objo{=}}

 \put(0,0){\obju{P}}
 \put(10,0){\obju{X}}
 \put(20,0){\obju{Q}}
 \put(35,0){\obju{B}}
 \put(60,0){\obju{P}}
 \put(70,0){\obju{X}}
 \put(80,0){\obju{Q}}
 \put(95,0){\obju{B}}
 \put(120,0){\obju{P}}
 \put(130,0){\obju{X}}
 \put(140,0){\obju{Q}}
 \put(157,0){\obju{B}}
 \egr
 since $B$ is $\C_0$-central in $\C$. Thus $\Sigma(f)$ is a
 $\C_0$-morphism.

 Now we construct the second map $\Pi$. Let $\varphi: \omega \tensor
\omega \=> \omega \tensor \omega \tensor M$. Define
$\lim\varphi(B_i,B_j) = \varphi(B,B): B \tensor B \=> B \tensor B
\tensor M$ as the uniquely determined morphism so that
 $$\bfig
 \putmorphism(0, 400)(1, 0)[B_i \tensor B_j`B_i \tensor B_j \tensor
M`\sst \varphi(B_i,B_j)]{900}1a
 \putmorphism(0, 0)(1, 0)[B \tensor B`B \tensor B \tensor M`\sst
\varphi(B,B)]{900}1a
 \putmorphism(0, 400)(0, -1)[``\sst \iota_i \tensor \iota_j]{400}1l
 \putmorphism(900, 400)(0, -1)[``\sst \iota_i \tensor \iota_j \tensor
1_M]{400}1r
 \efig$$
 commutes. Observe that $B \tensor B$ is the colimit of the $B_i
\tensor B_j$ since by assumption the tensor product preserves
colimits. Let
 $$\Pi(\varphi) := (\epsilon \tensor \epsilon \tensor
1_M)\varphi(B,B): B \tensor B \=> B \tensor B \tensor M \=> M.$$
 As in the proof of \ref{reccoalg} $\Pi(\varphi)$ is an $H$-comodule
morphism.

 We have $\Pi\Sigma(f) = (\epsilon \tensor \epsilon \tensor 1_M)(1_{B
\tensor B} \tensor f)(1_B \tensor \sigma_{B,B} \tensor 1_B)(\Delta_B
\tensor \Delta_B) = f(\epsilon \tensor \epsilon \tensor 1_{B \tensor
B})\Delta_{B \tensor B} = f$. Observe that $\delta_P: P \=> P \tensor
B$ is a $B$-comodule morphism with the $B$-structure on $P \tensor B$
coming from the one of $B$. So we have
 $$\begin{array}{l}
 \Sigma\Pi(\varphi)((P,\delta_P),(Q,\delta_Q))\\
 \hskip1cm = (1_{P \tensor Q} \tensor \epsilon \tensor \epsilon
\tensor 1_M)
   (1_{P \tensor Q} \tensor \varphi(B,B))
   (1_P \tensor \sigma_{B,Q} \tensor 1_B)
   (\delta_P \tensor \delta_Q)\\
 \hskip1cm = (1_{P \tensor Q} \tensor \epsilon \tensor \epsilon
\tensor 1_M)
   (1_{P \tensor Q} \tensor \lim\varphi(B_k,B_l))\\ \qquad \hskip1cm
   (1_P \tensor \sigma_{B,Q} \tensor 1_B)
   (\delta_P \tensor \delta_Q)\\
 \hskip1cm = (1_{P \tensor Q} \tensor \epsilon \tensor \epsilon
\tensor 1_M)
   \lim [(1_{P \tensor Q} \tensor \varphi(B_k,B_l))
   (1_P \tensor \sigma_{B_k,Q} \tensor 1_{B_l})]\\ \qquad \hskip1cm
   (1_P \tensor \iota_i \tensor 1_Q \tensor \iota_j)
   (\delta_{P,i} \tensor \delta_{Q,j})\\
 \hskip1cm = (1_{P \tensor Q} \tensor \epsilon \tensor \epsilon
\tensor 1_M)
   (1_{P \tensor Q} \tensor \iota_i \tensor \iota_j \tensor 1_M)\\
\qquad \hskip1cm
   (1_{P \tensor Q} \tensor \varphi(B_i,B_j))
   (1_P \tensor \sigma_{B_i,Q} \tensor 1_{B_j})]
   (\delta_{P,i} \tensor \delta_{Q,j})\\
 \hskip1cm = (1_{P \tensor Q} \tensor \epsilon\iota_i \tensor
\epsilon\iota_j \tensor 1_M)
   (1_P \tensor \sigma_{B_i,Q} \tensor 1_{B_j} \tensor 1_M)\\ \qquad
\hskip1cm
   (1_P \tensor \varphi(B_i,Q \tensor B_j))
   (\delta_{P,i} \tensor \delta_{Q,j})\\
 \hskip1cm = (1_P \tensor \epsilon\iota_i \tensor 1_Q \tensor
\epsilon\iota_j \tensor 1_M)
   \varphi(P \tensor B_i,Q \tensor B_j)
   (\delta_{P,i} \tensor \delta_{Q,j})\\
 \hskip1cm = (1_P \tensor \epsilon\iota_i \tensor 1_Q \tensor
\epsilon\iota_j \tensor 1_M)
   (\delta_{P,i} \tensor \delta_{Q,j} \tensor 1_M)
   \varphi(P,Q)\\
 \hskip1cm = \varphi((P,\delta_P),(Q,\delta_Q)).
 \end{array}$$
 Thus $\Sigma$ and $\Pi$ are inverses of each other.

 The multiplicative structure $\widetilde m_B$ of $B$ in $\C^H$ is now
the uniquely determined morphism which makes the diagram
 $$\bfig
 \putmorphism(0, 400)(1, 0)[P \tensor Q `P \tensor Q \tensor B \tensor
B `\sst \delta_2]{800}1a
 \putmorphism(0, 0)(1, 0)[P \tensor Q`P \tensor Q \tensor B`\sst
\delta_{P\tensor Q}]{800}1a
 \putmorphism(0,  400)(0, -1)[``\sst 1_{P\tensor Q}]{400}1l
 \putmorphism(800, 400)(0, -1)[``\sst 1_{P \tensor Q} \tensor
\widetilde m_B]{400}1r
 \efig$$
 commutative, since $\omega \tensor \omega = \omega(\x \tensor \x )
\buildrel \delta \over \longrightarrow \omega(\x \tensor \x ) \tensor
B$ is a $\C_0$-bimorphism of bifunctors. The unit is given by
$\lambda: I \=> I \tensor B = B$. The new multiplication can be
described by the braid diagram
 \bgr{80}{40}
 \put(20,10){\mult}
 \put(10,15){\braid}
 \put(0,25){\multmor{\delta}{10}}
 \put(20,25){\multmor{\delta}{10}}
 \put(50,25){\multmor{\delta}{10}}
 \put(70,25){\multmor{\delta}{10}}
 \put(70,5){\multmor{\widetilde m}{10}}
 \put(60,15){\brmor}
 \put(0,0){\idgr{25}}
 \put(5,35){\idgr{5}}
 \put(10,0){\idgr{15}}
 \put(25,0){\idgr{10}}
 \put(25,35){\idgr{5}}
 \put(30,15){\idgr{10}}
 \put(50,0){\idgr{25}}
 \put(55,35){\idgr{5}}
 \put(60,0){\idgr{15}}
 \put(75,0){\idgr{5}}
 \put(75,35){\idgr{5}}
 \put(80,15){\idgr{10}}
 \put(5,40){\objo{P}}
 \put(25,40){\objo{Q}}
 \put(55,40){\objo{P}}
 \put(75,40){\objo{Q}}
 \put(40,15){\objo{=}}
 \put(0,0){\obju{P}}
 \put(10,0){\obju{Q}}
 \put(25,0){\obju{B}}
 \put(50,0){\obju{P}}
 \put(60,0){\obju{Q}}
 \put(75,0){\obju{B}}
 \egr
 where the braid morphism on the left is the one in $\C$ and the braid
morphism one the right is the one in $\C^H$.

 The proof shows that the universal morphism is $(1_P \tensor
\sigma_{B,Q} \tensor 1_Q)(\delta_P \tensor \delta_Q):\omega \tensor
\omega \=> \omega \tensor \omega \tensor B \tensor B$.
 An analogous result holds for multifunctors $\omega \tensor \ldots
\tensor \omega = \omega^n$ and $\C_0$-morphisms $\varphi: \omega^n \=>
\omega^n \tensor M$. This proves that with a suitable element $\tau
\in B_n$ in the Artin braid group the morphism
 $$\delta_n:= \tau\delta^n: \omega^n \=> \omega^n \tensor B^n$$
 is the universal $\C_0$-morphism for all $n \in {\Bbb N}$, in
particular $\coend_{\C_0}(\omega^n) = B^n$.
  \end{pf}

 \subsubsection{}
 The proof of \ref{reccoalg} resp. \ref{recbialg} provides a proof for
the ``representability assumption for modules'' in (\cite{MAJ1} 3.2)
in a very general setting for the functor
$\Nat_{\C_0}(\omega^n,\omega^n \tensor \X )$ instead of the functor
$\Nat(\omega^n,\omega^n \tensor \X )$. A special case of Theorem
\ref{recbialg} ($\C = \Vek$) is \cite{MAJ2} Proposition A.4.

 The following is a generalization of \cite{PA2}, Corollary 6.4.

 \begin{cor}  \label{recC}
 Let $C$ be a $\C_0$-generated coalgebra in the braided monoidal
category $\C$ and let $\omega: \C_0^C \=> \C$ be the forgetful
functor. Then $\Nat_{\C_0}(\omega, \omega \tensor \X ) : \C \=> \Set$
is representable and $C = \coend_{\C_0}(\omega)$ as coalgebras in
$\C$.
 \end{cor}

 \begin{pf}
 Use $H=\kl$ in the above theorem.
 \end{pf}

 \begin{cor}\label{recI}
 Let $\omega: \C_0 \=> \C$ be the embedding functor. Then
$\Nat_{\C_0}(\omega, \omega \tensor \X ) : \C \=> \Set$ is
representable and $\coend_{\C_0}(\omega) = I$.

In particular $\coend_\C(\Id_\C) \iso I$ for any monoidal category
$\C$.
 \end{cor}

 \begin{pf}
 Use $C = \kl$ in the above Corollary.
 \end{pf}

 As we remarked after the definition of $\C_0$-generated coalgebras,
the condition that $C$ is $\C_0$-generated becomes vacuous in the case
$\C_0 = \C$.

 \begin{cor} \label{specrecbialg}
  Let $B$ be a $\C_0$-central bialgebra in $\C$ which is
 $\C_0$-generated as a coalgebra, let $\A := \C_0^B$, and let $\omega:
\C^B_0 \=> \C$ be the forgetful functor. Then
$\Nat_{\C_0}(\omega,\omega \tensor \X )$ is multirepresentable and
$\coend_{\C_0}(\omega) = B$ as bialgebras.
 \end{cor}

 By dualization of Proposition \ref{reccoalg} one gets

 \begin{cor}
 Let $A$ be a $\C_0$-generated algebra in $\C$, and let $\omega:
\C_{0A} \=> \C$ be the forgetful functor. Then $\Nat_{\C_0}(\omega
\tensor \X , \omega)$ is representable and $A = \rend_{\C_0}(\omega)$
as algebras in $\C$.
 \end{cor}

 \subsubsection{}
 Later on we will need the notion of cosmash products in $\A$. So we
define it here and show an important property. If $B$ is a bialgebra
and $C$ is a $B$-comodule-coalgebra in $\A$ then we can construct a
cosmash product $B \#^c C$ where the cosmash comultiplication is
defined by
 $$\begin{array}{c} \Delta: B \tensor C
 \buildrel \Delta_B \tensor \Delta_C \over \longrightarrow B \tensor B
\tensor C \tensor C
 \buildrel 1 \tensor \delta'_C \tensor 1 \over \longrightarrow B
\tensor B \tensor C \tensor B \tensor C \\
 \buildrel 1 \tensor \sigma \tensor 1 \over \longrightarrow B \tensor
C \tensor B \tensor B \tensor C
 \buildrel 1 \tensor m_B \tensor 1 \over \longrightarrow B \tensor C
\tensor B \tensor C.
 \end{array}$$
 It is easy to see that $B \#^c C$ is a coalgebra in $\A$.

 \subsubsection{} \label{twocosmashs}
 Let $H$ be a braided Hopf algebra in $\C$, $z : B \=> H$ be a
bialgebra morphism in $\C$ and $\widetilde B$ be the transmuted
bialgebra in $\C^H$ (as in Theorem \ref{recbialg}). Let $C$ be a
$\widetilde B$-comodule-coalgebra in $\A := \C^H$. Then we can form
the cosmash product $\widetilde B \widetilde \#^c C$ in $\C^H$.

 Since $B$ is a bialgebra in $\C$ and $C$ is a coalgebra in $\C$, $C$
is also a $B$-comodule-coalgebra by
 \bgr{120}{50}
 \put(0,15){\comult}
 \put(95,40){\bcomult{20}}
 \put(45,40){\bcomult{20}}
 \put(110,10){\mult}
 \put(100,15){\braid}
 \put(50,15){\brmor}
 \put(20,20){\twist{-5}{10}}
 \put(90,25){\multmor{\delta}{10}}
 \put(110,25){\multmor{\delta}{10}}
 \put(40,25){\multmor{\delta}{10}}
 \put(60,25){\multmor{\delta}{10}}
 \put(5,30){\multmor{\delta}{10}}
 \put(60,5){\multmor{\widetilde m}{10}}
 \put(90,0){\idgr{25}}
 \put(95,35){\idgr{5}}
 \put(100,0){\idgr{15}}
 \put(105,45){\idgr{5}}
 \put(115,0){\idgr{10}}
 \put(115,35){\idgr{5}}
 \put(40,0){\idgr{25}}
 \put(120,15){\idgr{10}}
 \put(45,35){\idgr{5}}
 \put(50,0){\idgr{15}}
 \put(55,45){\idgr{5}}
 \put(65,0){\idgr{5}}
 \put(65,35){\idgr{5}}
 \put(70,15){\idgr{10}}
 \put(0,0){\idgr{15}}
 \put(5,20){\idgr{10}}
 \put(10,0){\idgr{15}}
 \put(10,40){\idgr{10}}
 \put(20,0){\idgr{20}}
 \put(105,50){\objo{C}}
 \put(55,50){\objo{C}}
 \put(10,50){\objo{C}}
 \put(80,20){\objo{=}}
 \put(30,20){\objo{=}}
 \put(90,0){\obju{C}}
 \put(100,0){\obju{C}}
 \put(115,0){\obju{B}}
 \put(40,0){\obju{C}}
 \put(50,0){\obju{C}}
 \put(65,0){\obju{B}}
 \put(0,0){\obju{C}}
 \put(10,0){\obju{C}}
 \put(20,0){\obju{B}}
 \egr
 So there is a second way to define a cosmash product $B \# C$ this
time in $\C$. These two coalgebra structures on $B \tensor C$,
however, coincide as the following diagram shows
 \bgr{100}{45}
 \put(0,35){\comult}
 \put(60,35){\comult}
 \put(25,35){\bcomult{15}}
 \put(85,35){\bcomult{15}}
 \put(20,10){\mult}
 \put(10,15){\braid}
 \put(70,15){\brmor}
 \put(20,25){\multmor{\delta}{10}}
 \put(80,25){\multmor{\delta}{10}}
 \put(80,5){\multmor{\widetilde m}{10}}
 \put(0,0){\idgr{35}}
 \put(10,0){\idgr{15}}
 \put(10,25){\idgr{10}}
 \put(5,40){\idgr{5}}
 \put(25,0){\idgr{10}}
 \put(30,15){\idgr{10}}
 \put(32,40){\idgr{5}}
 \put(40,0){\idgr{35}}
 \put(90,15){\idgr{10}}
 \put(60,0){\idgr{35}}
 \put(65,40){\idgr{5}}
 \put(70,25){\idgr{10}}
 \put(70,0){\idgr{15}}
 \put(85,0){\idgr{5}}
 \put(92,40){\idgr{5}}
 \put(100,0){\idgr{35}}
 \put(5,45){\objo{B}}
 \put(32,45){\objo{C}}
 \put(65,45){\objo{B}}
 \put(92,45){\objo{C}}
 \put(50,20){\objo{=}}
 \put(0,0){\obju{B}}
 \put(10,0){\obju{C}}
 \put(25,0){\obju{B}}
 \put(40,0){\obju{C}}
 \put(60,0){\obju{B}}
 \put(70,0){\obju{C}}
 \put(85,0){\obju{B}}
 \put(100,0){\obju{\phantom{.}C.}}
 \egr

 \subsection{Finite reconstruction}

 In the previous section we started with an algebra $A$ or a coalgebra
$C$ in $\C$ and reconstructed them from $\omega: \C_A \=> \C$
resp.~$\omega: \C^C \=> \C$. If, however, an arbitrary $\C$-functor
$\omega: \B \=> \A$ is given it is not clear if $\Nat_\C(\omega,
\omega \tensor \X )$ is a representable functor or if
$\coend_\C(\omega)$ exists in $\C$. It is customary to call the
construction of $\coend_\C(\omega)$ also in this situation
``re''construction, although we do not start with an algebra or a
coalgebra in $\C$ and then reconstruct is from its category of
representations.

 In one particular situation the (restricted) reconstruction is
possible and well known, namely in the case of $\C = \Vek$ and a
functor $\omega: \B \=> \vek \subseteq \Vek$ into the category of
finite-dimensional vector spaces. Various generalizations of this
result are known. We will lift this result to braided monoidal
categories $\C$.

 Let $\C$ be a cocomplete braided monoidal category and $\C_0$ be the
full (rigid braided monoidal) subcategory of rigid objects.
Furthermore assume that the tensor product in $\C$ preserves arbitrary
colimits in both variables. Let $\omega: \B \=> \C_0 \subseteq \C$ be
a functor.

 \begin{thm} \label{finiterec}
 \begin{enumerate}
 \item  The functor $\Nat(\omega,\omega \tensor M)$ is
multirepresentable:
 $$\Nat(\omega,\omega \tensor M) \iso \C(\coend(\omega),M),$$
 \item If $\B$ is a $\C_0$-category and $\omega: \B \=> \C_0$ is a
$\C_0$-functor, then the functor $\Nat_{\C_0}(\omega,\omega \tensor
M)$ is representable:
 $$\Nat_{\C_0}(\omega,\omega \tensor M) \iso
\C(\coend_{\C_0}(\omega),M).$$
 If $B := \coend_{\C_0}(\omega)$ is $\C_0$-central then
$\Nat_{\C_0}(\omega,\omega \tensor M)$ is multirepresentable.
 \end{enumerate}
 \end{thm}

 \begin{pf}
 (1) The existence of a representing object for the functor
$\Nat(\omega, \omega \tensor M)$ is well known (see \cite{PA4}). We
recall the main steps of its construction, since they play a role in
the second part of the proof. The representing object $C =
\coend(\omega)$ is obtained as colimit of the diagram consisting of
all wedges for all morphisms $f: P \=> Q$ in $\B$:
 $$\bfig
 \putmorphism(900, 600)(-3, -1)[\omega(P)^* \tensor
\omega(P)`\omega(Q)^* \tensor \omega(P)`]{900}{-1}l
 \putmorphism(0, 300)(3, -1)[`\omega(Q)^* \tensor \omega(Q)`]{900}1l
 \putmorphism(900, 600)(-3, -1)[``\sst \omega(f)^* \tensor
1\quad]{600}0l
 \putmorphism(0, 300)(3, -1)[``\sst 1 \tensor \omega(f)\quad]{1200}0l
 \efig$$
 Any (cone-) morphism from this diagram to an object $M \in \C$
 $$\bfig
 \put(-1400, 300){\text{(1)}}
 \put(3200, 300){\phantom{\text{(1)}}}
 \putmorphism(900, 600)(-3, -1)[\omega(P)^* \tensor
\omega(P)`\omega(Q)^* \tensor \omega(P)`]{900}{-1}l
 \putmorphism(0, 300)(3, -1)[`\omega(Q)^* \tensor \omega(Q)`]{900}1l
 \putmorphism(900, 600)(-3, -1)[``\sst \omega(f)^* \tensor
1\quad]{600}0l
 \putmorphism(0, 300)(3, -1)[``\sst 1 \tensor \omega(f)\quad]{1200}0l
 \putmorphism(900, 600)(3, -1)[`M`]{900}1l
 \putmorphism(1800, 300)(-3, -1)[``]{900}{-1}l
 \putmorphism(900, 600)(3, -1)[``\sst \quad\psi(P)]{600}0r
 \putmorphism(1800, 300)(-3, -1)[``\sst \quad\psi(Q)]{1200}0r
 \efig$$
 is given by a natural transformation $\varphi: \omega \=> \omega
\tensor M$:
 $$\bfig
 \putmorphism(0, 400)(1, 0)[\omega(P)`\omega(P) \tensor M`\sst
\varphi(P)]{900}1a
 \putmorphism(0, 0)(1, 0)[\omega(Q)`\omega(Q) \tensor M`\sst
\varphi(Q)]{900}1a
 \putmorphism(0, 400)(0, -1)[``\sst \omega(f)]{400}1r
 \putmorphism(900, 400)(0, -1)[``\sst \omega(f) \tensor 1_M]{400}1r
 \efig$$
 The one-to-one correspondence between morphisms $\psi$ and morphisms
$\varphi$ is given, using the evaluation
 $\ev_{\omega(P)}: \omega(P)^* \tensor \omega(P) \=> I$
 and the dual basis
 $\db_{\omega(P)}: I\=> \omega(P) \tensor \omega(P)^*,$
 as
 $$\psi(P) = (\ev_{\omega(P)} \tensor 1_M)(1_{\omega(P)^*} \tensor
\varphi(P))$$
 and
 $$\varphi(P) := (1_{\omega(P)} \tensor \psi(P))(\db_{\omega(P)}
\tensor 1_{\omega(P)}).$$

 Given $\psi$ the morphisms $\varphi(P)$ form a natural transformation
since
 $$\bfig
 \putmorphism(0, 700)(1, 0)[\omega(P)`\omega(P) \tensor \omega(P)^*
\tensor \omega(P)`\sst \db_{\omega(P)} \tensor 1_{\omega(P)}]{1400}1a
 \putmorphism(1400, 700)(1, 0)[\phantom{\omega(P) \tensor \omega(P)^*
\tensor \omega(P)}`\omega(P) \tensor M`\sst 1_{\omega(P)} \tensor
\psi(P)]{1400}1a
 \putmorphism(700, 350)(1, 0)[\omega(Q) \tensor \omega(Q)^* \tensor
\omega(P)`\omega(Q) \tensor \omega(P)^* \tensor \omega(P)`\sst
]{1400}1a
 \putmorphism(700, 400)(1, 0)[``\sst 1_{\omega(Q)} \tensor \omega(f)^*
\tensor 1_{\omega(P)}]{1400}0a
 \putmorphism(0, 0)(1, 0)[\omega(Q)`\omega(Q) \tensor \omega(Q)^*
\tensor \omega(Q)`\sst \db_{\omega(Q)} \tensor 1_{\omega(Q)}]{1400}1b
 \putmorphism(1400, 0)(1, 0)[\phantom{\omega(Q) \tensor \omega(Q)^*
\tensor \omega(Q)}`\omega(Q) \tensor M`\sst 1_{\omega(Q)} \tensor
\psi(Q)]{1400}1b
 \putmorphism(0, 700)(0, -1)[``\sst \omega(f)]{700}1l
 \putmorphism(2800, 700)(0, -1)[``\sst \omega(f) \tensor 1_M]{700}1r
 \putmorphism(0, 700)(2, -1)[``\sst \db_{\omega(Q)} \tensor
1_{\omega(P)}]{700}1r
 \putmorphism(1400, 700)(2, -1)[``\sst \omega(f) \tensor
1_{\omega(P)^*} \tensor 1_{\omega(P)}]{700}1r
 \putmorphism(700, 350)(2, -1)[``\sst 1_{\omega(Q)} \tensor
1_{\omega(Q)^*} \tensor \omega(f)]{700}1l
 \putmorphism(2100, 350)(2, -1)[``\sst 1_{\omega(Q)} \tensor
\psi(P)]{700}1l
 \efig$$
 commutes where the left upper part of the diagram commutes by the
definition of the adjoint morphism $\omega(f)^*$. Conversely given
$\varphi: \omega \=> \omega \tensor M$ the diagram
 $$\bfig
 \putmorphism(400, 800)(1, 0)[\omega(P)^* \tensor
\omega(P)`\omega(P)^* \tensor \omega(P) \tensor M`\sst 1_{\omega(P)^*}
\tensor \varphi(P)]{1600}1a
 \putmorphism(0, 400)(1, 0)[\omega(Q)^* \tensor \omega(P)`\omega(Q)^*
\tensor \omega(P) \tensor M`\sst 1_{\omega(Q)^*} \tensor
\varphi(P)]{1600}1a
 \putmorphism(400, 0)(1, 0)[\omega(Q)^* \tensor \omega(Q)`\omega(Q)^*
\tensor \omega(Q) \tensor M`\sst 1_{\omega(Q)^*} \tensor
\varphi(Q)]{1600}1a
 \putmorphism(0, 400)(1, -1)[``\sst 1_{\omega(Q)^*} \tensor
\omega(f)]{400}1l
 \putmorphism(1600, 400)(1, -1)[``\sst 1_{\omega(Q)^*} \tensor
\omega(f) \tensor 1_M]{400}1l
 \putmorphism(2000, 800)(1, -1)[`M`\sst \ev_{\omega(P)} \tensor
1_M]{400}1r
 \putmorphism(400, 800)(-1, -1)[``\sst \omega(f)^* \tensor
1_{\omega(P)}]{400}{-1}l
 \putmorphism(2000, 800)(-1, -1)[``\sst \omega(f)^* \tensor
1_{\omega(P)} \tensor 1_M]{400} {-1}l
 \putmorphism(2400, 400)(-1, -1)[``\sst \ev_{\omega(Q)} \tensor
1_M]{400}{-1}r
 \efig$$
 commutes.

 Now we show that the functor $\Nat(\omega,\omega \tensor \X )$ is
multirepresentable. We restrict our attention just to the case $n =
2$. As before there is a bijective correspondence between the natural
transformations $\varphi(P,Q): \omega(P) \tensor \omega(Q) \=>
\omega(P) \tensor \omega(Q) \tensor M$ and cones $\psi(P,Q):
\omega(Q)^* \tensor \omega(P)^* \tensor \omega(P) \tensor \omega(Q)
\=> M$.

 Let $\delta: \omega \=> \omega \tensor \coend(\omega)$ be the
universal morphism and abbreviate $B := \coend(\omega)$. Let
$\theta(P): \omega(P)^* \tensor  \omega(P) \=> B$ be the induced
morphism. In the commutative diagram (colimit of a wedge in the sense
used above) induced by morphisms $f:P \=> R$ and $g: Q \=> S$ in $\B$
 $$\bfig
 \put(2800, 400){\text{(2)}}
 \put(-400, 400){\phantom{\text{(2)}}}
 \putmorphism(0, 800)(1, 0)[
   \omega(Q)^* \tensor \omega(P)^* \tensor \omega(P) \tensor
\omega(Q)`
   \omega(P)^* \tensor \omega(P) \tensor \omega(Q)^* \tensor
\omega(Q)`
   \sst    ]{1800}1a
 \putmorphism(0, 850)(1, 0)[ ` ` \sst \sigma_{\omega(Q)^*,\omega(P)^*
\tensor
   \omega(P)}^{-1} \tensor 1_{\omega(Q)} ]{1800}0a
 \putmorphism(0, 400)(1, 0)[\omega(S)^* \tensor \omega(R)^* \tensor
\omega(P)
   \tensor \omega(Q)``]{1600}0a
 \putmorphism(0, 0)(1, 0)[
   \omega(S)^* \tensor \omega(R)^* \tensor \omega(R) \tensor
\omega(S)`
   \omega(R)^* \tensor \omega(R) \tensor \omega(S)^* \tensor
\omega(S)`
   \sst ]{1800}1a
 \putmorphism(0, 50)(1, 0)[``\sst \sigma_{\omega(S)^*,\omega(R)^*
\tensor
   \omega(R)}^{-1} \tensor 1_{\omega(S)} ]{1800}0a
 \putmorphism(-100, 400)(1, -1)[``\sst 1_{\omega(S)^* \tensor
\omega(R)^*}
   \tensor \omega(f) \tensor \omega(g)]{400}1l
 \putmorphism(1600, 800)(1, -1)[``\sst\theta(P) \tensor
\theta(Q)]{400}1r
 \putmorphism(300, 800)(-1, -1)[``\sst \omega(g)^* \tensor \omega(f)^*
\tensor
   1_{\omega(P) \tensor \omega(Q)}]{400}{-1}l
 \putmorphism(2000, 400)(-1, -1)[B \tensor B``\sst \theta(R) \tensor
   \theta(S)]{400}{-1}r
 \efig$$
 $B \tensor B$ is a colimit with
 $$\theta(P,Q) = (\theta(P) \tensor \theta(Q))
(\sigma_{\omega(Q)^*,\omega(P)^* \tensor \omega(P)}^{-1} \tensor
1_{\omega(Q)})$$
 since tensor products preserve colimits. We have to show that the
induced morphism $\delta(\x ,\x ): \omega \tensor \omega \=> \omega
\tensor \omega \tensor B \tensor B$ is equal to $(1_\omega \tensor
\sigma_{B,\omega} \tensor 1_B)(\delta \tensor \delta): \omega^2 \=>
\omega^2 \tensor B \tensor B$. We use graphic calculus and observe
that the correspondence between the morphisms $\varphi$ and $\psi$ is
given by
 \bgr{160}{30}
 \put(30,20){\coeval}
 \put(140,5){\eval}
 \put(0,10){\multmor{\varphi}{10}}
 \put(40,10){\multmor{\psi}{10}}
 \put(110,10){\multmor{\psi}{10}}
 \put(150,10){\multmor{\varphi}{10}}
 \put(0,0){\idgr{10}}
 \put(0,20){\idgr{10}}
 \put(10,0){\idgr{10}}
 \put(30,0){\idgr{20}}
 \put(50,0){\idgr{10}}
 \put(50,20){\idgr{10}}
 \put(110,20){\idgr{10}}
 \put(120,0){\idgr{10}}
 \put(120,20){\idgr{10}}
 \put(140,10){\idgr{20}}
 \put(150,20){\idgr{10}}
 \put(160,0){\idgr{10}}
 \put(0,30){\objo{\omega}}
 \put(50,30){\objo{\omega}}
 \put(108,30){\objo{\omega^*}}
 \put(120,30){\objo{\omega}}
 \put(138,30){\objo{\omega^*}}
 \put(150,30){\objo{\omega}}
 \put(20,12){\objo{=}}
 \put(130,12){\objo{=}}
 \put(75,12){\objo{\mbox{and}}}
 \put(10,0){\obju{B}}
 \put(50,0){\obju{B}}
 \put(120,0){\obju{B}}
 \put(160,0){\obju{B}}
 \put(0,0){\obju{\omega}}
 \put(30,0){\obju{\omega}}
 \egr
 Then we get (writing $P$ and $Q$ instead of $\omega(P)$
resp.~$\omega(Q)$)
 \bgr{190}{70}
 \put(10,60){\coeval}
 \put(10,50){\coeval}
 \put(80,60){\coeval}
 \put(80,50){\coeval}
 \put(90,5){\eval}
 \put(120,5){\eval}
 \put(170,20){\braid}
 \put(20,40){\ibraid}
 \put(30,30){\ibraid}
 \put(90,40){\ibraid}
 \put(100,30){\ibraid}
 \put(0,50){\twist{10}{10}}
 \put(30,50){\twist{-10}{10}}
 \put(70,50){\twist{10}{10}}
 \put(100,50){\twist{-10}{10}}
 \put(120,20){\twist{-10}{10}}
 \put(20,10){\multmor{\theta}{10}}
 \put(40,10){\multmor{\theta}{10}}
 \put(100,10){\multmor{\delta}{10}}
 \put(130,10){\multmor{\delta}{10}}
 \put(160,30){\multmor{\delta}{10}}
 \put(180,30){\multmor{\delta}{10}}
 \put(0,0){\idgr{50}}
 \put(10,0){\idgr{50}}
 \put(20,20){\idgr{20}}
 \put(30,0){\idgr{10}}
 \put(30,20){\idgr{10}}
 \put(40,20){\idgr{10}}
 \put(40,40){\idgr{30}}
 \put(50,0){\idgr{10}}
 \put(50,20){\idgr{50}}
 \put(70,0){\idgr{50}}
 \put(80,0){\idgr{50}}
 \put(90,10){\idgr{30}}
 \put(100,20){\idgr{10}}
 \put(110,0){\idgr{10}}
 \put(110,40){\idgr{30}}
 \put(120,10){\idgr{10}}
 \put(130,20){\idgr{50}}
 \put(140,0){\idgr{10}}
 \put(160,0){\idgr{30}}
 \put(160,40){\idgr{30}}
 \put(170,0){\idgr{20}}
 \put(180,0){\idgr{20}}
 \put(180,40){\idgr{30}}
 \put(190,0){\idgr{30}}
 \put(40,70){\objo{P}}
 \put(110,70){\objo{P}}
 \put(160,70){\objo{P}}
 \put(50,70){\objo{Q}}
 \put(130,70){\objo{Q}}
 \put(180,70){\objo{Q}}
 \put(60,32){\objo{=}}
 \put(150,32){\objo{=}}
 \put(0,0){\obju{P}}
 \put(10,0){\obju{Q}}
 \put(30,0){\obju{B}}
 \put(50,0){\obju{B}}
 \put(70,0){\obju{P}}
 \put(80,0){\obju{Q}}
 \put(110,0){\obju{B}}
 \put(140,0){\obju{B}}
 \put(160,0){\obju{P}}
 \put(170,0){\obju{Q}}
 \put(180,0){\obju{B}}
 \put(190,0){\obju{\phantom{.}B.}}
 \egr

 (2) To describe the second isomorphism of the theorem the property of
$\C_0$-trans\-for\-ma\-tion for $\varphi: \omega \=> \omega \tensor M$
is given by the commutative diagram
 $$\bfig
 \putmorphism(0, 400)(1, 0)[\omega(X \tensor P)`\omega(X \tensor P)
\tensor M`\sst \varphi(X \tensor P)]{1200}1a
 \putmorphism(0, 0)(1, 0)[X \tensor \omega(P)`X \tensor \omega(P)
\tensor M`\sst 1_X \tensor \varphi(P)]{1200}1a
 \putmorphism(0, 400)(0, -1)[``\sst \zeta(X, P)]{400}1r
 \putmorphism(1200, 400)(0, -1)[``\sst \zeta(X, P) \tensor 1_M
]{400}1r
 \efig$$
 which translates into
 $$ \bfig
 \put(-1300, 400){\text{(3)}}
 \put(3100, 300){\phantom{\text{(1)}}}
 \putmorphism(0, 800)(1,0)[
  \omega(P)^* \tensor X^* \tensor \omega(X \tensor P)`
  \omega(X \tensor P)^* \tensor \omega(X \tensor P)`
  \sst \zeta(X, P)^* \tensor 1
  ]{1800}1a
 \putmorphism(0, 800)(1, 0)[``\sst \iso ]{1800}0b
 \putmorphism(0, 800)(0, -1)[``\sst 1_{\omega(P)^*} \tensor \zeta(X,P)
]{800}1l
 \putmorphism(0, 800)(0, -1)[``\sst \iso ]{800}0r
 \putmorphism(0, 0)(1, 0)[
  \omega(P)^* \tensor X^* \tensor X \tensor \omega(P)`
  \omega(P)^* \tensor \omega(P)`
  \sst 1_{\omega(P)^*} \tensor \ev_X \tensor 1_{\omega(P)}
  ]{1800}1a
 \putmorphism(1800, 800)(1, -1)[`M`\sst \psi(X \tensor P)]{400}1r
 \putmorphism(2200, 400)(-1, -1)[``\sst \psi(P)]{400}{-1}r
 \efig$$

 So the colimit $\coend_{\C_0}(\omega)$ exists and is described
(similar to the way given in \cite{PA4} Definition 2.2) as $\coprod_{P
\in \Ob(\B)} \omega(P)^* \tensor \omega(P)$ modulo the relations given
by all $f: P \=> Q$ as in the construction of $\coend(\omega)$ plus
the relations for any pair $(X,P)$ given above.

 In particular $\delta: \omega \=> \omega \tensor B$ with $B :=
\coend_{\C_0}(\omega)$ is a universal $\C_0$-morphism.

 Now we assume that $B = \coend_{\C_0}(\omega)$ is $\C_0$-central and
show that the functor $\Nat_{\C_0}(\omega,\omega \tensor \X )$ is
multirepresentable. We restrict our attention again just to the case
$n = 2$. Since $\delta_2 := (1_\omega \tensor \sigma_{B,\omega}
\tensor 1_{B})(\delta \tensor \delta): \omega^2 \=> \omega^2 \tensor
B^2$ is a $\C_0$-bimorphism -- $B$ is $\C_0$-central -- we can show
that $B \tensor B = \coend_{\C_0}(\omega \tensor \omega)$ with the
universal $\C_0$-bimorphism $\delta_2$. We first show that there is a
 one-to-one correspondence between $\C_0$-bimorphisms $\varphi(P,Q):
\omega(P) \tensor \omega(Q) \=> \omega(P) \tensor \omega(Q) \tensor M$
and cones $\psi(P,Q): \omega(Q)^* \tensor \omega(P)^* \tensor
\omega(P) \tensor \omega(Q) \=> M$ satisfying certain relations given
below. We saw earlier for two variables (see diagram (2)) that
$\varphi$ is a natural transformation iff $\psi$ is a cone. So we have
to translate the conditions for the $\C_0$-structure. It is an easy
exercise to show that under the correspondence between $\varphi$ and
$\psi$ the conditions
 \bgr{300}{30}
 \put(110,0){\braid}
 \put(160,20){\braid}
 \put(270,0){\braid}
 \put(220,20){\braid}
 \put(0,10){\multmor{\varphi}{30}}
 \put(60,10){\multmor{\varphi}{20}}
 \put(110,10){\multmor{\varphi}{30}}
 \put(170,10){\multmor{\varphi}{20}}
 \put(220,10){\multmor{\varphi}{30}}
 \put(280,10){\multmor{\varphi}{20}}
 \put(0,0){\idgr{10}}
 \put(10,0){\idgr{10}}
 \put(20,0){\idgr{10}}
 \put(30,0){\idgr{10}}
 \put(60,0){\idgr{10}}
 \put(70,0){\idgr{10}}
 \put(80,0){\idgr{10}}
 \put(130,0){\idgr{10}}
 \put(140,0){\idgr{10}}
 \put(170,0){\idgr{10}}
 \put(180,0){\idgr{10}}
 \put(190,0){\idgr{10}}
 \put(220,0){\idgr{10}}
 \put(230,0){\idgr{10}}
 \put(240,0){\idgr{10}}
 \put(250,0){\idgr{10}}
 \put(290,0){\idgr{10}}
 \put(300,0){\idgr{10}}
 \put(0,20){\idgr{10}}
 \put(10,20){\idgr{10}}
 \put(20,20){\idgr{10}}
 \put(60,20){\idgr{10}}
 \put(70,20){\idgr{10}}
 \put(110,20){\idgr{10}}
 \put(120,20){\idgr{10}}
 \put(130,20){\idgr{10}}
 \put(180,20){\idgr{10}}
 \put(240,20){\idgr{10}}
 \put(280,20){\idgr{10}}
 \put(290,20){\idgr{10}}
 \put(50,0){\idgr{30}}
 \put(160,0){\idgr{20}}
 \put(270,10){\idgr{20}}
 \put(0,30){\objo{X}}
 \put(10,30){\objo{P}}
 \put(20,30){\objo{Q}}
 \put(50,30){\objo{X}}
 \put(60,30){\objo{P}}
 \put(70,30){\objo{Q}}
 \put(110,30){\objo{P}}
 \put(120,30){\objo{X}}
 \put(130,30){\objo{Q}}
 \put(160,30){\objo{P}}
 \put(170,30){\objo{X}}
 \put(180,30){\objo{Q}}
 \put(220,30){\objo{X}}
 \put(230,30){\objo{P}}
 \put(240,30){\objo{Q}}
 \put(270,30){\objo{X}}
 \put(280,30){\objo{P}}
 \put(290,30){\objo{Q}}
 \put(40,10){\objo{=}}
 \put(150,10){\objo{=}}
 \put(260,10){\objo{=}}
 \put(33,0){(a)}
 \put(143,0){(b)}
 \put(253,0){(c)}
 \put(0,0){\obju{X}}
 \put(10,0){\obju{P}}
 \put(20,0){\obju{Q}}
 \put(30,0){\obju{M}}
 \put(50,0){\obju{X}}
 \put(60,0){\obju{P}}
 \put(70,0){\obju{Q}}
 \put(80,0){\obju{M}}
 \put(110,0){\obju{X}}
 \put(120,0){\obju{P}}
 \put(130,0){\obju{Q}}
 \put(140,0){\obju{M}}
 \put(160,0){\obju{X}}
 \put(170,0){\obju{P}}
 \put(180,0){\obju{Q}}
 \put(190,0){\obju{M}}
 \put(220,0){\obju{P}}
 \put(230,0){\obju{X}}
 \put(240,0){\obju{Q}}
 \put(250,0){\obju{M}}
 \put(270,0){\obju{P}}
 \put(280,0){\obju{X}}
 \put(290,0){\obju{Q}}
 \put(300,0){\obju{M}}
 \egr
 are equivalent to the following conditions
 \bgr{120}{30}
 \put(90,25){\eval}
 \put(0,10){\multmor{\psi}{50}}
 \put(70,10){\multmor{\psi}{50}}
 \put(0,20){\idgr{10}}
 \put(10,20){\idgr{10}}
 \put(20,20){\idgr{10}}
 \put(30,20){\idgr{10}}
 \put(40,20){\idgr{10}}
 \put(50,20){\idgr{10}}
 \put(70,20){\idgr{10}}
 \put(80,20){\idgr{10}}
 \put(110,20){\idgr{10}}
 \put(120,20){\idgr{10}}
 \put(25,0){\idgr{10}}
 \put(95,0){\idgr{10}}
 \put(0,30){\objo{Q^*}}
 \put(10,30){\objo{P^*}}
 \put(20,30){\objo{X^*}}
 \put(30,30){\objo{X}}
 \put(40,30){\objo{P}}
 \put(50,30){\objo{Q}}
 \put(70,30){\objo{Q^*}}
 \put(80,30){\objo{P^*}}
 \put(90,30){\objo{X^*}}
 \put(100,30){\objo{X}}
 \put(110,30){\objo{P}}
 \put(120,30){\objo{Q}}
 \put(60,10){\objo{=}}
 \put(53,0){(a)}
 \put(25,0){\obju{M}}
 \put(95,0){\obju{M}}
 \egr
 \bgr{270}{40}
 \put(90,25){\eval}
 \put(240,25){\eval}
 \put(100,30){\braid}
 \put(80,30){\ibraid}
 \put(230,30){\braid}
 \put(250,30){\ibraid}
 \put(0,10){\multmor{\psi}{50}}
 \put(70,10){\multmor{\psi}{50}}
 \put(150,10){\multmor{\psi}{50}}
 \put(220,10){\multmor{\psi}{50}}
 \put(0,20){\idgr{20}}
 \put(10,20){\idgr{20}}
 \put(20,20){\idgr{20}}
 \put(30,20){\idgr{20}}
 \put(40,20){\idgr{20}}
 \put(50,20){\idgr{20}}
 \put(70,20){\idgr{20}}
 \put(120,20){\idgr{20}}
 \put(150,20){\idgr{20}}
 \put(160,20){\idgr{20}}
 \put(170,20){\idgr{20}}
 \put(180,20){\idgr{20}}
 \put(190,20){\idgr{20}}
 \put(200,20){\idgr{20}}
 \put(220,20){\idgr{20}}
 \put(270,20){\idgr{20}}
 \put(80,20){\idgr{10}}
 \put(110,20){\idgr{10}}
 \put(230,20){\idgr{10}}
 \put(260,20){\idgr{10}}
 \put(25,0){\idgr{10}}
 \put(95,0){\idgr{10}}
 \put(175,0){\idgr{10}}
 \put(245,0){\idgr{10}}
 \put(0,40){\objo{Q^*}}
 \put(10,40){\objo{X^*}}
 \put(20,40){\objo{P^*}}
 \put(30,40){\objo{P}}
 \put(40,40){\objo{X}}
 \put(50,40){\objo{Q}}
 \put(70,40){\objo{Q^*}}
 \put(80,40){\objo{X^*}}
 \put(90,40){\objo{P^*}}
 \put(100,40){\objo{P}}
 \put(110,40){\objo{X}}
 \put(120,40){\objo{Q}}
 \put(150,40){\objo{Q^*}}
 \put(160,40){\objo{X^*}}
 \put(170,40){\objo{P^*}}
 \put(180,40){\objo{P}}
 \put(190,40){\objo{X}}
 \put(200,40){\objo{Q}}
 \put(220,40){\objo{Q^*}}
 \put(230,40){\objo{X^*}}
 \put(240,40){\objo{P^*}}
 \put(250,40){\objo{P}}
 \put(260,40){\objo{X}}
 \put(270,40){\objo{Q}}
 \put(60,15){\objo{=}}
 \put(210,15){\objo{=}}
 \put(53,0){(b)}
 \put(203,0){(c)}
 \put(25,0){\obju{M}}
 \put(95,0){\obju{M}}
 \put(175,0){\obju{M}}
 \put(245,0){\obju{M}}
 \egr
 In particular the induced cone $\theta_2: \omega(Q)^* \tensor
\omega(P)^* \tensor \omega(P) \tensor \omega(Q) \=> B \tensor B$
satisfies the conditions (a), (b), and (c). Observe that condition (b)
for $\varphi = \delta_2$ implies condition (c) for $\delta_2$ since
$B$ is $\C_0$-central. In fact (b) implies

 \bgr{220}{40}
 \put(60,10){\braid}
 \put(120,30){\braid}
 \put(190,10){\braid}
 \put(200,0){\braid}
 \put(60,0){\ibraid}
 \put(120,10){\ibraid}
 \put(0,20){\multmor{\delta_2}{40}}
 \put(60,20){\multmor{\delta_2}{40}}
 \put(130,20){\multmor{\delta_2}{30}}
 \put(180,20){\multmor{\delta}{10}}
 \put(210,20){\multmor{\delta}{10}}
 \put(0,0){\idgr{20}}
 \put(10,0){\idgr{20}}
 \put(20,0){\idgr{20}}
 \put(30,0){\idgr{20}}
 \put(40,0){\idgr{20}}
 \put(80,0){\idgr{20}}
 \put(90,0){\idgr{20}}
 \put(100,0){\idgr{20}}
 \put(140,0){\idgr{20}}
 \put(150,0){\idgr{20}}
 \put(160,0){\idgr{20}}
 \put(180,0){\idgr{20}}
 \put(220,0){\idgr{20}}
 \put(0,30){\idgr{10}}
 \put(10,30){\idgr{10}}
 \put(20,30){\idgr{10}}
 \put(60,30){\idgr{10}}
 \put(70,30){\idgr{10}}
 \put(80,30){\idgr{10}}
 \put(140,30){\idgr{10}}
 \put(180,30){\idgr{10}}
 \put(210,30){\idgr{10}}
 \put(120,0){\idgr{10}}
 \put(120,20){\idgr{10}}
 \put(130,0){\idgr{10}}
 \put(190,0){\idgr{10}}
 \put(200,20){\idgr{20}}
 \put(210,10){\idgr{10}}
 \put(0,40){\objo{P}}
 \put(10,40){\objo{X}}
 \put(20,40){\objo{Q}}
 \put(60,40){\objo{P}}
 \put(70,40){\objo{X}}
 \put(80,40){\objo{Q}}
 \put(120,40){\objo{P}}
 \put(130,40){\objo{X}}
 \put(140,40){\objo{Q}}
 \put(180,40){\objo{P}}
 \put(200,40){\objo{X}}
 \put(210,40){\objo{Q}}
 \put(50,20){\objo{=}}
 \put(110,20){\objo{=}}
 \put(170,20){\objo{=}}
 \put(0,0){\obju{P}}
 \put(10,0){\obju{X}}
 \put(20,0){\obju{Q}}
 \put(30,0){\obju{B}}
 \put(40,0){\obju{B}}
 \put(60,0){\obju{P}}
 \put(70,0){\obju{X}}
 \put(80,0){\obju{Q}}
 \put(90,0){\obju{B}}
 \put(100,0){\obju{B}}
 \put(120,0){\obju{P}}
 \put(130,0){\obju{X}}
 \put(140,0){\obju{Q}}
 \put(150,0){\obju{B}}
 \put(160,0){\obju{B}}
 \put(180,0){\obju{P}}
 \put(190,0){\obju{X}}
 \put(200,0){\obju{Q}}
 \put(210,0){\obju{B}}
 \put(220,0){\obju{B}}
 \egr
 hence
 \bgr{280}{50}
 \put(0,40){\braid}
 \put(60,40){\braid}
 \put(70,10){\braid}
 \put(80,0){\braid}
 \put(120,40){\braid}
 \put(140,0){\braid}
 \put(180,10){\braid}
 \put(200,10){\braid}
 \put(240,10){\braid}
 \put(130,10){\ibraid}
 \put(80,30){\twist{-10}{10}}
 \put(140,30){\twist{-10}{10}}
 \put(0,20){\multmor{\delta_2}{40}}
 \put(250,20){\multmor{\delta_2}{30}}
 \put(60,20){\multmor{\delta}{10}}
 \put(90,20){\multmor{\delta}{10}}
 \put(120,20){\multmor{\delta}{10}}
 \put(150,20){\multmor{\delta}{10}}
 \put(190,20){\multmor{\delta}{10}}
 \put(210,20){\multmor{\delta}{10}}
 \put(20,30){\idgr{20}}
 \put(90,30){\idgr{20}}
 \put(150,30){\idgr{20}}
 \put(190,30){\idgr{20}}
 \put(210,30){\idgr{20}}
 \put(250,30){\idgr{20}}
 \put(260,30){\idgr{20}}
 \put(0,0){\idgr{20}}
 \put(10,0){\idgr{20}}
 \put(20,0){\idgr{20}}
 \put(30,0){\idgr{20}}
 \put(40,0){\idgr{20}}
 \put(60,0){\idgr{20}}
 \put(100,0){\idgr{20}}
 \put(120,0){\idgr{20}}
 \put(160,0){\idgr{20}}
 \put(220,0){\idgr{20}}
 \put(260,0){\idgr{20}}
 \put(270,0){\idgr{20}}
 \put(280,0){\idgr{20}}
 \put(70,0){\idgr{10}}
 \put(130,0){\idgr{10}}
 \put(180,0){\idgr{10}}
 \put(190,0){\idgr{10}}
 \put(200,0){\idgr{10}}
 \put(210,0){\idgr{10}}
 \put(240,0){\idgr{10}}
 \put(250,0){\idgr{10}}
 \put(0,30){\idgr{10}}
 \put(10,30){\idgr{10}}
 \put(60,30){\idgr{10}}
 \put(80,20){\idgr{10}}
 \put(90,10){\idgr{10}}
 \put(120,30){\idgr{10}}
 \put(140,20){\idgr{10}}
 \put(150,10){\idgr{10}}
 \put(180,20){\idgr{30}}
 \put(240,20){\idgr{30}}
 \put(0,50){\objo{X}}
 \put(10,50){\objo{P}}
 \put(20,50){\objo{Q}}
 \put(60,50){\objo{X}}
 \put(70,50){\objo{P}}
 \put(90,50){\objo{Q}}
 \put(120,50){\objo{X}}
 \put(130,50){\objo{P}}
 \put(150,50){\objo{Q}}
 \put(180,50){\objo{X}}
 \put(190,50){\objo{P}}
 \put(210,50){\objo{Q}}
 \put(240,50){\objo{X}}
 \put(250,50){\objo{P}}
 \put(260,50){\objo{Q}}
 \put(50,20){\objo{=}}
 \put(110,20){\objo{=}}
 \put(170,20){\objo{=}}
 \put(230,20){\objo{=}}
 \put(0,0){\obju{P}}
 \put(10,0){\obju{X}}
 \put(20,0){\obju{Q}}
 \put(30,0){\obju{B}}
 \put(40,0){\obju{B}}
 \put(60,0){\obju{P}}
 \put(70,0){\obju{X}}
 \put(80,0){\obju{Q}}
 \put(90,0){\obju{B}}
 \put(100,0){\obju{B}}
 \put(120,0){\obju{P}}
 \put(130,0){\obju{X}}
 \put(140,0){\obju{Q}}
 \put(150,0){\obju{B}}
 \put(160,0){\obju{B}}
 \put(180,0){\obju{P}}
 \put(190,0){\obju{X}}
 \put(200,0){\obju{Q}}
 \put(210,0){\obju{B}}
 \put(220,0){\obju{B}}
 \put(240,0){\obju{P}}
 \put(250,0){\obju{X}}
 \put(260,0){\obju{Q}}
 \put(270,0){\obju{B}}
 \put(280,0){\obju{B}}
 \egr
 So condition (b) for $\psi = \theta_2$ implies condition (c) for
$\theta_2$. Now we show that $\theta_2: \omega(Q)^* \tensor
\omega(P)^* \tensor \omega(P) \tensor \omega(Q) \=> B \tensor B$ is
the injection morphism of a colimit. Then $\delta_2: \omega \tensor
\omega \=> \omega \tensor \omega \tensor B \tensor B$ is a universal
$\C_0$-morphism.

 We have already seen that diagram (2) is a tensor product of wedges
and cone morphisms of the type of diagram (1). Now we show that the
relations (a) and (b) come about as tensor products of relations for
$B$. We don't have to consider condition (c) which is automatically
satisfied. The diagrams
 \bgr{260}{60}
 \put(150,25){\eval}
 \put(230,55){\eval}
 \put(70,50){\ibraid}
 \put(80,40){\ibraid}
 \put(90,30){\ibraid}
 \put(100,20){\ibraid}
 \put(140,50){\ibraid}
 \put(150,40){\ibraid}
 \put(160,30){\ibraid}
 \put(170,20){\ibraid}
 \put(0,25){\multmor{\theta_2}{50}}
 \put(210,25){\multmor{\theta_2}{50}}
 \put(70,10){\multmor{\theta}{30}}
 \put(110,10){\multmor{\theta}{10}}
 \put(140,10){\multmor{\theta}{30}}
 \put(180,10){\multmor{\theta}{10}}
 \put(0,35){\idgr{25}}
 \put(10,35){\idgr{25}}
 \put(20,35){\idgr{25}}
 \put(30,35){\idgr{25}}
 \put(40,35){\idgr{25}}
 \put(50,35){\idgr{25}}
 \put(210,35){\idgr{25}}
 \put(220,35){\idgr{25}}
 \put(250,35){\idgr{25}}
 \put(260,35){\idgr{25}}
 \put(20,0){\idgr{25}}
 \put(30,0){\idgr{25}}
 \put(70,20){\idgr{30}}
 \put(80,20){\idgr{20}}
 \put(85,0){\idgr{10}}
 \put(90,20){\idgr{10}}
 \put(90,50){\idgr{10}}
 \put(100,40){\idgr{20}}
 \put(110,30){\idgr{30}}
 \put(115,0){\idgr{10}}
 \put(120,20){\idgr{40}}
 \put(140,20){\idgr{30}}
 \put(150,30){\idgr{10}}
 \put(155,0){\idgr{10}}
 \put(160,50){\idgr{10}}
 \put(170,40){\idgr{20}}
 \put(180,30){\idgr{30}}
 \put(185,0){\idgr{10}}
 \put(190,20){\idgr{40}}
 \put(230,0){\idgr{25}}
 \put(240,0){\idgr{25}}
 \put(0,60){\objo{Q^*}}
 \put(10,60){\objo{P^*}}
 \put(20,60){\objo{X^*}}
 \put(30,60){\objo{X}}
 \put(40,60){\objo{P}}
 \put(50,60){\objo{Q}}
 \put(70,60){\objo{Q^*}}
 \put(80,60){\objo{P^*}}
 \put(90,60){\objo{X^*}}
 \put(100,60){\objo{X}}
 \put(110,60){\objo{P}}
 \put(120,60){\objo{Q}}
 \put(140,60){\objo{Q^*}}
 \put(150,60){\objo{P^*}}
 \put(160,60){\objo{X^*}}
 \put(170,60){\objo{X}}
 \put(180,60){\objo{P}}
 \put(190,60){\objo{Q}}
 \put(210,60){\objo{Q^*}}
 \put(220,60){\objo{P^*}}
 \put(230,60){\objo{X^*}}
 \put(240,60){\objo{X}}
 \put(250,60){\objo{P}}
 \put(260,60){\objo{Q}}
 \put(60,25){\objo{=}}
 \put(130,25){\objo{=}}
 \put(200,25){\objo{=}}
 \put(20,0){\obju{B}}
 \put(30,0){\obju{B}}
 \put(85,0){\obju{B}}
 \put(115,0){\obju{B}}
 \put(155,0){\obju{B}}
 \put(185,0){\obju{B}}
 \put(230,0){\obju{B}}
 \put(240,0){\obju{B}}
 \egr
 and
 \bgr{260}{50}
 \put(170,25){\eval}
 \put(230,35){\eval}
 \put(240,40){\braid}
 \put(70,30){\ibraid}
 \put(80,20){\ibraid}
 \put(80,40){\ibraid}
 \put(90,30){\ibraid}
 \put(140,30){\ibraid}
 \put(150,20){\ibraid}
 \put(150,40){\ibraid}
 \put(160,30){\ibraid}
 \put(220,40){\ibraid}
 \put(0,20){\multmor{\theta_2}{50}}
 \put(210,20){\multmor{\theta_2}{50}}
 \put(70,10){\multmor{\theta}{10}}
 \put(90,10){\multmor{\theta}{30}}
 \put(140,10){\multmor{\theta}{10}}
 \put(160,10){\multmor{\theta}{30}}
 \put(0,30){\idgr{20}}
 \put(10,30){\idgr{20}}
 \put(20,30){\idgr{20}}
 \put(30,30){\idgr{20}}
 \put(40,30){\idgr{20}}
 \put(50,30){\idgr{20}}
 \put(210,30){\idgr{20}}
 \put(260,30){\idgr{20}}
 \put(20,0){\idgr{20}}
 \put(30,0){\idgr{20}}
 \put(230,0){\idgr{20}}
 \put(240,0){\idgr{20}}
 \put(75,0){\idgr{10}}
 \put(105,0){\idgr{10}}
 \put(145,0){\idgr{10}}
 \put(175,0){\idgr{10}}
 \put(70,20){\idgr{10}}
 \put(70,40){\idgr{10}}
 \put(100,20){\idgr{10}}
 \put(100,40){\idgr{10}}
 \put(110,20){\idgr{30}}
 \put(120,20){\idgr{30}}
 \put(140,20){\idgr{10}}
 \put(140,40){\idgr{10}}
 \put(170,40){\idgr{10}}
 \put(180,30){\idgr{20}}
 \put(190,20){\idgr{30}}
 \put(220,30){\idgr{10}}
 \put(250,30){\idgr{10}}
 \put(0,50){\objo{Q^*}}
 \put(10,50){\objo{X^*}}
 \put(20,50){\objo{P^*}}
 \put(30,50){\objo{P}}
 \put(40,50){\objo{X}}
 \put(50,50){\objo{Q}}
 \put(70,50){\objo{Q^*}}
 \put(80,50){\objo{X^*}}
 \put(90,50){\objo{P^*}}
 \put(100,50){\objo{P}}
 \put(110,50){\objo{X}}
 \put(120,50){\objo{Q}}
 \put(140,50){\objo{Q^*}}
 \put(150,50){\objo{X^*}}
 \put(160,50){\objo{P^*}}
 \put(170,50){\objo{P}}
 \put(180,50){\objo{X}}
 \put(190,50){\objo{Q}}
 \put(210,50){\objo{Q^*}}
 \put(220,50){\objo{X^*}}
 \put(230,50){\objo{P^*}}
 \put(240,50){\objo{P}}
 \put(250,50){\objo{X}}
 \put(260,50){\objo{Q}}
 \put(60,25){\objo{=}}
 \put(130,25){\objo{=}}
 \put(200,25){\objo{=}}
 \put(20,0){\obju{B}}
 \put(30,0){\obju{B}}
 \put(75,0){\obju{B}}
 \put(105,0){\obju{B}}
 \put(145,0){\obju{B}}
 \put(175,0){\obju{B}}
 \put(230,0){\obju{B}}
 \put(240,0){\obju{B}}
 \egr
 show that relations (a) and (b) are tensor products with the relation
 \bgr{80}{30}
 \put(60,25){\eval}
 \put(60,20){\twist{-10}{10}}
 \put(70,20){\twist{10}{10}}
 \put(0,10){\multmor{\theta}{30}}
 \put(60,10){\multmor{\theta}{10}}
 \put(0,20){\idgr{10}}
 \put(10,20){\idgr{10}}
 \put(20,20){\idgr{10}}
 \put(30,20){\idgr{10}}
 \put(15,0){\idgr{10}}
 \put(65,0){\idgr{10}}
 \put(0,30){\objo{P^*}}
 \put(10,30){\objo{X^*}}
 \put(20,30){\objo{X}}
 \put(30,30){\objo{P}}
 \put(50,30){\objo{P^*}}
 \put(60,30){\objo{X^*}}
 \put(70,30){\objo{X}}
 \put(80,30){\objo{P}}
 \put(45,12){\objo{=}}
 \put(15,0){\obju{B}}
 \put(65,0){\obju{B}}
 \egr
 used in the construction of $B$ (up to a preceding isomorphism).
Since we are now considering a tensor product of two diagrams and the
colimit thereof and since tensor products preserve colimits we have
proved the claimed result.
 \end{pf}

 \subsubsection{}
 Using the results of section 3 we obtain uniquely determined
coalgebra, bialgebra, and Hopf algebra structures (depending on the
given functor) on $\coend(\omega)$ and $\coend_{\C_0}(\omega)$.

 \section{Hidden symmetries}

 In section 4.1 we studied under which circumstances coalgebras and
bialgebras (possibly with a transmuted multiplication) can be
reconstructed from their categories of comodules and the functor
$\omega: \C^C \=> \C$. We saw that they are obtained as the
representing object $\coend_\C(\omega)$ of $\Nat_{\C}(\omega, \omega
\tensor \X ) : \C \=> \Set$. In this section we will see that this
reconstruction depends strongly on the choice of the control category
$\C$. If $\C$ is decreased to a category $\D$ then the representing
(reconstructed) object $\coend_\D(\omega)$ becomes larger. We will see
that under certain conditions the reconstructed object decomposes into
a cosmash product where one factor represents the ``hidden
symmetries''.

 \subsection{Functors of control categories}

 We consider of a braided monoidal functor $\F: \D \=> \C$ of control
categories.

 \subsubsection{}
 If $\B$ is a $\C$-category via $\tensor: \C \times \B \=> \B$, then
$\B$ becomes a $\D$-category by
 $$\tensor: \D \times \B \buildrel \F \times 1 \over \longrightarrow
\C \times \B \buildrel \tensor \over \longrightarrow \B$$
 with associativity morphism $\beta (\upsilon \tensor 1): (X \tensor
Y) \tensor P \=> X \tensor (Y \tensor P)$ and unary action $\pi
(\varsigma \tensor 1): I_\D \tensor P \=> P$ for $X,Y \in \D$ and $P
\in \B$.

 \subsubsection{}
 If $\chi: \B \=> \B'$ is a $\C$-functor, then $\chi$ becomes also a
$\D$-functor. If $\chi: \B \times \B' \=> \B''$ is a $\C$-bifunctor,
then $\chi$ becomes also a $\D$-bifunctor. In both cases the structure
morphisms $\zeta$ resp. $\tau$ remain unchanged.

 If $\chi, \chi': \B \=> \B'$ are $\C$-functors and $\zeta: \chi \=>
\chi'$ is a $\C$-morphism, then $\zeta$ is also a $\D$-morphism.

 \subsubsection{}
 If $\A$ is a $\C$-monoidal category, then it becomes also a
 $\D$-monoidal category.

 The above observations give immediately

 \begin{prop}
 If $\F: \D \=> \C$ is a braided monoidal functor, then it induces an
``underlying'' functor $\J(\F): \J(\C) \=> \J(\D)$.
 \end{prop}

 \subsubsection{}
 Let $(\B,\omega) \in \J(\C)$ and consider $\J(\F)(\B,\omega) =
(\B,\omega)$ with the induced structure morphisms. Assume that
$\Nat_\C(\omega,\omega \tensor \X )$ and $\Nat_\D(\omega,\omega
\tensor \X )$ are representable by $\coend_\C(\omega)$ resp.
$\coend_\D(\omega)$. Then $\A(\coend_C(\omega),M) \iso
\Nat_\C(\omega,\omega \tensor M) \subseteq \Nat_\D(\omega,\omega
\tensor M) \iso \A(\coend_D(\omega),M)$ as functors in $M \in \A$
hence there is an epimorphism of the representing objects
$\coend_\F(\omega): \coend_\D(\omega) \=> \coend_\C(\omega)$.

 \begin{thm} \label{indcoendmor}
 Let $\F:\D \=> \C$ be a braided monoidal functor. Let $\A$ be a
 $\C$-monoidal category, $\B$ a $\C$-category and $\omega: \B \=> \A$
a $\C$-functor. Assume that $\coend_\C(\omega)$ and $\coend_D(\omega)$
exist. Then there is an induced epimorphism of coalgebras
$\coend_\F(\omega): \coend_D(\omega) \=> \coend_\C(\omega)$ in $\C$.

 If in addition the comodule $(\coend_\C(\omega),\Delta)$ is liftable
along $\omega$ then $\coend_\F(\omega): \coend_D(\omega) \=>
\coend_\C(\omega)$ is a retraction of objects in $\C$.
 \end{thm}

 \begin{pf}
 Let $C := \coend_\C(\omega)$ and $D := \coend_\D(\omega)$. Let
$\delta: \omega \=> \omega \tensor C$ and $\partial: \omega \=> \omega
\tensor D$ be the universal morphisms. Write $C^2 := C \tensor C$,
$D^2 = D \tensor D$, and $f = \coend_\F(\omega)$. Then the
commutativity of
 $$\bfig
 \putmorphism(0, 1000)(1, 0)[\omega`\omega \tensor D`\sst
\partial]{800}1a
 \putmorphism(0, 200)(1, 0)[\omega \tensor D`\omega \tensor D^2`\sst
]{800}1a
 \putmorphism(0, 200)(1, 0)[``\sst \partial \tensor 1]{600}0a
 \putmorphism(400, 800)(1, 0)[\omega`\omega \tensor C`\sst
\delta]{800}1b
 \putmorphism(400, 0)(1, 0)[\omega \tensor C`\omega \tensor C^2`\sst
\delta \tensor 1]{800}1b
 \putmorphism(0, 1000)(2, -1)[``\sst =]{400}1l
 \putmorphism(800, 1000)(2, -1)[``\sst 1 \tensor f]{400}1r
 \putmorphism(0, 200)(2, -1)[``\sst ]{400}1l
 \putmorphism(0, 200)(2, -1)[``\sst 1 \tensor f]{500}0l
 \putmorphism(800, 200)(2, -1)[``\sst ]{400}1l
 \putmorphism(800, 200)(2, -1)[``\sst 1 \tensor f \tensor f]{500}0l
 \putmorphism(0, 1000)(0, -1)[``\sst \partial]{800}1r
 \putmorphism(400, 800)(0, -1)[``\sst \delta]{800}1r
 \putmorphism(800, 1000)(0, -1)[``\sst 1 \tensor \Delta]{800}1r
 \putmorphism(1200, 800)(0, -1)[``\sst 1 \tensor \Delta]{800}1r
 \efig$$
 and
  $$\bfig
 \putmorphism(0, 800)(1, 0)[\omega`\omega \tensor D`\sst
\partial]{1200}1a
 \putmorphism(0, 800)(3, -2)[`\omega \tensor C`\sst \delta]{600}1r
 \putmorphism(0, 800)(3, -4)[`\omega \tensor I`\sst \iso]{600}1l
 \putmorphism(1200, 800)(-3, -2)[``\sst 1 \tensor f]{600}1l
 \putmorphism(1200, 800)(-3, -4)[``\sst 1 \tensor \varepsilon]{600}1r
 \putmorphism(600, 400)(0, -1)[``\sst 1 \tensor \varepsilon]{400}1r
 \efig$$
 show that $f: D \=> C$ is a coalgebra morphism.

 Let $g: \omega(\widetilde C) \=> C$ be a $C$-comodule isomorphism.
Then the following diagram commutes
 $$\bfig
 \putmorphism(0, 800)(1, 0)[\omega(\widetilde C)`\omega(\widetilde C)
\tensor D`\sst \partial]{600}1a
 \putmorphism(600, 800)(1, 0)[\phantom{\omega(\widetilde C) \tensor
D}`I \tensor D \iso D`\sst \varepsilon g \tensor 1]{800}1a
 \putmorphism(0, 800)(3, -2)[`\omega(\widetilde C) \tensor C`\sst
\delta]{600}1l
 \putmorphism(600, 800)(0, -1)[``\sst 1 \tensor f ]{400}1r
 \putmorphism(1200, 800)(0, -1)[``\sst 1 \tensor f]{800}1l
 \putmorphism(1600, 800)(0, -1)[``\sst f]{800}1r
 \putmorphism(0, 800)(0, -1)[`C`\sst g ]{400}1l
 \putmorphism(600, 400)(0, -1)[``\sst g \tensor 1]{400}1l
 \putmorphism(0, 400)(3, -2)[``\sst \Delta]{600}1l
 \putmorphism(600, 0)(1, 0)[C \tensor C`I \tensor C \iso C.`\sst
\varepsilon \tensor 1]{800}1a
 \efig$$
 The lower morphism of the diagram is the identity hence $f$ is a
retraction in $\A$.
 \end{pf}

 \subsubsection{}
 Observe that the morphism $\coend_\F(\omega): \coend_\D(\omega) \=>
\coend_\C(\omega)$ is the uniquely defined morphism such that
$(1_\omega \tensor \coend_\F(\omega)) \circ \partial = \delta$.

 \subsubsection{}
 The preceding theorem shows that the reconstructed coalgebra $D$
w.r.t. $\D$ is larger than $C$. We consider the additional part in $D$
as hidden symmetries in the sense described in the introduction. It is
responsible for $\D$-morphisms $\varphi: \omega \=> \omega \tensor M$
which are not $\C$-morphisms or certain elements in $\Nat_\D(\omega,
\omega \tensor M)$ which are not contained in $\Nat_\C(\omega, \omega
\tensor M)$. As we have seen this part of $D$ tends to split off.

 An example of a hidden symmetry can be obtained for superalgebra
representations. This is dual to the above considerations. Given an
algebra $A$ considered as a superalgebra $(A,0)$. Consider $\A = \C$,
the category of super vector spaces ($\kl{\Bbb Z}_2$-Comod), the
category $\B = \C_A$ of super $A$-modules, and the forgetful functor
$\omega: \C_A \=> \C$. Let $\F : \D = \Vek \=> \C$ be the functor
which sends each vector space $V$ to the super vector space $(V,0)$.
Any $\D$-morphism $\varphi: \omega \=> \omega$ is described by its
image under $\varphi \in \Nat_\D(\omega \tensor I, \omega) \iso \A(I,
\rend_\D(\omega))$.

 The natural transformation $\varphi:P \=> P$ given by $(p_0,p_1)
\mapsto (p_0,-p_1)$ is a symmetry for all representations of $A$ (a
natural automorphism of $\omega$), which is not induced by the
multiplication with any element of $A$. A multiplication with an
element $a = (a,0) \in A$ on $A$-modules $(P_0,P_1)$ in $\C_A$ would
have to satisfy $(p_0,p_1)a = (p_0a, p_1a) = (p_0,-p_1)$ for all
choices of $(p_0,p_1)$ which is not possible. The natural
transformation $\varphi$ is, however, a $\D$-morphism and thus
comes from multiplication with an element $b \in \rend_\D(\omega)$, in
fact from the element $e_1-e_t \in (\kl{\Bbb Z}_2)^* \subseteq
\rend_\D(\omega)$ where $e_1,e_t$ is the dual basis to $1,t \in
\kl[t]/(t^2-1) = \kl{\Bbb Z}_2$.

 \subsubsection{} \label{symmgroups}
 A special case of the preceding example occurs in representation
theory of groups as discussed in the introduction. If we consider
representations of a group $G$ in vector spaces over a field $\kl$,
i.e. the category $\M_{\kl G}$, then each element $g \in G$ induces a
$\C$-monoidal automorphism $\varphi_g: \omega \=> \omega$ where
$\omega: \M_{\kl G} \=> \M = \Vek = \C$ (observe that any natural
transformation of functors into $\C$ is a $\C$-morphism by Theorem
\ref{nattransisC}). Conversely given any $\C$-monoidal automorphism
$\varphi: \omega \=> \omega$ there is precisely one $g \in G$ with
$\varphi = \varphi_g$. Thus $G$ can be reconstructed from its
representations, i.e. from $\omega: \M_{\kl G} \=> \M$.

 To consider representations of $G$ in super vector spaces over $\kl$
let $\C = \M^{\kl {\Bbb Z}_2} = \A$, $\F: \D = \Vek \=> \C = \M^{\kl
{\Bbb Z}_2}$, and $\omega: \A_{\kl G} \=> \A$. We may consider $\kl G$
as a Hopf algebra $(\kl G,0)$ in $\A$ and have $(p_0,p_1)g =
(p_0g,p_1g)$ with a suitable $G$-structure on $P_0$ and $P_1$
separately. Then each element $g \in G$ induces a $\C$-monoidal
automorphism $\varphi_g: \omega \=> \omega$. For any $\C$-monoidal
automorphism there is precisely one $g \in G$ with $\varphi =
\varphi_g$. For the $\D$-monoidal automorphism $\varphi: \omega \=>
\omega$ with $\varphi(P_0,P_1)(p_0,p_1) := (p_0,-p_1)$ there is,
however, no $g \in G$ with $\varphi = \varphi_g$.

 \subsubsection{}
 We consider now the situation of a morphism $[\chi,\zeta]:
(\B,\omega) \=> (\B',\omega')$ in $\J(\C)$ together with a braided
monoidal functor $\F: \D \=> \C$. Assume the universal objects and
morphisms $\delta: \omega \=> \omega \tensor C$, $\partial: \omega \=>
\omega \tensor D$, $\delta': \omega' \=> \omega' \tensor C'$, and
$\partial': \omega' \=> \omega' \tensor D'$ exist. Then by
\ref{coalgmor} we get induced morphisms $z: C \=> C'$ and $y: D \=>
D'$ such that $(\zeta \tensor z) \circ \delta = \delta' \chi \circ
\zeta$ and $(\zeta \tensor y) \circ \partial = \partial' \chi \circ
\zeta$. Furthermore by Theorem \ref{indcoendmor} there are induced
morphisms $f := \coend_\F(\omega): D \=> C$ and  $f' :=
\coend_\F(\omega'): D' \=> C'$ such that $(1_\omega \tensor f) \circ
\partial = \delta$ and $(1_{\omega'} \tensor f') \circ \partial' =
\delta'$. Hence by the universal property of $\partial$ the diagram
 $$\bfig
 \putmorphism(0, 1000)(1, 0)[\omega`\omega \tensor D`\sst
\partial]{600}1a
 \putmorphism(0, 400)(1, 0)[\omega'\chi`\omega'\chi \tensor D'`\sst
\partial'\chi]{600}1a
 \putmorphism(0, 1000)(0, -1)[``\sst \zeta]{600}1l
 \putmorphism(600, 1000)(0, -1)[``\sst \zeta \tensor y]{600}1l
 \putmorphism(1200, 600)(0, -1)[\omega \tensor C`\omega' \chi \tensor
C'`\sst \zeta \tensor z]{600}1r
 \putmorphism(600, 1000)(3, -2)[``\sst 1 \tensor f]{600}1r
 \putmorphism(600, 400)(3, -2)[``\sst 1 \tensor f']{600}1r
 \putmorphism(0, 1000)(3, -1)[``\sst \delta]{1200}1r
 \putmorphism(0, 400)(3, -1)[``\sst \delta'\chi]{1200}1r
 \efig$$
 commutes and from $\zeta \tensor zf = (\zeta \tensor z)(1 \tensor f)
= (1 \tensor f')(\zeta \tensor y) = \zeta \tensor f' y$ and the
uniqueness of induced morphisms we get a commutative diagram of
coalgebra morphisms
 $$\bfig
 \putmorphism(0, 400)(1, 0)[D`C`\sst f]{400}1a
 \putmorphism(0, 0)(1, 0)[D'`C'`\sst f']{400}1a
 \putmorphism(0, 400)(0, -1)[``\sst y]{400}1l
 \putmorphism(400, 400)(0, -1)[``\sst z]{400}1r
 \efig$$ or
 $$\coend_\C([\chi,\zeta]) \coend_\F(\omega) = \coend_\F(\omega')
\coend_\D([\chi,\zeta]).$$
 In particular we have proved

 \begin{thm}
 Let $\F: \D \=> \C$ be a braided monoidal functor. Assume that the
functors $\coend_\C: \J(\C) \=> \A\coalg$ and $\coend_\D: \J(\C) \=>
\A\coalg$ exist. Then $\coend_\F: \coend_\D \=> \coend_\C$ is a
natural epimorphism of functors from $\J(\C)$ to $\A\coalg$.
 \end{thm}

 \subsubsection{}
 Assume now that $\B$ is $\C$-monoidal, that $\omega: \B \=> \A$ is a
$\C$-monoidal functor and that $\Nat_\C(\omega, \omega \tensor \X )$
and $\Nat_\D(\omega, \omega \tensor \X )$ are multirepresentable. Then
$z = \coend_\C([\chi,\zeta]): D \=> C$ is a bialgebra morphism. The
multiplicativity follows from the commutative diagram
 $$\bfig
 \putmorphism(0, 1600)(1, 0)[\omega \tensor \omega`\omega \tensor
\omega \tensor D \tensor D`\sst \partial_2]{800}1a
 \putmorphism(0, 400)(1, 0)[\omega(\tensor)`\omega(\tensor) \tensor
D`\sst \partial]{800}1a
 \putmorphism(0, 1600)(0, -1)[``\sst \upsilon^{-1}]{1200}1l
 \putmorphism(800, 1600)(0, -1)[`\omega \tensor \omega \tensor D`\sst
1 \tensor m_D]{600}1l
 \putmorphism(800, 1000)(0, -1)[``\sst \upsilon^{-1} \tensor 1]{600}1l
 \putmorphism(1600, 1200)(0, -1)[\omega \tensor \omega \tensor C
\tensor C` \omega \tensor \omega \tensor C`\sst 1 \tensor m_C]{600}1r
 \putmorphism(1600, 600)(0, -1)[`\omega \tensor C`\sst \upsilon^{-1}
\tensor 1]{600}1r
 \putmorphism(800, 1600)(2, -1)[``\sst 1 \tensor f \tensor f]{800}1r
 \putmorphism(800, 1000)(2, -1)[``\sst 1 \tensor f]{800}1r
 \putmorphism(800, 400)(2, -1)[``\sst 1 \tensor f]{800}1r
 \putmorphism(0, 1600)(4, -1)[``\sst \delta_2]{1600}1l
 \putmorphism(0, 400)(4, -1)[``\sst \delta]{1600}1l
 \efig$$
 (where $\omega(\tensor)(P,Q) := \omega(P \tensor Q)$) and the unary
property is proved similarly.

 If the bialgebras $C'$ and $D'$ are Hopf algebras then by Corollary
\ref{adjcomodalg} $C$ is a $C'$-comodule coalgebra by the coadjoint
coaction w.r.t. the induced coalgebra morphism $z: C \=> C'$ and $D$
is a $D'$-comodule coalgebra by the coadjoint coaction w.r.t. the
induced coalgebra morphism $y: D \=> D'$. Furthermore $f':D' \=> C'$
is an epimorphism and a bialgebra (Hopf algebra) morphism.

 \subsection{Hidden symmetries of the base category}

 Consider a braided monoidal functor $\F: \D \=> \C$ and a morphism
$[\chi,\zeta]: (\B,\omega) \=> (\B',\omega')$ in $\J(\C)$. Assume that
$\B'$ is a $\C$-monoidal category and $\omega'$ is a $\C$-monoidal
functor. Let $\delta: \omega \=> \omega \tensor D$ be a universal
$\D$-morphism. Let $D'$ be a bialgebra in $\A$ and let $\delta':
\omega' \=> \omega' \tensor D'$ be a $\D$-morphism compatible with the
bialgebra structure of $D'$ (e.g. $\delta'$ is a universal
 $\D$-morphism). Let $E$ be a coalgebra in $\B'$ and let $\mu: \chi
\=> \chi \tensor E$ be a $\C$-morphism compatible with the structure
of $E$ (e.g. a universal $\C$-morphism).

 \begin{thm} \label{smashprd}
 In the setup given above $D' \tensor \omega' E$ carries the structure
of a cosmash product $D' \#^c \omega' E$ and there is a canonical
coalgebra morphism $f: D \=> D' \#^c \omega' E$.
 \end{thm}

 \begin{pf}
 We first observe that $\omega' E$ is a coalgebra in $\A$ since
$\omega'$ is a monoidal functor. Furthermore $\omega' E$ is a
 $D'$-comodule coalgebra by the morphism $\delta' E: \omega' E \=>
\omega' E \tensor D'$.

 Consider the following induced morphism $f: D \=> D' \#^c \omega' E$
defined by
 $$\bfig
 \putmorphism(0, 800)(1, 0)[\omega`\omega \tensor D`\sst
\delta]{2000}1a
 \putmorphism(0, 0)(1, 0)[\omega'(\chi \tensor E)`\omega'\chi \tensor
\omega' E`\sst \upsilon]{800}1a
 \putmorphism(800, 0)(1, 0)[\phantom{\omega'\chi \tensor \omega'
E}`\omega'\chi \tensor D' \tensor \omega' E.`\sst \delta'\chi \tensor
1]{1200}1a
 \putmorphism(0, 800)(0, -1)[`\omega' \chi`\sst \zeta]{400}1l
 \putmorphism(0, 400)(0, -1)[``\sst \omega'\mu]{400}1l
 \putmorphism(2000, 800)(0, -1)[`\omega \tensor D' \tensor \omega'
E`\sst 1 \tensor f]{400}1r
 \putmorphism(2000, 400)(0, -1)[``\sst \zeta \tensor 1]{400}1r
 \efig$$
 Since the composition along the lower edge of the square is a
 $\D$-morphism $\omega \=> \omega \tensor D' \tensor \omega' E$ the
morphism $f$ is uniquely determined. We show that $f$ is a coalgebra
morphism with $D' \tensor \omega' E = D' \#^c \omega' E$ the cosmash
product. For this purpose define a morphism
 $$\tau_0: D' \tensor \omega' E
 \buildrel 1 \tensor \delta' E \over \longrightarrow D' \tensor
\omega' E \tensor D'
 \buildrel \sigma \tensor 1 \over \longrightarrow \omega' E \tensor D'
\tensor D'
 \buildrel 1 \tensor m_{D'} \over \longrightarrow \omega' E \tensor
D'.$$
 Then the following diagram of $\D$-morphisms commutes
  $$\bfig
 \putmorphism(0, 1200)(1, 0)[\omega' \chi` \omega'(\chi \tensor
E)`\sst \omega' \mu]{1000}1a
 \putmorphism(1000, 1200)(1, 0)[\phantom{\omega'(\chi \tensor
E)}`\omega' \chi \tensor \omega' E`\sst \upsilon]{800}1a
 \putmorphism(1800, 1200)(1, 0)[\phantom{\omega' \chi \tensor \omega'
E}`\omega' \chi \tensor D' \tensor \omega' E`\sst \delta' \chi \tensor
1 ]{1000}1a
 \putmorphism(0, 0)(1, 0)[\omega' \chi \tensor D'`\omega'( \chi
\tensor E) \tensor D'`\sst \omega'\mu \tensor 1]{1000}1a
 \putmorphism(1000, 0)(1, 0)[\phantom{\omega'( \chi \tensor E) \tensor
D'}`\omega' \chi \tensor \omega' E \tensor D'`\sst \upsilon \tensor 1
]{1800}1a
 \putmorphism(0, 1200)(0, -1)[``\sst \delta' \chi]{1200}1l
 \putmorphism(1000, 1200)(0, -1)[``\sst \delta' (\chi \tensor
E)]{1200}1r
 \putmorphism(2800, 1200)(0, -1)[`\omega' \chi \tensor D' \tensor
\omega' E \tensor D'`\sst 1 \tensor \delta' E]{400}1r
 \putmorphism(2800, 800)(0, -1)[`\omega' \chi \tensor \omega' E
\tensor D' \tensor D'`\sst 1 \tensor \sigma \tensor 1]{400}1r
 \putmorphism(2800, 400)(0, -1)[``\sst 1 \tensor m_{D'}]{400}1r
 \efig$$
 by the very fact that the bialgebra structure of $D'$ is compatible
with $\delta'$. Hence with suitable identifications we get the
commutative diagram
 $$\bfig
 \putmorphism(0, 800)(1, 0)[\omega' \chi` \omega' \chi \tensor \omega'
E`\sst \omega' \mu]{1000}1a
 \putmorphism(0, 0)(1, 0)[\omega' \chi \tensor D'`\omega' \chi \tensor
\omega' E \tensor D'.`\sst \omega'\mu \tensor 1]{1000}1a
 \putmorphism(0, 800)(0, -1)[``\sst \delta' \chi]{800}1l
 \putmorphism(1000, 800)(0, -1)[`\omega' \chi \tensor D' \tensor
\omega' E`\sst \delta' \chi \tensor 1 ]{400}1r
 \putmorphism(1000, 400)(0, -1)[``\sst 1 \tensor \tau_0]{400}1r
 \efig$$
 Now consider the commutative diagram
 $$\bfig
 \putmorphism(0, 1400)(1, 0)[\omega' \chi`\omega' \chi \tensor D`\sst
\delta]{1200}1a
 \putmorphism(1200, 1400)(1, 0)[\phantom{\omega' \chi \tensor
D}`\omega'\chi \tensor D \tensor D`\sst \delta \tensor 1]{1200}1a
 \putmorphism(0, 1000)(1, 0)[\omega' \chi \tensor \omega' E` \omega'
\chi \tensor D' \tensor \omega' E`\sst \delta' \chi \tensor 1]{1100}1a
 \putmorphism(1800, 800)(1, 0)[\omega' \chi \tensor \omega' E \tensor
D`\omega' \chi \tensor D' \tensor \omega' E \tensor D`\sst \delta'
\chi \tensor 1]{1200}1a
 \putmorphism(1500, 400)(1, 0)[\omega' \chi \tensor \omega' E \tensor
D' \tensor \omega' E`\omega' \chi \tensor D' \tensor \omega' E \tensor
D' \tensor \omega' E`\sst \delta' \chi \tensor 1]{1600}1a
 \putmorphism(1300, 0)(1, 0)[\omega' \chi \tensor D' \tensor \omega' E
\tensor \omega' E `\omega' \chi \tensor D' \tensor D' \tensor \omega'
E \tensor  \omega' E`\sst \delta' \chi \tensor 1]{1700}1a
 \putmorphism(0, 0)(1, 0)[\omega' \chi \tensor \omega' E \tensor
\omega' E` \phantom{\omega' \chi \tensor D' \tensor \omega' E \tensor
\omega' E}`\sst \delta' \chi \tensor 1]{1300}1a

 \putmorphism(0, 1400)(0, -1)[``\sst \omega' \mu]{400}1l
 \putmorphism(0, 1000)(0, -1)[``\sst \omega' \mu \tensor 1]{1000}1r
 \putmorphism(1200, 1400)(0, -1)[``\sst 1 \tensor f]{400}1l
 \putmorphism(1800, 800)(0, -1)[``\sst 1 \tensor f]{400}1r
 \putmorphism(3000, 800)(0, -1)[``\sst 1 \tensor 1 \tensor f]{400}1r

 \putmorphism(1200, 1400)(1, -1)[``\sst \omega' \mu \tensor 1]{600}1r
 \putmorphism(1200, 1000)(1, -1)[``\sst \omega' \mu \tensor 1]{600}1l
 \putmorphism(1800, 400)(-3, -2)[``\sst 1 \tensor \tau_0 \tensor
1]{600}{-1}r
 \putmorphism(3000, 400)(-3, -2)[``\sst 1 \tensor \tau_0 \tensor
1]{600}{-1}r
 \putmorphism(2400, 1400)(1, -1)[``\sst 1 \tensor f \tensor 1]{600}1r
 \efig$$
 From it we get that all morphisms $\omega' \chi \=> \omega' \chi
\tensor D' \tensor \omega' E \tensor D' \tensor \omega' E$ in the
following diagram are equal
 $$\bfig
 \putmorphism(0, 1400)(1, 0)[\omega' \chi`\omega' \chi \tensor D`\sst
\delta]{1200}1a
 \putmorphism(1200, 1400)(1, 0)[\phantom{\omega' \chi \tensor
D}`\omega'\chi \tensor D \tensor D`\sst ]{1200}0a
 \putmorphism(1200, 1380)(1, 0)[\phantom{\omega' \chi \tensor
D}`\phantom{\omega'\chi \tensor D \tensor D}`\sst \delta \tensor
1]{1200}1b
 \putmorphism(1200, 1420)(1, 0)[\phantom{\omega' \chi \tensor
D}`\phantom{ \omega'\chi \tensor D \tensor D}`\sst 1 \tensor
\Delta_D]{1200}1a
 \putmorphism(0, 1000)(1, 0)[\omega' \chi \tensor \omega' E` \omega'
\chi \tensor D' \tensor \omega' E`\sst \delta' \chi \tensor 1]{1100}1a
 \putmorphism(1800, 800)(1, 0)[\omega' \chi \tensor \omega' E \tensor
D`\omega' \chi \tensor D' \tensor \omega' E \tensor D`\sst \delta'
\chi \tensor 1]{1200}1a
 \putmorphism(1500, 400)(1, 0)[\omega' \chi \tensor \omega' E \tensor
D' \tensor \omega' E`\omega' \chi \tensor D' \tensor \omega' E \tensor
D' \tensor \omega' E`\sst \delta' \chi \tensor 1]{1600}1a
 \putmorphism(1300, 0)(1, 0)[\omega' \chi \tensor D' \tensor \omega' E
\tensor \omega' E `\omega' \chi \tensor D' \tensor D' \tensor \omega'
E \tensor  \omega' E`\sst ]{1700}0a
 \putmorphism(1300, 20)(1, 0)[\phantom{\omega' \chi \tensor D' \tensor
\omega' E \tensor \omega' E} `\phantom{\omega' \chi \tensor D' \tensor
D' \tensor \omega' E \tensor  \omega' E} `\sst \delta' \chi \tensor
1]{1700}1a
 \putmorphism(1300, -20)(1, 0)[\phantom{\omega' \chi \tensor D'
\tensor \omega' E \tensor \omega' E} `\phantom{ \omega' \chi \tensor
D' \tensor D' \tensor \omega' E \tensor  \omega' E}`\sst 1 \tensor
\Delta_{D'} \tensor 1]{1700}1b
 \putmorphism(0, 0)(1, 0)[\omega' \chi \tensor \omega' E \tensor
\omega' E` \phantom{\omega' \chi \tensor D' \tensor \omega' E \tensor
\omega' E}`\sst \delta' \chi \tensor 1]{1300}1a

 \putmorphism(0, 1400)(0, -1)[``\sst \omega' \mu]{400}1l
 \putmorphism(20, 1000)(0, -1)[``\sst \omega' \mu \tensor 1]{1000}1r
 \putmorphism(-20, 1000)(0, -1)[``\sst 1 \tensor \Delta_{\omega'
E}]{1000}1l
 \putmorphism(900, 1000)(0, -1)[``\sst 1 \tensor \Delta_{\omega'
E}]{1000}1l
 \putmorphism(1200, 1400)(0, -1)[``\sst 1 \tensor f]{400}1l
 \putmorphism(1800, 800)(0, -1)[``\sst 1 \tensor f]{400}1r
 \putmorphism(3000, 800)(0, -1)[``\sst 1 \tensor 1 \tensor f]{400}1r

 \putmorphism(1200, 1400)(1, -1)[``\sst \omega' \mu \tensor 1]{600}1r
 \putmorphism(1200, 1000)(1, -1)[``\sst \omega' \mu \tensor 1]{600}1l
 \putmorphism(1800, 400)(-3, -2)[``\sst 1 \tensor \tau_0 \tensor
1]{600}{-1}r
 \putmorphism(3000, 400)(-3, -2)[``\sst 1 \tensor \tau_0 \tensor
1]{600}{-1}r
 \putmorphism(2400, 1400)(1, -1)[``\sst 1 \tensor f \tensor 1]{600}1r
 \efig$$
 We define
 $$\Delta_{D' \tensor \omega' E} : D' \tensor \omega' E
 \buildrel \Delta_{D'} \tensor \Delta_{\omega' E} \over
\longrightarrow D' \tensor D' \tensor \omega' E \tensor \omega' E
 \buildrel 1 \tensor \tau_0 \tensor 1 \over \longrightarrow D' \tensor
\omega' E \tensor D' \tensor \omega' E$$
 and observe that $\delta: \omega'\chi \=> \omega'\chi \tensor D$ is a
universal $\D$-morphism. Hence the diagram of induced morphisms
 $$\bfig
 \putmorphism(0, 400)(1, 0)[D`D \tensor D`\sst \Delta_D]{1200}1a
 \putmorphism(0, 0)(1, 0)[D' \tensor \omega' E`D' \tensor \omega' E
\tensor D' \tensor \omega' E`\sst \Delta_{D' \tensor \omega'
E}]{1200}1a
 \putmorphism(0, 400)(0, -1)[``\sst f]{400}1r
 \putmorphism(1200, 400)(0, -1)[``\sst f \tensor f]{400}1r
 \efig$$
 commutes. It is now easy to see that $D' \tensor \omega' E$ with
$\Delta_{D' \tensor \omega' E}$ and $\varepsilon: D' \tensor \omega' E
\buildrel \varepsilon \tensor \omega'\varepsilon \over \longrightarrow
I$ is a coalgebra, the cosmash product $D' \#^c \omega' E$, and that
$f: D \=> D' \tensor \omega' E$ is a coalgebra morphism.
 \end{pf}

 A special application of the theorem is the following

 \begin{cor}
 Let $(\B,\omega)$ be in $\J(\C)$ and let $\F: \D \=> \C$ be a braided
monoidal functor. If $\coend_\D(\omega)$, $\coend_\C(\omega)$, and
$\coend_\D(\id_\A)$ (with $\Nat_\D(\id_\A, \id_\A \tensor \X )$
multirepresentable) exist then there is a canonical coalgebra morphism
 $$f : \coend_\D(\omega) \=> \coend_\D(\id_\A) \#^c
\coend_\C(\omega).$$
 \end{cor}

 In certain cases the canonical morphism of the preceding corollary is
an isomorphism. If this is the case then we have identified the hidden
symmetries of the functor $\omega: \B \=> \A$ as the component $D' =
\coend_\D(\id_\A)$ in the cosmash product.

 \begin{thm}
 Let $\D$ be a braided monoidal category and $H$ be a braided Hopf
algebra in $\D$. Let $\A = \C = \D^H$. Let $C$ be a coalgebra in $\A$
and $\omega: \A^C \=> \A$ be the forgetful functor. Then with
$\coend_\D(\id_\A) \widetilde \#^c \coend_\C(\omega)$ the cosmash
product in $\A$
 $$f: \coend_\D(\omega) \=> \coend_\D(\id_\A) \widetilde \#^c
\coend_\C(\omega)$$
 is an isomorphism of coalgebras in $\A$.
 \end{thm}

 \begin{pf}
 We consider the diagram
 $$\bfig
 \putmorphism(0, 400)(1, 0)[\A^C`\A`\sst \omega]{600}1a
 \putmorphism(0, 400)(3, -4)[`\A`\sst \omega]{300}1l
 \putmorphism(600, 400)(-3, -4)[``\sst \id]{300}1r
 \efig$$
 where $\omega: \A^C \=> \A$ is the forgetful functor together with
the functor $\F: \D \=> \C$ induced by $u:I \=> H$. It is easy to
check that $\A^C = (\D^H)^C$ and $\D^{H \#^c C}$ are isomorphic
 $\D$-categories with the cosmash product formed in $\D$ (for cosmash
products and transmutation see also \ref{twocosmashs}) by sending each
object $(P, \mu: P \=> P \tensor C)$ in $\A^C$ to the object $(P,
\delta: P \buildrel \mu \over \longrightarrow P \tensor C \buildrel
\delta' \tensor 1 \over \longrightarrow P \tensor H \tensor C)$ in
$\D^{H \#^c C}$ (see \ref{cosmashcomods}), where the cosmash
comultiplication is defined by
 $$\begin{array}{c}
 \Delta: H \tensor C
 \buildrel \Delta_H \tensor \Delta_C \over \longrightarrow H \tensor H
\tensor C \tensor C
 \buildrel 1 \tensor \delta'_C \tensor 1 \over \longrightarrow H
\tensor H \tensor C \tensor H \tensor C \\
 \buildrel 1 \tensor \sigma \tensor 1 \over \longrightarrow H \tensor
C \tensor H \tensor H \tensor C
 \buildrel 1 \tensor m_H \tensor 1 \over \longrightarrow H \tensor C
\tensor H \tensor C.
 \end{array}$$
 The morphism $z = 1 \tensor \varepsilon_C: H \#^c C \=> H$  is a
coalgebra morphism. It induces a $\D$-functor $\D^z: \D^{ H \#^c C}
\=> \D^H$ which can be identified with $\omega: \A^C \=> \A$.

 By Theorem \ref{reccoalg} we get $\coend_\D(\omega) = H \#^c C$, the
cosmash product defined in $\D$ as above.

 From Theorem \ref{recbialg} we get $\coend_\D(\id_\A) = H$ as a
coalgebra but with the new multiplication $\widetilde m_H: H \tensor H
\=> H$ as defined in Proposition \ref{fuprbialgebra} by the braid
diagram
 \bgr{80}{40}
 \put(20,10){\mult}
 \put(10,15){\braid}
 \put(0,25){\multmor{\delta}{10}}
 \put(20,25){\multmor{\delta}{10}}
 \put(50,25){\multmor{\delta}{10}}
 \put(70,25){\multmor{\delta}{10}}
 \put(70,5){\multmor{\widetilde m}{10}}
 \put(60,15){\brmor}
 \put(0,0){\idgr{25}}
 \put(5,35){\idgr{5}}
 \put(10,0){\idgr{15}}
 \put(25,0){\idgr{10}}
 \put(25,35){\idgr{5}}
 \put(30,15){\idgr{10}}
 \put(50,0){\idgr{25}}
 \put(55,35){\idgr{5}}
 \put(60,0){\idgr{15}}
 \put(75,0){\idgr{5}}
 \put(75,35){\idgr{5}}
 \put(80,15){\idgr{10}}
 \put(5,40){\objo{P}}
 \put(25,40){\objo{Q}}
 \put(55,40){\objo{P}}
 \put(75,40){\objo{Q}}
 \put(40,15){\objo{=}}
 \put(0,0){\obju{P}}
 \put(10,0){\obju{Q}}
 \put(25,0){\obju{H}}
 \put(50,0){\obju{P}}
 \put(60,0){\obju{Q}}
 \put(75,0){\obju{H}}
 \egr
 In particular $\widetilde m_H: H \tensor H \=> H$ is a morphism of
$H$-comodules under the coadjoint coaction. (This morphism has been
studied in \cite{MAJ3} under the notion of transmutation.)

 Furthermore we have $\coend_\C(\omega) = C$ as coalgebras in $\A$
again by Theorem \ref{reccoalg}.

 By Theorem \ref{smashprd} we thus get a canonical coalgebra morphism
$f: H \#^c C \=> H \widetilde \#^c C$ where the first cosmash product
was described above and the second cosmash product comes from the
transmutation multiplication on $H$ and the braiding in $\A = \D^H$.
As observed in \ref{twocosmashs} these two cosmash products are the
same.

 With these coalgebras the canonical morphism $f$ is defined as in the
proof of Theorem \ref{smashprd} by
 $$\bfig
 \putmorphism(0, 400)(1, 0)[\omega`\omega \tensor H \#^c C`\sst
\delta]{800}1a
 \putmorphism(0, 0)(1, 0)[\omega \tensor C`\omega \tensor H \tensor
C`\sst \delta'\omega \tensor 1]{800}1a
 \putmorphism(0, 400)(0, -1)[``\sst \mu]{400}1l
 \putmorphism(800, 400)(0, -1)[``\sst 1 \tensor f]{400}1r
 \efig$$
 with $\delta = (\delta'\omega \tensor 1)\mu$ as above, so that the
uniquely determined morphism $f$ under the given identifications is
the identity.
 \end{pf}

 \begin{cor}
 Let $H$ be a coquasitriangular Hopf algebra over the field $\kl$ and
let $C$ be an $H$-comodule coalgebra. Let $\omega : (\M^H)^C \=> \M^H$
be the forgetful functor. Then
 $$\coend(\omega) \iso H \#^c C.$$
 \end{cor}

 \begin{pf}
 Use Proposition \ref{nattransisC} to show $\coend(\omega) =
\coend_\M(\omega)$.
 \end{pf}

 The last corollary shows that the hidden symmetries as given in the
examples are represented by the Hopf algebra $H$ (with the
transmutation multiplication), i.e. $H = \coend_{\M^H}(\id: \M^H \=>
\M^H)$.

 \section{Appendix on $\kl$-additive categories and $\C$-categories}

 Let $\kl$ be a commutative ring and let ${\cal C} := \kl\modp$ be the
category of finitely generated projective $\kl$-modules. Then $\C$ is
a symmetric monoidal $\kl$-abelian category.

 Let $\A$ be a category with splitting idempotents, i.e.~for a
morphism $f: X \=> X$ with $f^2 = f$ there are morphisms $q: X \=> P$
and $j: P \=> X$ with $jq = f$ and $qj = 1_P$. Then $q: X \=> P$ is a
coequalizer of $(1_X, f)$ and $q$ and $j$ are unique up to
isomorphisms of $P$.

 \begin{lma} \label{splitting}
 Let $f:X \=> X$ and $g: X \=> X$ be idempotents with $fg = gf$ and
splittings $(P,q,j)$ of $f$ and $(P',q',j')$ of $g$. Let $(R,q'',j'')$
be a splitting of the idempotent $jgq$. Then $(R,q''q,jj'')$ is a
splitting of $fg: X \=> X$.
 \end{lma}

 \begin{pf}
 $qgj$ is idempotent since $qgjqgj = qgfgj = qggfj = qgjqj = qgj$, so
a splitting $(R,q'',j'')$ exists with $qgj = j''q''$ and $q''j'' = 1$.
Then $q''qjj'' = q''j'' = 1_R$ and $jj''q''q = jqgjq = fgf = ffg =
fg$.
 \end{pf}

The following theorem is a generalization of \cite{SCH1} Lemma 2.2.2.

 \begin{thm} \label{addto Ccat}
 Let $\A$ be a $\kl$-additive category with splitting idempotents.
Then $\A$ is a $\C$-category.
 \end{thm}

 \begin{pf}
 Let $\A$ be a $\kl$-additive category. Then the following hold for
$\alpha \in \kl$ and $f,g$ morphisms in $\A$: $\alpha(f + g) = \alpha
f + \alpha g$, $\alpha (fg) = (\alpha f)g = f (\alpha g)$ and $\alpha
(f \oplus g) = (\alpha f) \oplus (\alpha g)$.

 We define the functor $(\kl \tensor \X : \A \=> \A) := (\Id_\A: \A
\=> \A)$. This is a $\kl$-additive functor. For $n \geq 0$ we define
$(\kl^n \tensor \X : \A \=> \A) := (\Id^n: \A \=> \A)$ (so for an
object $P \in \A$ we have $\kl^n \tensor P = P^n$) which again is a
$\kl$-additive functor.

 Let $f: \kl^m \=> \kl^n$ be a morphism in $\C$. Then $f = \sum_{ij}
\alpha_{ij} e_{ij}$ where the $e_{ij}$ are given by the compositions
$e_{ij} := \kl^m \buildrel p_i \over \longrightarrow \kl \buildrel
\iota_j \over \longrightarrow \kl^n$, the canonical basis of the
matrix space $\Hom_\kk(\kl^m,\kl^n)$.

 Since $\A$ is additive, we have corresponding natural transformations
 $$\bar e_{ij}(P): P^m \buildrel {\bar p_i} \over \longrightarrow P
\buildrel {\bar \iota_j} \over \longrightarrow P^n.$$
 For $f: \kl^m \=> \kl^n$ we define $f \tensor P: P^m \=> P^n$ by $f
\tensor P := \sum_{ij}\alpha_{ij} \bar e_{ij}$. Then it is easy to
verify, that $f \tensor \x : \kl^m \tensor \X \=> \kl^n \tensor \X $
is a natural transformation. Furthermore it is easy to see that $fg
\tensor \x = (f \tensor \x )(g \tensor \x )$ and $\id \tensor \x =
\id$. If $\C^-$ is the full subcategory of $\C$ with the objects
$\kl^n$, then $\tensor : \C^- \times \A \=> \A$ is a $\kl$-bilinear
bifunctor.

 Now let $X$ be a finitely generated projective $\kl$-module. Then
there are homomorphisms $j: X \=> \kl^n$ and $q: \kl^n \=> X$ for some
$n \in {\Bbb N}$ with $qj = 1_X$. For $P \in \A$ the morphism $jq
\tensor P: \kl^n \tensor P \=> \kl^n \tensor P$ is an idempotent $(jq
\tensor P)^2 = jq \tensor P$ so there is a splitting $q \tensor P:
\kl^n \tensor P \=> X \tensor P$  and $j \tensor P: X\tensor P \=>
\kl^n \tensor P$ (thus defining $q \tensor P$, $j\tensor P$ and
$X\tensor P$) with
 $$(j \tensor P)(q \tensor P) = jq \tensor P \mbox{ and }
 (q \tensor P)(j \tensor P) = 1_{X \tensor P}.$$
 In particular $q \tensor P: \kl^n \tensor P \=> X \tensor P$ is a
cokernel of $(1-jq) \tensor P: \kl^n \tensor P \=> \kl^n \tensor P$
and we have a commutative diagram
 $$\bfig
 \putmorphism(0, 400)(1, 0)[\kl^n \tensor P`\kl^n \tensor P`\sst
 (1-jq) \tensor P]{1000}1a
 \putmorphism(1000, 400)(1, 0)[\phantom{\kl^n \tensor P}`X \tensor
P`\sst q \tensor P ]{600}1a
 \putmorphism(0, 0)(1, 0)[\kl^n \tensor P`\kl^n \tensor P`\sst (1-jq)
\tensor P]{1000}1a
 \putmorphism(1000, 0)(1, 0)[\phantom{\kl^m \tensor P}`X \tensor
P`\sst q \tensor P ]{600}1a
 \putmorphism(0, 400)(0, -1)[``\sst jq \tensor P]{400}1r
 \putmorphism(1000, 400)(0, -1)[``\sst jq \tensor P]{400}1l
 \putmorphism(1600, 400)(0, -1)[`` = ]{400}1r
 \putmorphism(1600, 400)(-3, -2)[``\sst j \tensor P]{600}1r
 \efig$$
 Let $j: X \=> \kl^n$, $q: \kl^n \=> X$ and $j': X \=> \kl^m$ and $q':
\kl^m \=> X$ be two choices of a representation of $X$ as a direct
summand of a free $\kl$-modules. Then the following diagram commutes
 $$\bfig
 \putmorphism(400, 800)(1, 0)[\kl^n \tensor P`\kl^n \tensor P`\sst
 (1-jq) \tensor P]{1000}1a
 \putmorphism(1400, 800)(1, 0)[\phantom{\kl^n \tensor P}`X \tensor
P`\sst q \tensor P ]{600}1a
 \putmorphism(0, 400)(1, 0)[\kl^m \tensor P`\kl^m \tensor P`\sst
 (1-j'q') \tensor P]{1000}1a
 \putmorphism(1000, 400)(1, 0)[\phantom{\kl^m \tensor P}`[X \tensor
P]`\sst q' \tensor P ]{600}1a
 \putmorphism(400, 0)(1, 0)[\kl^n \tensor P`\kl^n \tensor P`\sst
 (1-jq) \tensor P]{1000}1a
 \putmorphism(1400, 0)(1, 0)[\phantom{\kl^n \tensor P}`X \tensor
P`\sst q \tensor P ]{600}1a
 \putmorphism(400, 400)(0, -1)[``\sst ]{400}1r
 \putmorphism(400, 800)(0, -1)[``\sst jq \tensor P]{350}1r
 \putmorphism(1400, 400)(0, -1)[``\sst ]{400}1r
 \putmorphism(1400, 800)(0, -1)[``\sst jq \tensor P]{350}1r
 \putmorphism(2000, 800)(0, -1)[``\sst 1_{X \tensor P}]{800}1r
 \putmorphism(400, 800)(-1, -1)[``\sst j'q \tensor P]{400}1l
 \putmorphism(1400, 800)(-1, -1)[``\sst j'q \tensor P]{400}1l
 \putmorphism(2000, 800)(-1, -1)[``\sst \cong]{400}1r
 \putmorphism(0, 400)(1, -1)[``\sst jq' \tensor P]{400}1l
 \putmorphism(1000, 400)(1, -1)[``\sst jq' \tensor P]{400}1l
 \putmorphism(1600, 400)(1, -1)[``\sst \cong]{400}1r
 \efig$$
 and the isomorphisms between $X \tensor P$ and $[X \tensor P]$ (the
corresponding object for the second representation of $X$) arise from
this and the symmetric diagram with $(j,q,\kl^n)$ and $(j',q',\kl^m)$
interchanged.

 Since the morphism $(1 - jq) \tensor P$ is a natural transformation
it commutes with morphisms $f: P \=> Q$ and thus induces uniquely
determined morphisms on the cokernels $X \tensor f: X \tensor P \=> X
\tensor Q$, so that the following diagram commutes:
  $$\bfig
 \putmorphism(0, 400)(1, 0)[\kl^n \tensor P`\kl^n \tensor P`\sst
 (1-jq)\tensor P]{1000}1a
 \putmorphism(1000,400)(1, 0)[\phantom{\kl^n \tensor P}`X \tensor
P`\sst ]{600}1a
 \putmorphism(0, 0)(1, 0)[\kl^n \tensor Q`\kl^n \tensor Q`\sst
 (1-jq)\tensor Q]{1000}1a
 \putmorphism(1000,0)(1, 0)[\phantom{\kl^n \tensor Q}`X \tensor
Q.`\sst ]{600}1a
 \putmorphism(0,400)(0, -1)[``\sst \kl^n \tensor f]{400}1r
 \putmorphism(1000, 400)(0, -1)[``\sst \kl^n \tensor f]{400}1r
 \putmorphism(1600, 400)(0, -1)[``\sst X \tensor f]{400}1r
 \efig$$
 In particular one sees that $X \tensor \X : \A \=> \A$ is a
 $\kl$-additive functor.

 Now let $f: X \=> X'$ be a homomorphism. Choose $j: X \=> \kl^n$, $q:
\kl^n \=> X$, $j': X' \=> \kl^m$, and $q': \kl^m \=> X'$ with $qj =
1_X$ and $q'j' = 1_{X'}$. Then by the cokernel property there is a
unique morphism $f \tensor P$ which makes the following diagram
commutative:
 $$\bfig
 \putmorphism(0, 400)(1, 0)[\kl^n \tensor P`\kl^n \tensor P`\sst
 (1-jq) \tensor P]{1000}1a
 \putmorphism(1000, 400)(1, 0)[\phantom{\kl^n \tensor P}`X \tensor
P`\sst ]{600}1a
 \putmorphism(0, 0)(1, 0)[\kl^m \tensor P`\kl^m \tensor P`\sst
 (1-j'q') \tensor P]{1000}1a
 \putmorphism(1000, 0)(1, 0)[\phantom{\kl^m \tensor P}`X' \tensor
P.`\sst ]{600}1a
 \putmorphism(0, 400)(0, -1)[``\sst j'fq \tensor P]{400}1r
 \putmorphism(1000, 400)(0, -1)[``\sst j'fq \tensor P]{400}1r
 \putmorphism(1600, 400)(0, -1)[``\sst f \tensor P]{400}1r
 \efig$$
 By the universal property of the cokernel we get, that $f \tensor P:
X \tensor P \=> X' \tensor P$ is a natural transformation. Furthermore
we get $fg \tensor \x = (f \tensor \x )(g \tensor \x )$ and $\id
\tensor \x = \id$. In particular we have that $\tensor : \C \times \A
\=> \A$ is a $\kl$-bilinear bifunctor.

 We sketch the construction of the associativity morphism $\beta: (X
\tensor Y) \tensor P \=> X \tensor (Y \tensor P)$. We first observe
that $\kl^n \tensor (\kl^m \tensor P) \iso P^{nm} \iso (\kl^n \tensor
\kl^m) \tensor P$, which defines $\beta$ in the free case. Now
consider representations $q: \kl^n \=> X$, $j: X \=> \kl^n$ and $q':
\kl^m \=> Y$, $j': Y \=> \kl^m$. Then
 $$f : = jq \tensor 1 \tensor 1: \kl^n \tensor \kl^m \tensor P \=> X
\tensor (\kl^n \tensor P) \=> \kl^n \tensor \kl^m \tensor P$$
 $$g : = 1 \tensor j'q' \tensor 1: \kl^n \tensor \kl^m \tensor P \=>
\kl^n \tensor (Y \tensor P) \=> \kl^n \tensor \kl^m \tensor P$$
 are idempotents with $fg = (jq \tensor 1 \tensor 1)(1 \tensor j'q'
\tensor 1) = jq \tensor j'q' \tensor 1 = gf$. So we can apply Lemma
\ref{splitting}. Since $(1 \tensor q' \tensor 1)(q \tensor 1 \tensor
1): \kl^n \tensor \kl^m \tensor P \=> X \tensor (\kl^m \tensor P) \=>
X \tensor (Y \tensor P)$ and $(j \tensor 1 \tensor 1)(1 \tensor j'
\tensor 1): X \tensor (Y \tensor P) \=> X \tensor (\kl^m \tensor P)
\=> \kl^n \tensor \kl^m \tensor P$ is a splitting of $(j \tensor 1
\tensor 1)g(q \tensor 1 \tensor 1)$, we get a splitting $\kl^n \tensor
\kl^m \tensor P \=> X \tensor (Y \tensor P) \=> \kl^n \tensor \kl^m
\tensor P$ which defines up to unique isomorphism the object $(X
\tensor Y) \tensor P$, so $\beta: (X \tensor Y) \tensor P \iso X
\tensor (Y \tensor P)$. By the uniqueness this isomorphism is a
natural transformation and coherent.

 Hence $\A$ is a $\C$-category.
 \end{pf}

 \begin{prop}
 Let $\A$ and $\B$ be $\kl$-additive categories with splitting
idempotents with the $\C$-structure derived in Theorem \ref{addto
Ccat} and let $\omega: \B \=> \A$ be a $\kl$-additive functor. Then
$\omega$ is a $\C$-functor.

 If $\omega$ and $\omega'$ are $\kl$-additive functors from $\B$ to
$\A$ and $\varphi: \omega \=> \omega'$ is a natural transformation
then $\varphi$ is a $\C$-morphism.
 \end{prop}

 \begin{pf}
 An object in $\C$ is given as above by the splitting $j: X \=> \kl^n$
and $q: \kl^n \=> X$. Then the tensor product with $X$ is given as the
splitting $j \tensor 1: X \tensor P \=> \kl^n \tensor P$ with $q
\tensor 1: \kl^n \tensor P \=> X \tensor P$. We define $\xi: \omega(X
\tensor P) \iso X \tensor \omega(P)$ as the uniquely defined morphism
by the isomorphic splittings
 $$\bfig
 \putmorphism(0, 400)(1, 0)[\omega(X \tensor P)`X \tensor
\omega(P)`\sst \xi]{800}1a
 \putmorphism(0, 0)(1, 0)[\omega(\kl^n \tensor P)`\kl^n \tensor
\omega(P)`\sst \iso]{800}1a
 \putmorphism(0, 400)(0, -1)[``\sst ]{400}1r
 \putmorphism(40, 400)(0, -1)[``\sst ]{400}{-1}r
 \putmorphism(760, 400)(0, -1)[``\sst ]{400}1r
 \putmorphism(800, 400)(0, -1)[``\sst ]{400}{-1}r
 \efig$$
 We leave it to the reader to check naturality and coherence of $\xi$.

 The following commutative diagram shows that $\varphi$ is a
 $\C$-morphism
 $$\bfig
 \putmorphism(600, 600)(1, 0)[\omega(X \tensor P)`X \tensor
\omega(P)`\sst \xi]{800}1a
 \putmorphism(600, 200)(1, 0)[\omega(\kl^n \tensor P)`\kl^n \tensor
\omega(P)`\sst \iso]{800}1a
 \putmorphism(1000, 400)(1, 0)[\omega'(X \tensor P)`X \tensor
\omega'(P)`\sst ]{800}1a
 \putmorphism(1000, 0)(1, 0)[\omega'(\kl^n \tensor P)`\kl^n \tensor
\omega'(P)`\sst ]{800}1a
 \putmorphism(0, 200)(1, 0)[\omega(P^n) =``\sst ]{600}0a
 \putmorphism(400, 0)(1, 0)[\omega'(P^n) =``\sst ]{600}0a
 \putmorphism(1400, 200)(1, 0)[`= \omega(P)^n`\sst ]{700}0a
 \putmorphism(1800, 0)(1, 0)[`= \omega'(P)^n.`\sst ]{700}0a
 \putmorphism(600, 600)(0, -1)[``\sst ]{400}1r
 \putmorphism(640, 600)(0, -1)[``\sst ]{400}{-1}r
 \putmorphism(1040, 400)(0, -1)[``\sst ]{400}1r
 \putmorphism(1080, 400)(0, -1)[``\sst ]{400}{-1}r
 \putmorphism(1400, 600)(0, -1)[``\sst ]{400}1r
 \putmorphism(1440, 600)(0, -1)[``\sst ]{400}{-1}r
 \putmorphism(1800, 400)(0, -1)[``\sst ]{400}1r
 \putmorphism(1840, 400)(0, -1)[``\sst ]{400}{-1}r
 \putmorphism(600, 600)(2, -1)[``\sst \varphi(X \tensor P)]{400}1r
 \putmorphism(1400, 600)(2, -1)[``\sst X \tensor \varphi(P)]{400}1r
 \putmorphism(0, 200)(2, -1)[``\sst \varphi(P^n)]{400}1l
 \putmorphism(600, 200)(2, -1)[``\sst ]{400}1r
 \putmorphism(1400, 200)(2, -1)[``\sst ]{400}1r
 \putmorphism(2100, 200)(2, -1)[``\sst \varphi(P)^n]{400}1r
 \efig$$
 \end{pf}

 \begin{thm}\label{nattransisC}
 Let $\C$ be a full monoidal subcategory of $\M = \kl\Mod$ and $\A$ be
a $\C$-category. Let $\omega, \omega': \A \=> \C$ be $\C$-functors.
Then every natural transformation $\varphi: \omega \=> \omega'$ is a
$\C$-morphism.
 \end{thm}

 \begin{pf}
 Since
 $$\bfig
 \putmorphism(0, 800)(1, 0)[\omega(\kl \tensor P)`\omega'(\kl \tensor
P)`\sst \varphi(\kl \tensor P)]{900}1a
 \putmorphism(0, 400)(1, 0)[\omega(P)`\omega'(P)`\sst
\varphi(P)]{900}1a
 \putmorphism(0, 0)(1, 0)[\kl \tensor \omega(P)`\kl \tensor
\omega'(P)`\sst 1 \tensor \varphi(P)]{900}1a
 \putmorphism(0, 800)(0, -1)[``\sst \omega(\pi_P)]{400}1r
 \putmorphism(0, 400)(0, -1)[``\sst \lambda_P^{-1} ]{400}1r
 \putmorphism(900, 800)(0, -1)[``\sst \omega'(\pi_P)]{400}1r
 \putmorphism(900, 400)(0, -1)[``\sst \lambda_P^{-1}]{400}1r
 \efig$$
 commutes and the vertical morphisms are by coherence the canonical
morphisms $\xi: \omega(\kl \tensor P) \=> \kl \tensor \omega(P)$ resp.
$\xi': \omega'(\kl \tensor P) \=> \kl \tensor \omega'(P)$. For $x \in
X \in \C$ let $f_x:\kl \=> X$ be the homomorphism with $f_x(1) = x$.
Then the following diagram commutes, possibly with the exception of
the front face
 $$\bfig
 \putmorphism(0, 1200)(1, 0)[\omega(\kl \tensor P)`\omega'(\kl \tensor
P)`\sst \varphi(\kl \tensor P)]{1100}1a
 \putmorphism(0, 300)(1, 0)[\kl \tensor \omega(P)`\kl \tensor
\omega'(P)`\sst 1_\kl \tensor \varphi(P)]{1100}1b
 \putmorphism(300, 900)(1, 0)[\omega(X \tensor P)`\omega'(X \tensor
P)`\sst \varphi(X \tensor P)]{1100}1a
 \putmorphism(300, 0)(1, 0)[X \tensor \omega(P)`X \tensor
\omega'(P).`\sst 1_X \tensor \varphi(P)]{1100}1b
 \putmorphism(0,1200)(0, -1)[``\sst \xi ]{900}1l
 \putmorphism(300,900)(0, -1)[``\sst \xi ]{900}1r
 \putmorphism(1100, 1200)(0, -1)[``\sst \xi']{900}1l
 \putmorphism(1400, 900)(0, -1)[``\sst \xi']{900}1r
 \putmorphism(0, 1200)(1, -1)[``\sst \omega(f_x \tensor 1)]{300}1r
 \putmorphism(0, 300)(1, -1)[``\sst f_x \tensor 1]{300}1l
 \putmorphism(1100, 1200)(1, -1)[``\sst \omega'(f_x \tensor 1)]{300}1r
 \putmorphism(1100, 300)(1, -1)[``\sst f_x \tensor 1]{300}1l
 \efig$$
 So for $q \in \omega(P)$ we get
 $\varphi(X \tensor P)\xi^{-1}(x \tensor q)
 = \varphi(X \tensor P)\xi^{-1}(f_x \tensor 1)(1 \tensor q)
 = {\xi'}^{-1} (1_X \tensor \varphi(P))(f_x \tensor 1)(1 \tensor q)
 = {\xi'}^{-1} (1_X \tensor \varphi(P))(x \tensor q)$.
 This holds for all $x \in X$ and all $q \in \omega(P)$ so that
$\varphi(X \tensor P)\xi^{-1} = ({\xi'}^{-1} \tensor 1) (1_X \tensor
\varphi(P))$ and $\varphi$ is a $\C$-morphism.
 \end{pf}

\end{document}